\documentclass[unsortedaddress,twocolumn,eqsecnum,aps,floatfix]{revtex4}
\usepackage[dvips]{graphicx}
\usepackage{color}

\begin{document}

\title{Effect of Varying the TD-lc-DFTB Range-Separation Parameter on
  Charge and Energy Transfer in a Model Pentacene/Buckminsterfullerene Heterojunction}
\author{Ala Aldin M. H. M. Darghouth}
\email[e-mail: ]{aladarghouth@uomosul.edu.iq}
\affiliation{
Department of Chemistry, College of Science, University of Mosul, Iraq
}
\author{Mark E. Casida,} 
\email[e-mail: ]{mark.casida@univ-grenoble-alpes.fr}
\affiliation{
Laboratoire de Spectrom\'etrie, Interactions et Chimie Th\'eorique (SITh), D\'epartement de Chimie Mol\'eculaire (DCM),
Institut de Chimie Mol\'eculaire de Grenoble (ICMG), Universit\'e Grenoble-Alpes,
301 rue de la Chimie, CS 40700, 38058 Grenoble Cedex 9, France}
\author{Xi Zhu 
\includegraphics[scale=0.80]{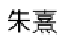})}
\email[Present address: {T}he {C}hinese {U}niversity of {H}ong {K}ong, {S}henzhen,
                {N}o.~2001 {L}ongxiang {B}lvd., {L}onggang {D}ist., {S}henzhen,
                {G}uangdong, {C}hina, 518172]{}
\author{Bhaarathi Natarajan}
\author{Haibin Su 
\includegraphics[scale=0.60]{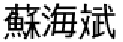}),}
\email[Present address: {D}epartment of {C}hemistry, {T}he {H}ong {K}ong
                {U}niversity of {S}cience and {T}echnology, {H}ong {K}ong, {C}hina; e-mail: ]{hbsu@ntu.edu.sg}
\affiliation{Institute of Advanced Studies, Nanyang Technological University, 60 Nanyang View,
639673 Singapore}
\author{Alexander Humeniuk}
\author{Evgenii Titov} 
\email[e-mail: ]{evgenii.v.titov@gmail.com}
\author{Xincheng Miao 
\includegraphics[scale=0.40]{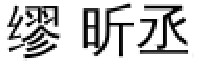})}
\email[e-mail: ]{xincheng.miao@uni-wuerzburg.de}
\author{Roland Mitri\'c}
\email[e-mail: ]{roland.mitric@uni-wuerzburg.de}
\affiliation{
  Institut f\"ur Physikalische und Theoretische Chemie,
 Julius-Maximilians-Universit\"at W\"urzburg,
 Emil-Fischer-Stra{\ss}e 42,
   D-97074 W\"urzburg, Germany}
 -------------------------------------------------------------------------
\begin{abstract}
Atomistic modeling of energy and charge transfer at the heterojunction of organic solar
cells is an active field with many remaining outstanding questions owing, in part, to the 
difficulties of performing reliable photodynamics calculations on very large systems.
One approach to being able to overcome these difficulties is to design and apply an
appropriate simplified method.  Density-functional tight binding (DFTB) has become
a popular form of approximate density-functional theory (DFT) based upon a minimal
valence basis set and neglect of all but two center integrals.  We report the results
of our tests of a recent long-range correction (lc) [{\em J. Chem. Phys.} {\bf 143}, 134120 (2015)]
for time-dependent (TD) lc-DFTB by carrying out TD-lc-DFTB fewest switches surface hopping (FSSH) 
calculations of energy and charge transfer times using the relatively new {\sc DFTBaby}
[{\em Comp. Phys. Comm.} {\bf 221}, 174 (2017)] program.  An advantage of this method is
the ability to run enough trajectories to get meaningful ensemble averages.  Our 
interest in the present work is less in determining exact energy and charge transfer rates than in understanding 
how the results of these calculations vary with the value of the range-separation parameter 
($R_{lc}=1/\mu$) for a model organic solar cell heterojunction consisting of a van der 
Waals complex {\bf P}/{\bf F} made up of single pentacene ({\bf P}) molecule together with a single 
buckminsterfullerene ({\bf F}) molecule.  The default value of $R_{lc} = 3.03 \, a_0$ is 
found to be much too small as neither energy nor charge transfer is observed until 
$R_{lc} \approx 10 \, a_0$.  Tests at a single geometry show that best agreement with 
high-quality {\em ab-initio} spectra is obtained in the limit of no lc (i.e., very large $R_{lc}$.)  
A plot of energy 
and charge transfer rates as a function of $R_{lc}$ is provided which suggests that a value of 
$R_{lc} \approx 15 \, a_0$ yields the typical literature charge transfer time of about 
100 fs.  However energy and charge transfer times become as high as $\sim$300 fs for 
$R_{lc} \approx 25 \, a_0$.  A closer examination of the charge transfer process 
{\bf P}$^*$/{\bf F} $\rightarrow$ {\bf P}$^+$/{\bf F}$^-$ shows that the initial electron
transfer is accompanied by a partial delocalization of the {\bf P} hole onto {\bf F} 
which then relocalizes back onto {\bf P}, consistent with a polaron-like picture in which the
nuclei relax to stabilize the resultant redistribution of charges.
\end{abstract}
\maketitle
\section{Introduction}
\label{sec:intro}

\begin{figure}
\includegraphics[width=\columnwidth]{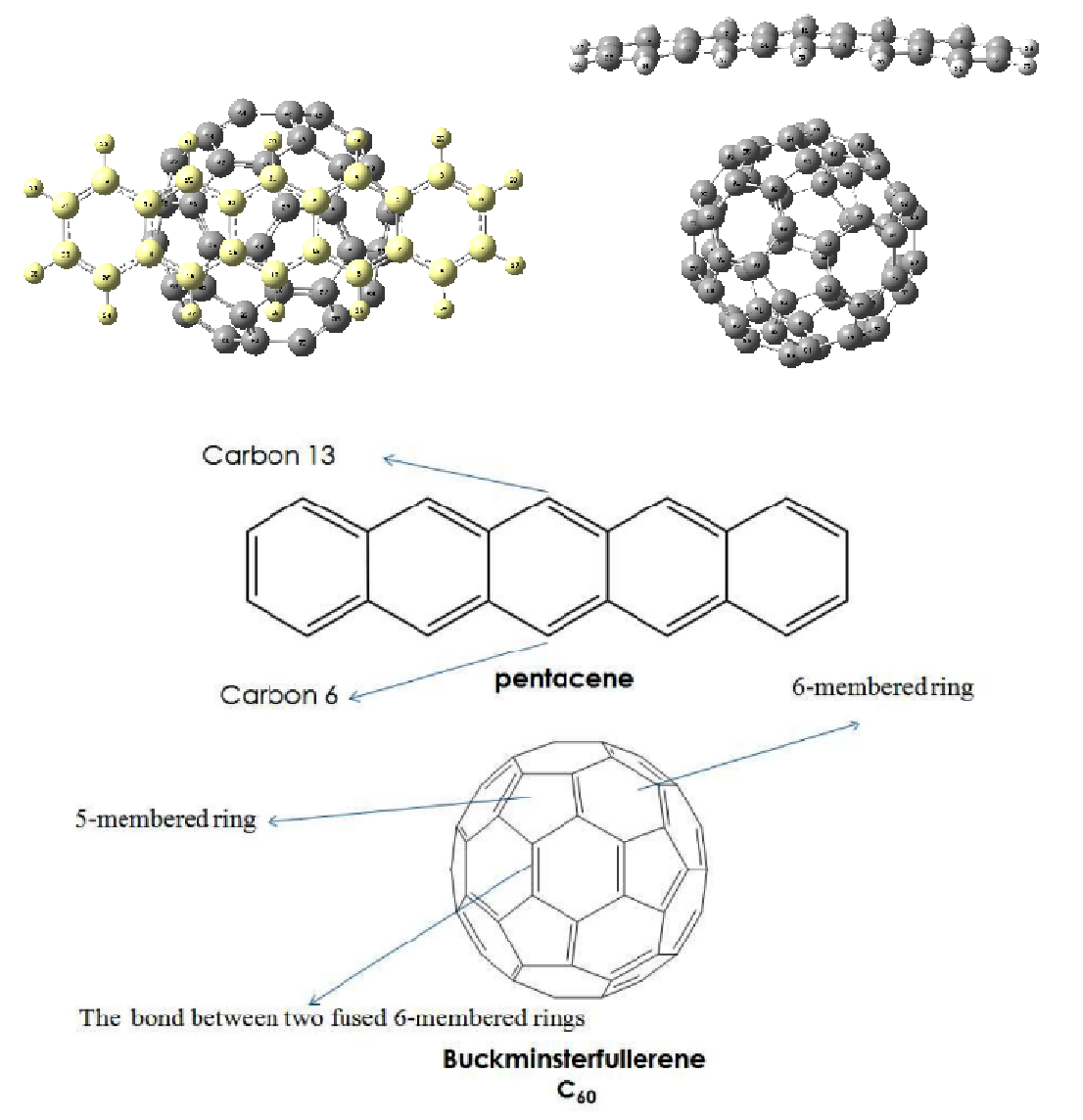}
\caption{\label{fig:2a}
The model system used in this work to study the behavior of TD-lc-DFTB
calculations as $R_{lc}$ is varied.  Some very rough dimensions have been
given for comparison with $R_{lc}$.  Bottom row: Buckminsterfullerene ({\bf F}),
roughly 7 {\AA} (13.2 $a_0$) in diameter.  Middle row: Pentacene ({\bf P}), 
roughly about 14 {\AA} (26.4 $a_0$) long.  Top row: two views of the same {\bf P}/{\bf F}
van der Waals complex, the distance of closest approach between the two molecules
is roughly 3.0 {\AA} (5.7 $a_0$.) Note that all of these dimensions are subject to
change during geometry optimizations and during dynamics calculations.}
\end{figure}
\begin{figure}
\includegraphics[width=\columnwidth]{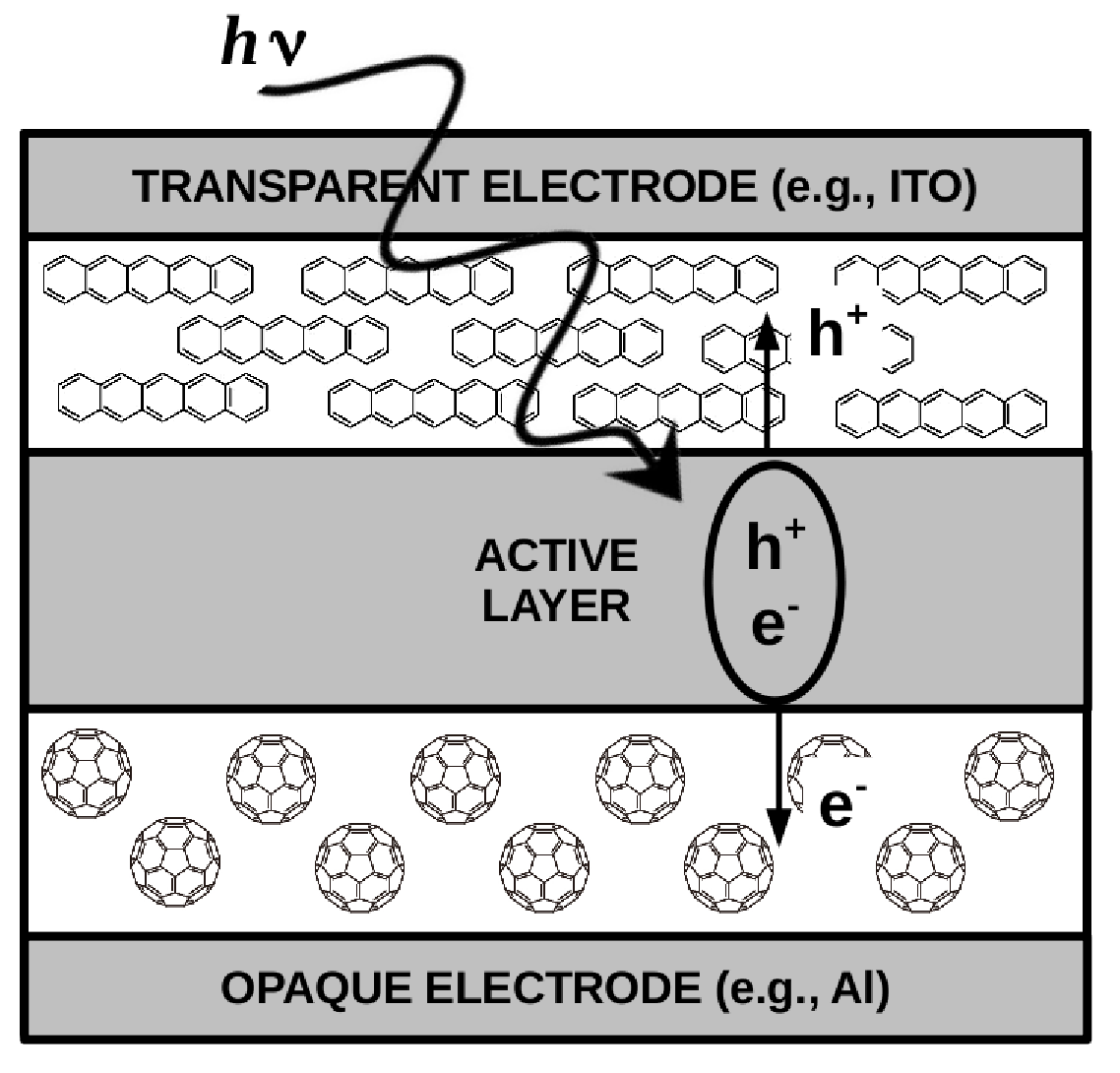}
\caption{\label{fig:SolarCell}
Cartoon of an idealized Tang-type \cite{T86} {\bf P}/{\bf F} solar cell.  ITO is a common
abbreviation for indium titanium oxide, a transparent conducting oxide. (A realistic
working {\bf P}/{\bf F} is described in Ref.~\cite{YDK04}.)
}
\end{figure}

Charge transfer (CT) and energy transfer (ET) are major topics in
physical chemistry/chemical physics \cite{MK00} that enter into such diverse problems as, for 
example,
photobiology \cite{AGV00} and organic electronics \cite{PS99}.  Time and distance scales for CT
and ET may differ over many orders of magnitude for different systems, necessitating the 
development and use of different theoretical models.  Here we are interested in a method 
able to treat phenomena occurring on time scales of the order of 100 fs, such as CT 
occurring at the heterojunction of an organic solar cell.  In general, proper 
atomistic modeling of such a system involves nonadiabatic photodynamics where electronic states
are delocalized over several molecules.  The number of atoms involved and the need to 
calculate ensemble-averaged properties means that efficiency is important.  At the same time, 
the methods used must be realistic enough to give at least a qualitatively-correct
description of the phenomena involved.  
In the present article, we consider the use of time-dependent (TD) 
\cite{NSD+01,FSE+02,HNWF07,N09,DAF+13}
long-range-corrected (lc) \cite{HM15,ND12,LAN15,VKK+18}
density-functional tight binding (DFTB) \cite{PFK+95,KM09,ES14,EPJ+98} 
fewest switches surface hopping (FSSH) photodynamics \cite{HT94,T90}
for atomistic calculations of CT and ET times in model organic solar cell heterojunctions.
The particular version of TD-lc-DFTB used in this paper is that implemented in the relatively
recent program {\sc DFTBaby} \cite{HM15,HM17,DFTBaby}.
(Recent applications of {\sc DFTBaby} include \cite{HSH+17,DCJ+18,THM18,TPL+19}.)
Unlike conventional TD-DFTB, TD-lc-DFTB contains a range-separation parameter $R_{lc} = 1/\mu$ 
which has a large effect on CT energetics and so can also be expected to have a large effect on 
CT times. It also turns out to have a large effect on ET times.  It is thus important to understand
how these times vary with $R_{lc}$ in order to be able to make rational choices for particular
problems.  Here we investigate how the choice of $R_{lc}$ affects CT and ET in a model 
{\bf P}/{\bf F} junction ({\bf P} = pentacene, {\bf F} = buckminsterfullerene, see 
{\bf Fig.~\ref{fig:2a}}.)
It is to be emphasized that our interest at this time is not in obtaining quantitatively 
correct estimates of ET and CT times but rather is on understanding which values of 
$R_{lc}$ lead to qualitatively 
correct behavior.  Indeed our model is 
most likely too small to be the basis for a quantitatively correct comparison with ET and CT
times at the heterojunction in a real {\bf P}/{\bf F} solar cell 
({\bf Fig.~\ref{fig:SolarCell}}).  
But the calculations that we are reporting are useful before applications to larger, 
more realistic models are carried out, and may already provide some new insight into this well-studied system.

\begin{table}
\squeezetable
\begin{tabular}{ll}
\hline \hline
\multicolumn{2}{c}{Six-Step Model} \\
\hline
(i)   & Exciton formation via photon absorption \\
(ii)  & Excition diffusion to the heterojunction\footnotemark[1] \\
(iii) & Exciton dissociation into closely-bound \\
      & charge-transfer (CT) states at the heterojunction\footnotemark[2] \\
(iv)  & Dissociation of these CT states into charge-separated (CS) \\
      & states composed of free mobile charges\footnotemark[3] \\
(v)   & Charge transport away from the heterojunction\footnotemark[4] \\
(vi)  & Charge collection at the electrodes \\
\hline \hline
\end{tabular}
\footnotetext[1]{Diffusion length $\leq$ about 10 nm \cite{T86}.}
\footnotetext[2]{CT time $\sim$100 fs (Table~\ref{tab:incohCTtimes}.)}
\footnotetext[3]{Over a time scale on the order a picosecond and over a distance scale on the
      order of a nanometer \cite{VDA+13}.}
\footnotetext[4]{This is a mesoscopic to macroscopic step with time scales from
      nanoseconds to milliseconds and distance scales on the order of
      millimeters.}
\caption{
\label{tab:SixStepModel}
The generally accepted model for organic heterojunction solar cells \cite{LZY12}.
}
\end{table}

Indeed, although
{\bf P}/{\bf F} solar cells may be the most heavily studied and best characterized organic solar cells, 
both at the experimental and at the theoretical level, the fact that this solar cell continues
to be studied is also a good indication that plenty of things remain to be understood.
{\bf Table~\ref{tab:SixStepModel}} shows the well-accepted six-step model for the physics of
organic solar cells.
To this, we could add that singlet fission also occurs in {\bf P}/{\bf F} solar cells \cite{AP14}, 
though we shall not discuss singlet fission any further in the present work.  Our study concerns step (iii) of the
six-step model.  {\bf Tables~\ref{tab:cohCTtimes}} and {\bf \ref{tab:incohCTtimes}} provide experimental and theoretical times for this
CT step obtained via various methods for a few organic solar cells, including but not limited
to {\bf P}/{\bf F}.  The first thing to notice is the presence of two different
CT times.  Coherent CT (Table~\ref{tab:cohCTtimes}) refers to rapid charge transfer back and forth between {\bf P} and {\bf F}
and may go by additional names, such as Rabi or St\"uckelberg oscillations depending upon how
the oscillations are explained.  Incoherent CT (Table~\ref{tab:incohCTtimes}) refers to a longer time scale non-oscillatory
charge transfer.  Caution is in order when studying the literature as sometimes 
the distinction between coherent and incoherent CT does not seem to be very clear.
Here we are primarily only interested in incoherent CT.  
The table shows that literature numbers for this quantity vary from as low as about 40 fs 
to as high as about 700 fs with the most common estimates being around 100 fs.  We will also find
an incoherent CT time of this order of magnitude from our TD-lc-DFTB calculations with
an appropriate choice of $R_{lc}$.

\begin{table}
\begin{tabular}{ccc}
\hline \hline
Donor/Acceptor & CT time & Reference \\ 
\hline 
\multicolumn{3}{c}{Experiment} \\
P3HT/PCBM  & 25 fs\footnotemark[1] & \cite{FRB+14} \\ 
p-DTS(FBTTh$_2$)$_2$/PC$_{71}$BM & $<$40 fs\footnotemark[2] 
           & \cite{GRK+14} \\ 
P3HT/PCBM   &  20 fs\footnotemark[3] & \cite{SCP+14} \\ 
\multicolumn{3}{c}{Theory} \\
4T/{\bf F}  & 25 fs\footnotemark[4] & \cite{FRB+14} \\ 
{\bf P}/{\bf F}  & 25 fs\footnotemark[5] & \cite{JRB17} \\ 
{\bf P}/{\bf F}  &$\sim$15 fs\footnotemark[6] & PW \\
\hline \hline
\end{tabular}
\footnotetext[1]{High time-resolution pump-probe spectroscopy 
and time-dependent density functional theory.}
\footnotetext[2]{Transient absorption spectroscopy.}
\footnotetext[3]{Two-dimensional electronic spectroscopy.}
\footnotetext[4]{Ehrenfest TDLDA calculations.}
\footnotetext[5]{TD-DFT FSSH.}
\footnotetext[6]{CIS/AM1.}
\caption{
\label{tab:cohCTtimes}
This is a sampling of some organic heterojunction coherent CT times collected from the literature.
Except for {\bf P}/{\bf F}, the reader is referred to cited references for the definitions 
of the various acronyms used to denote the donor and acceptors.  Our own results have also
been included in the table with the reference ``PW'' (present work.)  
}
\end{table}  

\begin{table}
\begin{tabular}{ccc}
\hline \hline
Donor/Acceptor & CT time & Reference \\ 
\hline 
\multicolumn{3}{c}{Experiment} \\
P3HT/PCBM & $<$100 fs\footnotemark[1] & \cite{HMML10}\\ 
APFO3/PCBM & 200 fs\footnotemark[2] & \cite{DPM+07} \\   
MDMO-PPV/PC$_{70}$BM & $\leq$ 100 fs\footnotemark[3] & \cite{BRP+12} \\ 
PCPDTBT/PC$_{70}$BM  & $\leq$ 100 fs\footnotemark[3] & \cite{BRP+12} \\ 
p-DTS(FBTTh$_2$)$_2$/PC$_{71}$BM & 82 fs &  \cite{GRK+14} \\
{\bf P}/{\bf F} & 110 fs\footnotemark[4] & \cite{CLJ+11} \\ 
\multicolumn{3}{c}{Theory} \\
4T/{\bf F}  & 97 fs\footnotemark[5] & \cite{FRB+14} \\ 
{\bf P}/{\bf F}  & 100 fs\footnotemark[6] & \cite{YCB09} \\ 
{\bf P}/{\bf F}  & 714 fs\footnotemark[7] & \cite{LKCH14} \\ 
{\bf P}/{\bf F}  & 40 fs\footnotemark[8] & \cite{AP14} \\ 
{\bf P}/{\bf F}  & 164 fs\footnotemark[9]  & PW \\ 
{\bf P}/{\bf F}  & $\sim$300 fs\footnotemark[10]  & PW \\ 
\hline \hline
\end{tabular}
\footnotetext[1]{Transient absorption spectroscopy and 
quasi-steady-state photoinduced absorption spectroscopy.}
\footnotetext[2]{Transient absorption spectroscopy.}
\footnotetext[3]{Infrared pump-probe spectroscopy.}
\footnotetext[4]{Femtosecond time-resolved two-photon photoemission spectroscopy.}

\footnotetext[5]{Ehrenfest TDLDA calculations.}
\footnotetext[6]{Figure~7 of the cited reference, Marcus theory, 
dimer parallel bond-alignment geometry {\bf 2A}, 
$^1B_{1u}^P \otimes ^1A_{g}^{C60}$, $d_i= \mbox{ 3.5 {\AA} }$, 
fastest decay time for geometries with
$d_{int} \leq \mbox{ 2.0 {\AA}}$.}
\footnotetext[7]{From Table I of the cited reference, Marcus theory, 
bulk singlet CT, quoted result is for the aligned structure 
without interfacial effects.}
\footnotetext[8]{TD-DFT FSSH (simplified scheme.)}
\footnotetext[9]{TD-lc-DFTB FSSH $R_{lc}=15 \, a_0$.}
\footnotetext[10]{TD-lc-DFTB FSSH $R_{lc}=25 \, a_0$.}
\caption{
\label{tab:incohCTtimes}
This is a sampling of some organic heterojunction incoherent CT times collected from the literature.
Except for {\bf P}/{\bf F}, the reader is referred to cited references for the definitions 
of the various acronyms used to denote the donor and acceptors.  Our own results have also
been included in the table with the reference ``PW'' (present work.)  
}
\end{table}  

\begin{figure}
\includegraphics[width=0.5\columnwidth]{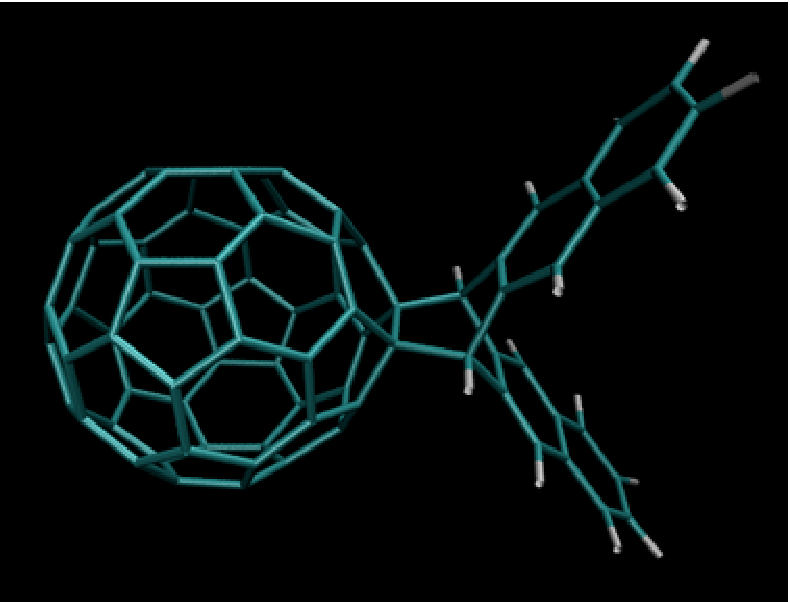}
\caption{Electrocyclic addition product.
\label{fig:electrocyclicproduct}
}
\end{figure}
Of the many theoretical studies of the {\bf P}/{\bf F} system reported in the literature,
a few stand out as overlapping in one way or another with the present work.  The group of
Br\'edas have a series of papers where they have established the van der Waals complex
of a single {\bf P} with a single {\bf F} molecule as one of their model systems of choice 
\cite{YCB09,YYZ+14,ZSY+14,JRB17}.  (See also Ref.~\cite{CRL+16}.) They considered many 
orientations of one molecule relative to the other molecule.  The orientation used
in the present work was somewhat arbitrarily 
chosen because of its relation to the known thermal electrocyclic reaction of the two 
molecules whose product is shown in {\bf Fig.~\ref{fig:electrocyclicproduct}} \cite{MM97,MKFK99}.
[Though no indication of such a reaction was found in our TD-lc-DFTB photochemical 
simulations, we did see it in a DFTB thermal simulation. See the Supplementary Information (SI).]
It is essentially the same as geometry {\bf 2A} in the Marcus theory study of CT and charge
recombination times in Ref.~\cite{YCB09}.  In Ref.~\cite{ZSY+14}, they tuned the $\omega$B97X-D
functional \cite{CH08} for describing charge transfer in this system.  (Note that
the range-separation parameter is called $\omega$ in this functional rather than $\mu$.
In fact, both notations are used in the literature.) Inverting the resultant value 
$\omega=0.137 \, a_0^{-1}$ gives $R_{lc}=7.30 \, a_0$, which is about equal to the radius
of {\bf F}, is about a quarter the length of {\bf P}, and is a little longer than
the closest distance between {\bf P} and {\bf F} (Fig.~\ref{fig:2a}.)  
Three years later they 
carried out
TD-$\omega$B97X-D FSSH calculations to gain insight into coherent CT times \cite{JRB17}.  
Note that only seven trajectories were included in these calculations, probably because 
these calculations rapidly become very time and resource intensive. 

Another notable study comes from the group of Prezhdo \cite{AP14}.  This is an approximate
periodic TD-DFT FSSH calculation of a {\bf P}/{\bf F} interface in which the repeated unit cell
consisted of two {\bf P} and one {\bf F}.  This is not a true TD-DFT FSSH calculation, but rather involves
some simplifications:  Precomputed ground state trajectories are used instead of
true FSSH trajectories, excluding applications where photochemical reactions occur which
are different from those in the ground state.  Also, the nonadiabatic couplings are calculated 
assuming excited-state wave functions of single determinantal form \cite{CDP05,AP13} rather 
than using the correct multideterminantal form \cite{M06}.  And excitation energies are 
approximated as orbital energy differences \cite{AP13}.  These approximations are one way to 
increase the efficiency of an approximate TD-DFT FSSH calculation.  
The TD-lc-DFTB FSSH approach explored in this paper represents a different way to increase
the efficiency of approximate TD-DFT FSSH calculations and one which we believe is more faithful
to the spirit of TD-DFT FSSH calculations.

The remainder of this paper is organized
as follows: The next section reviews those aspects of TD-lc-DFTB FSSH theory which are important
for this paper and explains how we define ET and CT.  Computational details are provided in 
Sec.~\ref{sec:details}. Results are presented in Sec.~\ref{sec:results}.  Section~\ref{sec:conclude}
contains our concluding discussion.

\section{Theoretical methods}
\label{sec:theory}

The purpose of this section is to describe the methodology in this paper which might
be considered either to be novel or at least nonstandard. 

\subsection{TD-lc-DFTB}

We now explain the need for and difficulty of formulating TD-lc-DFTB.
The explanation assumes a certain familiarity with DFTB and TD-DFTB, but
Appendix~\ref{sec:DFTB} provides a minimal review which should make this
section comprehensible for those new to the field.
It is now well-established that TD-DFT with conventional density-functional
approximations works best for
localized low-energy excitations without too much CT.  On the other
hand, CT excitations may be underestimated by one or two electron volts
\cite{TAH+99,DWH03}.  The best solution to date for fixing the problem of underestimated
CT excitation energies is to use TD-lc-DFT \cite{TTY+04,TTH06,VS06,PTS+07,LB07,LB08}.
In long-range corrected (lc) functionals, the electron repulsion is separated into a 
short-range (sr) part and a long range (lr) part,
\begin{eqnarray}
  \frac{1}{r_{12}} & = & \left( \frac{1}{r_{12}} \right)_{sr} + 
                      \left( \frac{1}{r_{12}} \right)_{lr} \nonumber \\
  \left( \frac{1}{r_{12}} \right)_{sr} & = & \frac{\mbox{erfc}(\mu r_{12})}{r_{12}}
            \nonumber \\
  \left( \frac{1}{r_{12}} \right)_{lr} & = & \frac{\mbox{erf}(\mu r_{12})}{r_{12}}
  \, ,
  \label{eq:theory.11}
\end{eqnarray}
where erf is the error function and erfc is the complementary error function and
$\mu=1/R_{lc}$ is the range-separation parameter.
The sr part is described with DFT using special functionals based only on the form of the
short-range part of the electron repulsion, and the lr part is described using the
appropriate long-range exact exchange. 

Work on lc-DFTB began with work by Niehaus and Della~Sala \cite{ND12}, but has become
more active in the last five years \cite{HM15,HM17,LAN15,VKK+18}.  Three years ago it was
extended to TD-lc-DFTB FSSH \cite{HM17}.

The TD-lc-DFTB scheme \cite{HM15,HM17} used in the present work is based upon the simplest 
type of range-separated functional, namely the  lc scheme of Iikura, Tsuneda, Yanai, and 
Hirao \cite{ITYH01} which was later generalized to TD-lc-DFT by Tawada, Tsuneda, Yanagisa, 
Yanai, and Hirao \cite{TTY+04}.  In this scheme, the range separation 
is applied only to exchange so that the new exchange-correlation energy is,
\begin{equation}
  E_{xc}^{lc}(\mu) = E_x^{sr,\text{GGA}}(\mu) + E_x^{lr,\text{HF}}(\mu) + E_c \, .
  \label{eq:theory.12}
\end{equation}

Now suppose we want to do this in the context of DFTB.  A minor point is that DFTB
is based upon certain separability assumptions which are rigorous for local and
semi-local density functionals, but which are lost when including Hartree-Fock
exchange.  This is compensated by long experience gained with approximating
Hartree-Fock exchange in the context of traditional semi-empirical quantum 
chemistry techniques \cite{BJ05}.
But there is also a major problem which is the need (at least in principle)
for extensive reparameterization each time a new density-functional approximation 
is considered.  {\em A priori} lc-DFTB should involve three terms,
\begin{equation}
  E(\mu) = E_{\text{BS}}(\mu) + E_{\text{rep}}(\mu) + E_{\text{coul}}(\mu) \, .
  \label{eq:theory.13}
\end{equation}
This is avoided in the method of Humeniuk and Mitri\'c (used in
the present article) where a long-range correction for exchange is added with
\begin{equation}
  \gamma_{I,J}^{lr}(R_{I,J}) = \text{erf}(\mu R_{I,J}) \gamma_{I,J}(R_{I,J}) 
  \label{eq:theory.14}
\end{equation}
and no reparameterizations.  Furthermore Humeniuk and Mitri\'{c} neglect the lr contribution
to the BS energy on the grounds that the zero-order system ``already accounts for all
interactions between electrons in the neutral atoms'' \cite{HM17}, which means that
we are actually using an expression of the form,
\begin{equation}
  E^{\text{lc-DFTB}}(\mu) = E_{\text{BS}} + E_{\text{rep}} + E_{coul}(\mu) \, .
  \label{eq:theory.15}
\end{equation}
While it often seems to be true that a good value of $\mu$ to use in this formulation of
TD-lc-DFTB is close to that found to be good to use in TD-lc-DFT, we must be aware that
this need not always be the case.  

It should be clear from the above very rapid review that the largest differences between
TD-lc-DFTB and TD-lc-DFT are in the lc part.  Developing lc-DFTB is still an active area
of research and much testing of the new methodology is needed.  The present work on ET and CT
in our {\bf P}/{\bf F} van der Waals complex investigates one particularly challenging case.

\subsection{TD-lc-DFTB FSSH}

Appendix~\ref{sec:FSSH} contains a very brief review of the TD-DFT FSSH method.  This 
allows us to emphasize what is either particularly important or what is new in the 
TD-lc-DFTB FSSH used here.  We use the version of TD-lc-DFTB FSSH implemented in the 
{\sc DFTBaby} program \cite{HM17,DFTBaby}.  This solves the full TD-lc-DFTB equation, 
but calculates the non-adiabatic coupling (NAC) elements in the same way as did Tapavicza {\em et al.} \cite{TTR07} 
using Casida's {\em Ansatz} \cite{C95}.  Three additions were added to the {\sc DFTBaby} 
code specifically for the project reported here. 

\subsubsection{Choice of Initial State}
\label{sec:initial}
In order to be as consistent as possible with the conventional model of an organic solar
cell, we do {\em not} start with an adiabatic wavefunction as is usually done in FSSH.
Instead, we imagine the formation of the lowest excited state of {\bf P} which is known to
be the singlet spin projected highest occupied molecular orbital (HOMO)
$\rightarrow$ lowest unoccupied molecular orbital (LUMO) of {\bf P} \cite{DCJ+18}.
We simulate the arrival of this {\bf P}$^*$ excited state at our model {\bf P}/{\bf F}
``heterojunction'' by projecting the molecular {\bf P}$^*$ excited state onto the
van der Waals complex to form a {\bf P}$^*$/{\bf F} diabatic state whose wavefunction
is then projected onto the basis of {\bf P}/{\bf F} adiabatic excited states to obtain
the initial wavefunction as a superposition of electronic states.  
The initial adiabatic surface is selected stochastically with the probability
equal to the modulus squared of the corresponding component in the electronic
wavevector.
Note that this was
only done for our TD-lc-DFTB FSSH calculations and not for the CIS/AM1 FSSH calculations.
In the latter case, it was judged that the initial adiabatic excited state was close enough
to a pure {\bf P}$^*$/{\bf F} excitation.  

\subsubsection{Incorporation of a Decoherence Correction}  
\label{sec:decohere}
We will be using Casida's {\em Ansatz} wavefunction to obtain a quantitative measure of ET and CT.
Before doing this, we need to first discuss the issue of how observables should be
calculated in the FSSH method.  It turns out that this is less trivial than might at
first be believed because the FSSH method is constantly switching back and forth
between a quantum description of the evolution of the electronic wave function and
the classical propagation of nuclei on adiabatic potential energy surfaces.  The classical
part of the algorithm represents a measurement of the state of the quantum system which
collapses the quantum wave packet in the sense of fixing the adiabatic electronic state.  
On the other hand, the quantum part of the calculation ignores this collapse and continues
on with the time-dependent propagation of the wave function in the field of the moving
nuclei.  This raises the question of whether observables should be calculated from the
classical populations of the different adiabatic states or quantum mechanically from the
time-dependent wave function.  Following Landry, Falk, and Subotnik \cite{LFS13},
we call the first method the ``surface method'' while the second method is called the
``electronic wavevector method.''  It was recognized early on that the surface and electronic
wavevector methods gave different expectation values and that the surface method usually
agreed better with the results of fully quantum calculations.  This may be a little surprising
as the basic assumption of Tully's FSSH is that the two methods should agree in the limit
of an ensemble average over a very large number of trajectories.  It turns out that the
primary reason for the discrepancy is that the FSSH method is overcoherent.  Put simply
there is a problem because the electronic wave function can have nonnegligeable probability
density for finding the state on different adiabatic potential energy surfaces even once
the classical trajectories have left the region of significant surface hopping probability.
The result is the appearance of artifactual St\"uckelberg-like oscillations \cite{T90},
improper scaling of electron transfer rates in Marcus theory \cite{LS11,LS12}, and disagreement
between the surface and electronic wavevector methods for calculating 
observables \cite{LFS13}.  These problems are solved by the inclusion of proper decoherence 
corrections.  Reference~\cite{BLS05} contains an excellent review of decoherence corrections in
trajectory-based photodynamics modeling.  For the present study an energy-based decoherence 
correction \cite{ZNJT04,GP07} has been implemented in {\sc DFTBaby}.  
Our method of calculating ET and CT requires access to an electronic wavefunction.  This 
wavefunction could be an adiabatic wavefunction (surface method) or the time-dependent
wavefunction (wavevector method.)  We usually use the wavevector method.  However, as we shall
see the decoherence correction makes this choice of wavefunction largely irrelevant.

\subsubsection{Particle-hole definition of charge and energy transfer}  
Defining ET and CT is actually less trivial than might at first be assumed.  We take
a very direct approach here by looking at the charges due to excited electrons (particles)
and holes on each fragment molecule.  However, it should be noted that this is not the
only way to define ET and CT.  For example, Ref.~\cite{DCJ+18} treats ET and CT in 
stacks of ethylene and pentacene molecules using a generalization of Kasha's exciton 
model \cite{KRE65}.  It is shown there that CT excitations may be identified whose energy is 
underestimated by TD-DFT and by TD-DFTB but corrected by TD-lc-DFT and TD-lc-DFTB even 
though no actual charge is transferred.  ET and CT in the sense of Kasha's exciton
model could be treated using the methodology developed by Plasser and Lischka \cite{PL12,P20}.
However, we have preferred in the present work to use the simpler, more direct, 
definition of ET and CT in terms of the transfer (or nontransfer) of particle and hole 
charges between fragments.

Appendix~\ref{sec:charges} explains how
to calculate four nonnegative numbers---namely the population $q_h^P$ of holes 
on {\bf P}, the population $q_p^P$ of excited electrons (``particles'') on {\bf P}, the 
population $q_h^F$ of holes on {\bf F}, and the population $q_p^F$ of excited electrons 
(``particles'') on {\bf F}.  
These could be gathered into a single matrix,
\begin{equation}
   {\bf q} = \left[ \begin{array}{cc} q_h^P & q_p^P \\ 
                                      q_h^F & q_p^F \end{array} \right]
   \, ,
   \label{eq:theory.20} 
\end{equation}
if so desired, though this would be more for aesthetic than for practical reasons. 
These will allow us to quantify ET and CT.
Notice that the symbol $q$ has been used which usually refers
to charge, but that these charges are all nonnegative: $q_h^P \geq 0$,
$q_p^P \geq 0$, $q_h^F \geq 0$, and $q_p^F \geq 0$.  Nevertheless, by
conservation of charge, we have that,
\begin{equation}
  q_h^F + q_h^P = q_p^F + q_p^P = 1 \, .
  \label{eq:theory.21} 
\end{equation}
It is easy to see that the amount of charge transferred from pentacene to
buckminsterfullerene ({\bf P} $\rightarrow$ {\bf F}) is,
\begin{equation}
  \mbox{CT } = q_h^P - q_p^P = q_p^F - q_h^F \, .
  \label{eq:theory.22}
\end{equation}
[Notice that the two different ways to define CT are equivalent because
of Eq.~(\ref{eq:theory.21}).] With this definition, CT = +1 for {\bf P}$^+$/{\bf F}$^-$
and CT = -1 for {\bf P}$^-$/{\bf F}$^+$.  It is also possible for a neutral excitation
to be localized either on {\bf P} or on {\bf F} or partially on both at the same time.
In order to quantify the amount of neutral excitation transferred from
pentacene to buckminsterfullerene, we define,
\begin{eqnarray}
  \mbox{ET } & = &  \left( q^F_h + q^F_p \right) -1
             = 1 - \left( q_h^P + q_p^P \right) \nonumber \\
             & = & \frac{q_h^F+ q^F_p}{2} - \frac{q_h^P + q_p^P}{2} \, .
  \label{eq:theory.23}  
\end{eqnarray}
Then ET = -1 for {\bf P}$^*$/{\bf F} and ET = +1 for {\bf P}/{\bf F}$^*$.  Note that 
CT and ET are independent in the sense that these two parameters were created from
the four components of the charge matrix ${\bf q}$ by making use of the
two dependence relations given in Eq.~(\ref{eq:theory.21}).

\section{Computational Details}
\label{sec:details}

Four different programs were used to carry out the calculations reported
in this paper:  
(A) {\sc Gaussian} versions 09 \cite{Gaussian09} and
16 \cite{Gaussian16} were used to construct start geometries and to carry out
some single point spectra calculations.  
(B) {\sc TurboMole} \cite{ABH+89} versions 6.5 \cite{TURBOMOLE13} 
and 7.0 \cite{TURBOMOLE15} were used to carry out second-order coupled 
cluster \cite{CKJ95} (CC2) and second-order algebraic diagrammatic 
construction
[ADC(2), also known as strict ADC(2) or ADC(2)-s, Eq.~(53c)
of Ref.~\cite{S82}] single point spectra calculations.
(C) {\sc DFTBaby} \cite{HM17} was used to carry out 
Tully-type TD-lc-DFTB/classical trajectory surface hopping calculations.
The energies, gradients, and nonadiabatic couplings produced by the 
(D) {\sc MNDO2005} program \cite{T05} 
were used as input for a local program  ({\sc Field\_Hopping})
to carry out AM1/CIS FSSH dynamics.

\subsection{\sc Gaussian}
\noindent
was used to generate start geometries and for some single-point spectra calculations.
Start geometries for the individual {\bf P} and {\bf F}  molecules were obtained by 
gas-phase optimization of initial crystal geometries taken from the Crystallography 
Open Database (COD) \cite{COD,SHDN07,DM94} and then optimized at the
B3LYP/6-31G(d,p) level --- that is, with the B3LYP 
functional (i.e., Becke's B3P functional \cite{B93b} with Perdew's correlation 
generalized gradient approximation (GGA) replaced with the Lee-Yang-Parr 
GGA \cite{LYP88} without further optimization \cite{B3LYP94})
\cite{B93a,B93b} using the 6-31G(d,p) basis set \cite{HDP72,HP73}.

The initial geometry for the {\bf P}/{\bf F} complex was generated as follows:
A first geometry, called {\bf A}, was obtained from the
minimum of the potential energy curve for the unrelaxed molecules as a function of 
the intermolecular distance using the orientation shown in the upper left-hand 
corner of Fig.~\ref{fig:2a}.  These curves were calculated at the CAM-B3LYP-D3/6-31G(d,p)
level --- that is, the range-separated CAM-B3LYP \cite{YTH04} was supplemented 
with Grimme's semi-empirical van der Waals correction \cite{GAEK10}.
A second geometry, called {\bf C}, was then obtained by completely relaxing 
geometry {\bf A} at the same level of theory.  We only report spectra calculated
at geometry {\bf C} in the present work.  (Spectra calculated with geometry {\bf A}
were found to be very similar to those calculated with geometry {\bf C}.)
The $(x,y,z)$-coordinates of geometry {\bf C} may 
be found in the Supplementary Information (SI.)

\subsection{\sc TurboMole} 
\label{sec:TurboMole}
\noindent
was used to carry out single point spectra calculations at
the CC2 and ADC(2) levels of theory.  All these calculations were carried out using the
cc-pVDZ basis set \cite{D89}.  The resolution-of-the-identity (RI) approximation was 
employed with the standard {\sc TurboMole} cc-pVDZ-RI auxiliary basis 
set \cite{WKH02,H05}.  
Tight convergence criteria were employed ({\tt thrdiis = 4}, {\tt thrpreopt  = 5} 
in {\tt \$excitations}).  Oscillator strengths have been calculated using the length 
gauge 
({\tt operators=diplen}).  The {\tt freeze} option was used to freeze out the core orbitals 
from the active space in our CC2 and ADC(2) calculations.
Besides CC2 and ADC(2) calculations, two variants were also tried using default parameters.  
These are spin-component scaled \cite{G03b,HGH08} (SCS) CC2 and ADC(2)({\tt cos=1.2 css=0.33333})
and scaled opposite-spin (SOS) \cite{JLDH04,HGH08} CC2 and ADC(2) ({\tt SOS: cos=1.3}).
In particular, Lischka and co-workers have found SOS-ADC(2) to be an improvement
over ADC(2) for charge-transfer complexes \cite{ABN+14}.

\subsection{\sc DFTBaby}
\noindent
was used to carry out the initial calculation of the absorption spectrum
and was used to carry out mixed TD-lc- DFTB/classical trajectory surface hopping using
the FSSH algorithm. 

DFTB uses an underlying atomic basis set whose definition requires atomic
calculations on both the free atoms and for atoms confined within a 
quadratic potential.  These calculations were carried out with the PBE
functional \cite{PBE96,PBE97}.  The atomic confinement radius used for hydrogen
was $r_0 (\mbox{H}) = 1.757 \, a_0$.  The confinement radius used for carbon
has varied between different versions of {\sc DFTBaby}: $r_0 (\mbox{C}) = 2.657 \, a_0$
was used in earlier versions, while $r_0 (\mbox{C}) = 4.309 \, a_0$ is used in the 
present work.

A full active space was used to calculate the absorption spectrum at a fixed geometry.  
The dynamics calculations are more involved. For each value of $R_{lc}$, the ground
state ($S_0$) geometry was reoptimized starting from geometry {\bf C}. A thermal ensemble
was then generated for this geometry by choosing velocities at random with a 300 K 
Maxwell-Boltzmann distribution.  The electronic wavevector was initialized by 
projecting the diabatic {\bf P}*/{\bf F} excited state onto the adiabatic basis 
and the initial adiabatic surface for the classical propagation of the nuclei 
was drawn randomly as described above in Sec.~\ref{sec:initial}.

Calculations were carried out with a nuclear time step of 0.1 fs. 
All trajectories were at least 500 fs long.
The active space used in the FSSH calculations was restricted to 40 occupied orbitals 
and 40 unoccupied orbitals.  At least ten excited states, in addition
to the ground state, have been followed during each trajectory calculation.  
(For $R_{lc} = 20 \, a_0$ and $R_{lc} = 25 \, a_0$, twenty excited states were followed.
This increased number of excited states was needed only for these values of the 
range-separation parameter as it was found that the {\bf P}$^*$/{\bf F} diabatic
state had significant amplitude for some of the higher excited states.)
Excited state gradients were calculated analytically (Appendix B of 
Ref.~\cite{HM17}.) Adiabatic energies and scalar non-adiabatic couplings were 
interpolated linearly when integrating the electronic Schr\"odinger equation between 
nuclear time steps.  As explained in Sec.~\ref{sec:decohere}, a decoherence correction
was added to {\sc DFTBaby} for the present work.  Hops from a lower
to a higher state were rejected if the kinetic energy was less than the energy gap 
between the states so as not to violate the principle of conservation of energy.  
Velocities were uniformly scaled after an allowed hop so that the total (kinetic 
plus potential) energy was conserved.  It is known that artifacts can occur when
integrating the electronic Schr\"odinger equation in an adiabatic basis near a 
photochemical funnel.  This problem was avoided using a locally diabatic 
basis~\cite{GPT01}
which avoids numerical instabilities due to ``trivial crossings.''
Unlike some previous implementations of DFTB which may not have been suitable for
some dynamics calculations because they were not parameterized for all interatomic
distances, {\sc DFTBaby} is well adapted for dynamics calculations as it is parameterized 
for every interatomic distances from zero to infinity.

In a few of our calculations, {\sc DFTBaby} was used to run ground state dynamics
at 300 K without the lc option.  [Although just a parenthesis to the present work,
we note that the electrocyclic addition product (Fig.~\ref{fig:electrocyclicproduct}) was
found after a few hundred fs in some of our DFTB+D3 thermal dynamics calculations.] 

\subsection{\sc MNDO2005}
\noindent
was used to generate energies, gradients,
and nonadiabatic couplings which were then used together with {\sc Field\_Hopping},
a local program in W\"urzburg group, to do CIS/AM1 photochemical dynamics.
{\sc Field\_Hopping} contains the decoherence correction 
[Eq. (17) of Ref.~\cite{GP07} with $C = \mbox{1 Ha}$].
AM1 is described in Ref.~\cite{DZHS85} and its CIS variant is described in 
Ref.~\cite{KBT03}.  A description of the method for calculating nonadiabatic coupling 
elements is given in Ref.~\cite{FKT08}.  Our AM1/CIS calculations used an active 
space of 12 occupied and 12 virtual orbitals.  
Initial state preparation began with ground-state molecular dynamics
at the PBE/def2-SVP level with {\sc Turbomole 6.5} interfaced with 
{\sc metaFALCON} \cite{LSRM19}.  Equilibration to 300 K used the Berendsen
thermostat and a time constant of 100 fs.  The ground-state dynamics was propagated
with a time-step of 0.5 fs for 20 ps.  The temperature was found to be 
stable after 5 ps.  After that time 100 initial configurations were selected
equally spaced in time for starting the FSSH dynamics.
Excitation was to  the lowest lying excited state in the first 19 excited states
having greater than 60\% {\bf P}$^*$/{\bf F} character.  

\section{Results}
\label{sec:results}

Although physics suggests a rough physical value for
the range-separation parameter $R_{lc} = 1/\mu$ to be used in TD-lc-DFT,
this parameter is often optimized in practice in order to fit known physical
quantities from experiment or from the results of a more exact theory.  
As the long-range correction (lc) in TD-lc-DFTB affects only some of the terms 
that the lc affects in TD-lc-DFT, there is no clear reason why the same 
range-separation parameter should be used in the two theories.  Thus a separate
investigation of the effect of varying $R_{lc}$ is needed for TD-lc-DFTB.
Our results are presented in this section. 

Note that optimization by fitting to experiment is not really available to us here 
for two reasons.  The first is that our particular physical system {\bf P}/{\bf F}, 
consisting of a van der Waals complex of pentacene ({\bf P}) and of 
buckminsterfullerene ({\bf F}) is unlikely to exist in nature, except possibly in 
outer space.  Secondly, although we do not doubt that {\bf P}/{\bf F} might be made, for 
example in a molecular beam, it would most likely involve a larger mixture 
of configurations than we consider here.  

As will be seen below we can compare against the results of high-quality {\em ab-intio}
calculations of spectra at a single geometry and that is done below to find the range
of values of $R_{lc}$ which appear to give the most physical results.  Along the way,
we also explore how some simpler theories that resemble TD-lc-DFTB do when compared
to the results of high-quality {\em ab-initio} results.

TD-lc-DFTB FSSH calculations of charge transfer (CT) and energy transfer (ET) are
carried out specifically to see how these properties vary with $R_{lc}$.  This provides
additional information which supports and extends what we learned by looking at spectra
calculated at only a single geometry.

However, before doing any of this, we find it useful to review the importance of having
good statistics and what types of FSSH results we get from another semi-empirical theory,
namely CIS/AM1.  This will form a different type of comparison against which to judge
TD-lc-DFTB FSSH calculations.

\subsection{Importance of Statistics}

It is worth emphasizing the importance of ensemble averaging and hence
of good statistics.  It has been pointed out that, ``the dynamics of 
individual FSSH trajectories are not physical, and it is unclear how many 
trajectories in the swarm must be averaged before observables can be 
calculated'' \cite{BLS05}.  While this statement seems overly pessimistic 
in the sense that individual FSSH trajectories are often thought to reflect 
different possible photochemical pathways, it is also true that the most 
rigorous interpretation of FSSH observables requires taking ensemble 
averages over many trajectories and that great care must be taken to 
obtain the necessary statistics.  

We have tried TD-DFT FSSH calculations and have found them to be quite resource 
intensive compared to semi-empirical FSSH calculations for our problem.  
However we mentioned one previous TD-DFT FSSH study of the {\bf P}/{\bf F} 
system \cite{JRB17}.  As the objective of that study was to study coherent
CT and the primary subject of the present study is incoherent CT, the procedure
followed was rather different.
Calculations in Ref.~\cite{JRB17} were started in the lowest singlet state.  
This state has dominant {\bf P}$^+$/{\bf F}$^-$ character. Only seven trajectories 
were carried out, presumably
because these are computational resource intensive to do.  The charge on {\bf F}
was studied as a function of time.  The excited electron (particle) always
began on {\bf F} but then moved back and forth between {\bf P} and {\bf F} with an apparently
regular frequency.  The results were deemed consistent with 
previous theoretical \cite{GRK+14,SCP+14} and experimental \cite{FRB+14,AP14} 
work indicating the presence of charge oscillations occurring with periods of 
some 20-25 fs.  

\begin{figure}
\includegraphics[width=1.0\linewidth]{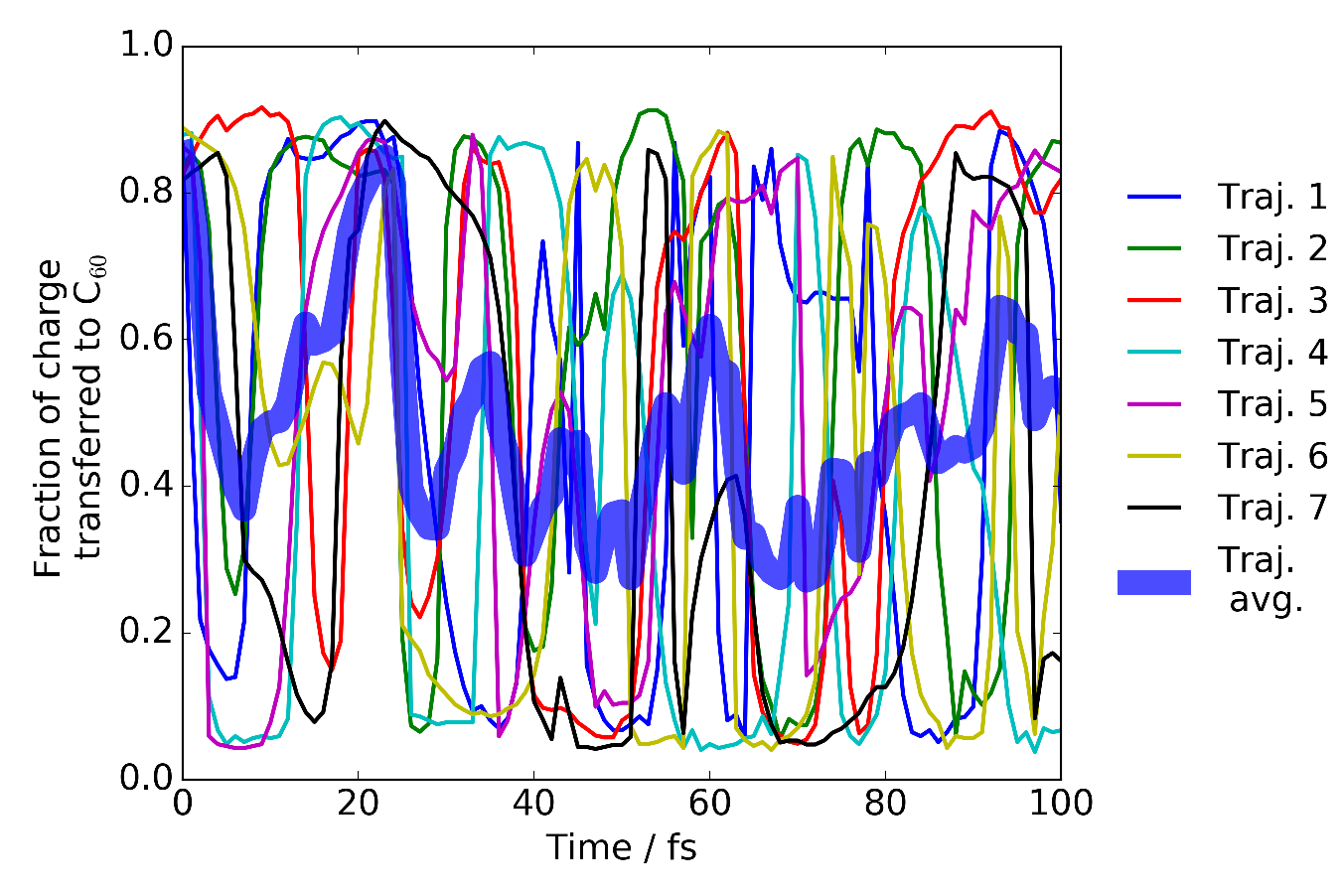} 
\caption{
Fraction of an electron on {\bf F} in the {\bf P}/{\bf F} studied in Ref.~\cite{JRB17}:
All seven trajectories and their ensemble average.
\label{fig:JRB17ave}
}
\end{figure}

Now let us take a different point of view and look at ensemble averages rather than
individual trajectories.  When studying coherent CT, ensemble averaging has the possible
disadvantage of averaging away at least some of the coherence, but we are interested
in incoherent CT whose study demands ensemble averaging.  
As the authors of Ref.~\cite{JRB17} provided graphs of the charge on {\bf F} 
for all seven trajectories, we may take their data and use it to illustrate the
effect of ensemble averaging.  Chemical intuition might have suggested that the electron 
should remain on {\bf F} but {\bf Fig.~\ref{fig:JRB17ave}} is consistent with the
above-mentioned physical picture of the charge moving back and forth
between {\bf F} and {\bf P}.  As the oscillations are not the same for all seven
trajectories the averaging process leads to the picture of the decay
of the initial {\bf P}$^+$/{\bf F}$^-$ charge separation, with on average
the electron spending half of its time on {\bf P} and half of its time on {\bf F}.
Of course, the statistics from averaging over only seven trajectories 
are far from satisfactory if our goal is to calculate an ensemble property.
It is, for example, impossible to extract ensemble-based relaxation times from these
data.

Semi-empirical calculations are much less resource intensive than DFT
calculations. We are thus in the presence of a trade-off between the
accuracy of the electronic structure calculation and the ability to carry out
significantly more trajectories in order to carry out more reliable 
ensemble averages.  We illustrate this concept by showing the ensemble
averages from our own CIS/AM1 FSSH calculations. 
Our minimum objective is to calculate a CT relaxation time.
As it happens, out of 100 initial trajectories, all but four ran to 
completion to that ensemble averages are reported over 96 trajectories.  

\begin{figure}
\includegraphics[width=1.0\linewidth]{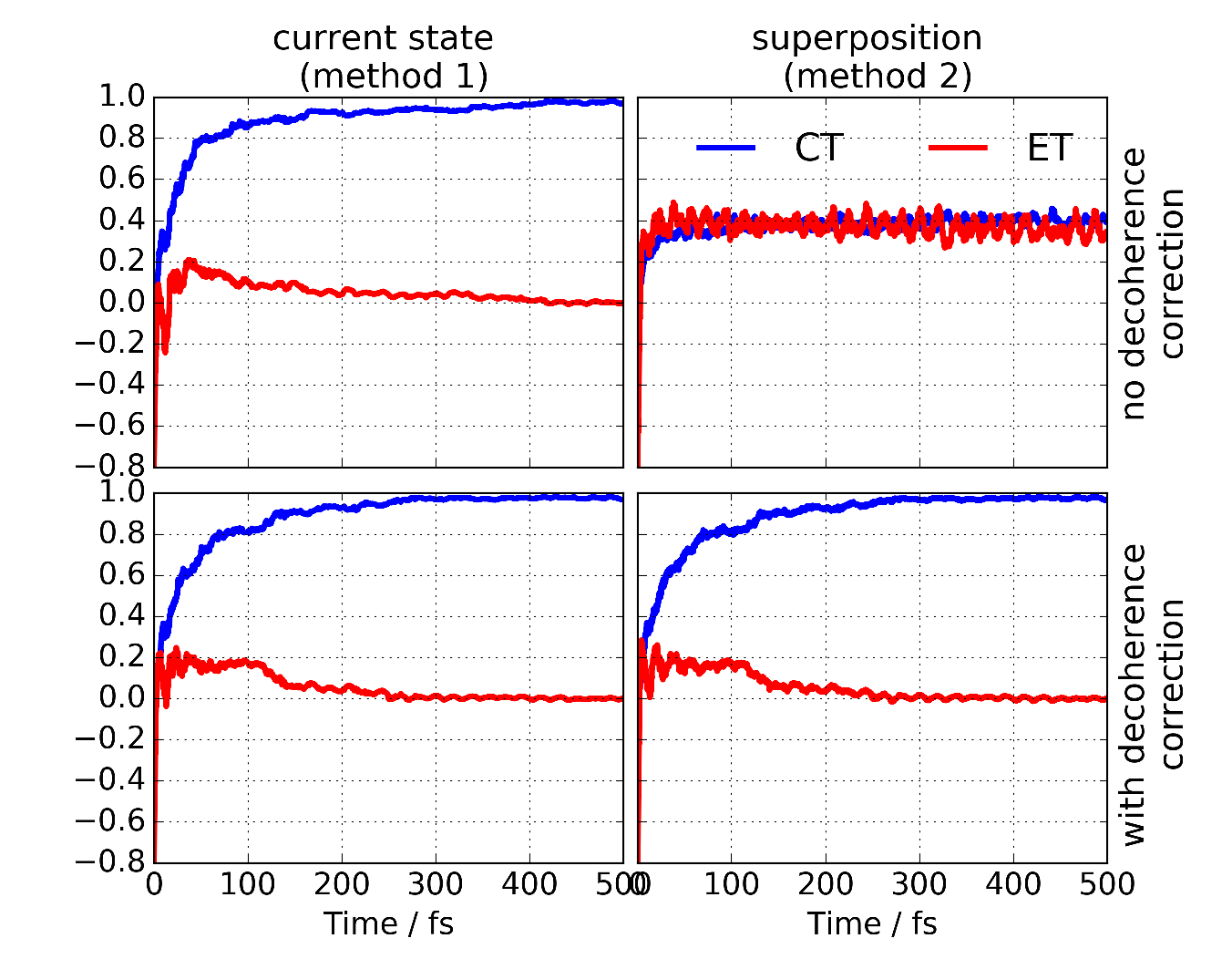} 
\caption{
CIS/AM1 FSSH calculations: 
``current state'' is the same as the surface method; ``superposition'' is the same as
the wavevector method.
\label{fig:Xincheng1}
}
\end{figure}

As discussed in Sec.~\ref{sec:theory}, there are two main ways to 
calculate expectation values in the FSSH method.
Tully's FSSH method was designed so that the fraction of running trajectories
for any given state (method 1, surfaces \cite{LFS13}) should equal the 
diagonal element of the electronic density matrix for that state (method 2, 
electronic wavevector \cite{LFS13}), provided that averaging occurs over 
a sufficiently large number of trajectories.  Unfortunately, this does not 
happen for the original FSSH algorithm with a reasonable number of 
trajectories. 
As shown in {\bf Fig.~\ref{fig:Xincheng1}}, 
inclusion of the decoherence correction of 
Ref.~\cite{GP07} brings CT and ET observables calculated with the two
different methods into nearly perfect agreement.  
It is now possible to fit the CT curve to either a single exponential or
to a linear combination of two exponentials and extract {\bf P}$^*$/{\bf F} $\rightarrow$ 
{\bf P}$^+$/{\bf F}$^-$ relaxation times.  A single exponential gives a CT time of 
$\tau_{\mbox{CT}} = 1/k_{\mbox{CT}} = \mbox{ 164 fs}$.  A fit with a linear combination
of two exponentials gives two relaxation times, a fast one 
$\tau_{\mbox{CT}}^{\text{fast}} \approx \mbox{ 16 fs}$ and a slow one 
$\tau_{\mbox{CT}}^{\text{slow}} \approx \mbox{ 91 fs}$.

\begin{figure} 
\includegraphics[width=0.8\linewidth]{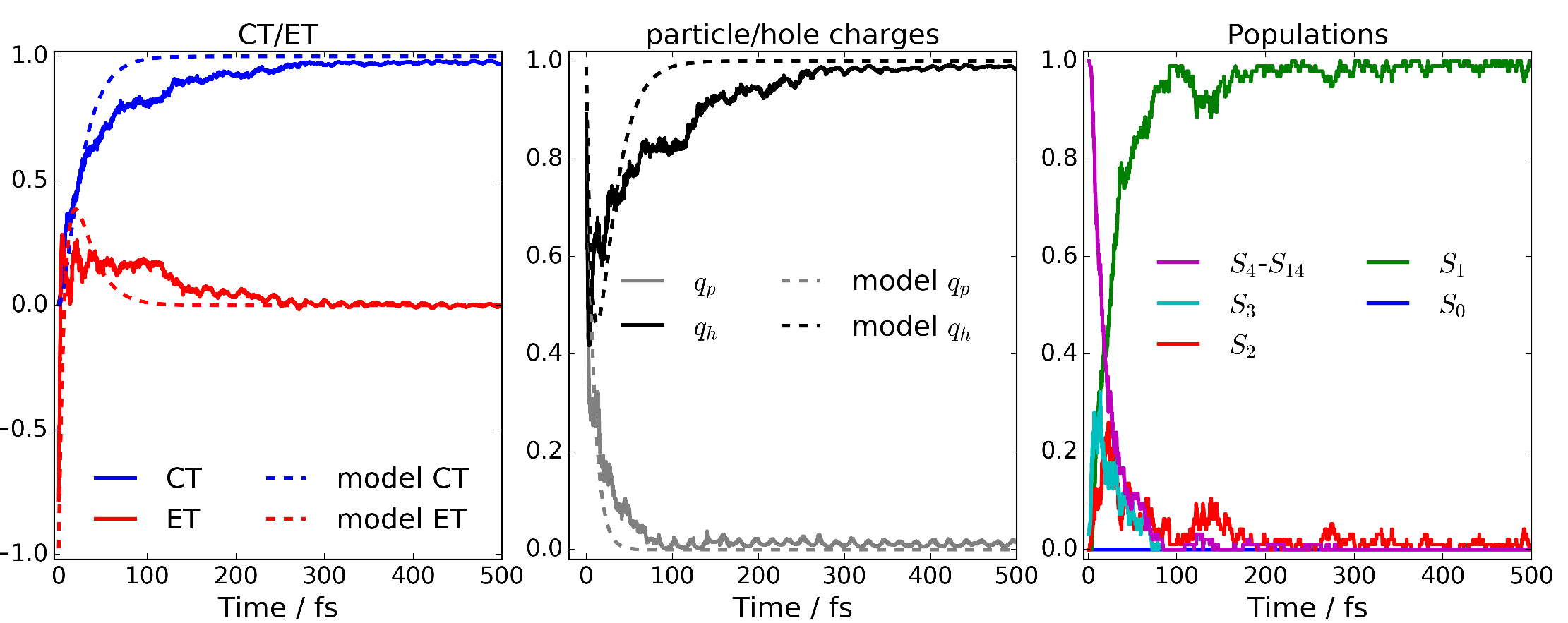}
\caption{
CIS/AM1 FSSH calculations of particle and hole charges along with state populations.
\label{fig:Xincheng5}
}
\end{figure}

It is very interesting to convert these ET and CT plots into particle and hole charge
plots as in {\bf Fig.~\ref{fig:Xincheng5}}.
Small oscillations reminiscent of the coherent charge 
oscillations of Ref.~\cite{JRB17} or of St\"uckelberg oscillations \cite{S32,T90} appear with 
a period of about 15 fs.  However more interesting to us is that we observe 
that the transfer of the excited electron 
(particle) from {\bf P} to {\bf F} induces a {\em partial} delocalization of the hole
from {\bf P} to {\bf F} consistent with the idea that the pairing of opposite charges
should be energetically favorable.  In the present case, this process only lasts about
5 fs before the hole relocalizes back onto {\bf P}.  Such a mechanism is consistent
with the solid-state concept of a polaron, namely a charge defect which is localized by
a lattice deformation. 
\onecolumngrid

\newpage
\twocolumngrid

The reader is reminded that a complicated process is taking place which
requires nuclear motion in order for there to be nonadiabatic hopping between different
electronic states. Nevertheless, the particle and hole dynamics suggest that qualitative 
(perhaps even semi-quantitative) results may be obtained from the very simple ``textbook'' 
model,
\begin{equation}
  \begin{array}{ccccc}  & k_{\text{ET}} &   & k_{\text{CT}} &   \\ 
  \mbox{{\bf P}$^*$/{\bf F}} & \longrightarrow & \mbox{{\bf P}/{\bf F}$^*$} 
        & \longrightarrow & \mbox{{\bf P}$^+$/{\bf F}$^-$}
  \end{array} \, .
  \label{eq:results.1}
\end{equation}
The well-known solution is,
\begin{eqnarray}
  [\mbox{{\bf P}$^*$/{\bf F}}] & = & e^{-k_{\text{ET}} t}   \nonumber \\
  \mbox{ [{\bf P}/{\bf F}$^*$] } & = & 
  \frac{ k_{\text{ET}} } {k_{\text{CT}}-k_{\text{ET}}} 
  \left( e^{-k_{\text{ET}} t} - e^{-k_{\text{CT}} t} \right)   \nonumber \\
  \mbox{[{\bf P}$^+$/{\bf F}$^-$]} & = &  1 +
   \frac{ 
          k_{ \text{ET} } 
                         e^{-k_{\text{CT}} t} 
        - k_{ \text{CT} } 
                         e^{-k_{\text{ET}} t} 
        } 
   {
          k_{ \text{CT} }
         -k_{ \text{ET} }
   }
  \label{eq:results.2}
\end{eqnarray}
assuming that $[\mbox{{\bf P}$^*$/{\bf F}}]_0=1$.  Since
\begin{eqnarray}
  q_p & = & q_p^P = [\mbox{{\bf P}$^*$/{\bf F}}] \nonumber \\
  q_h & = & q_h^P = [\mbox{{\bf P}$^*$/{\bf F}}] + [\mbox{{\bf P}$^+$/{\bf F}$^-$}] \, ,
  \label{eq:results.3}
\end{eqnarray}
then
\begin{eqnarray}
  q_p & = & e^{-k_{\text{ET}} t} \nonumber \\
  q_h & = & 1+\frac{k_{\text{ET}}}{k_{\text{CT}}-k_{\text{ET}}} 
       \left( e^{-k_{\text{CT}} t} - e^{-k_{\text{ET}} t} \right) \, .
  \label{eq:results.4}
\end{eqnarray}
The easiest way to find values for $k_{\text{ET}}$ and for $k_{\text{CT}}$ is to
first obtain $k_{\text{ET}}$ from a fit of $q_p$ and then to adjust $k_{\text{CT}}$
until the fit of $q_h$ looks reasonable.  This was done for $0 < t < \mbox{50 fs}$
and then simply applied at longer times (RHS.)  
CT and ET may then be obtained from,
\begin{eqnarray}
  \mbox{CT} & = & q_h - q_p \nonumber \\
  \mbox{ET} & = & 1 - (q_h + q_p) \, .
  \label{eq:results.5}
\end{eqnarray}
The result of the fit is shown in 
Fig.~\ref{fig:Xincheng5}.
The fit is by no means quantitative but it is at least qualitative
(if not semi-quantitative.)  This is even more impressive when it is
realized that the fit was done using only the data from the first 50 fs 
and then simply applied to the entire 500 fs run.  The resultant CT time is
$\tau_{\text{CT}} = 1/k_{\text{CT}} = \mbox{ 20 fs}$ and 
$\tau_{\text{ET}} = 1/k_{\text{ET}} = \mbox{ 8.3 fs}$. 
It is interesting to note how different this is from the above-mentioned 
$\tau_{\text{CT}} = \mbox{ 164 fs}$ extracted from the same data but using
a different kinetics model.  However the 20 fs CT time does resemble 
$\tau_{\text{CT}}^{\text{fast}} \approx \mbox{ 16 fs}$ obtained from the
double exponential fit.  This simply means that great care must be taken when specifying
relaxation times to also specify the fitting method.  Finally we note that
it is possible to obtain even better fits with a more elaborate kinetic 
model (SI.)

\subsection{Importance of Spectra}

As emphasized above, our choice of a semi-empirical method is governed by the need to
make a compromise between accuracy and being able to carry out a large number of 
trajectories.  We would therefore like to benchmark TD-lc-DFTB for our system 
against high-quality {\em ab-initio} methods in order to establish an ``exact'' result
against which to compare.  At the same time, we would also like to establish
two other benchmarks: The first is TD-DFT as TD-DFTB is supposed to be designed 
to behave like TD-DFT, rather than {\em ab-initio} methods.  And the secondd
is the semi-empirical CIS/AM1 method as TD-DFTB shares many aspects in common
with semi-empirical theories. 
As the {\em ab-initio} methods considered here are fairly resource intensive, this
subsection is restricted to calculations at a single geometry (geometry {\bf C}).

Since our initial model geometry is a purely theoretical construct, there are no 
experimental results against which to compare.  Nevertheless, we can make a common 
sense crude estimate of the position of the lowest CT excitation using the formula,
\begin{equation}
  \Delta E_{\text{CT}} \approx I_{\text{P}} - A_{\text{F}} - \frac{e^2}{R} \, ,
  \label{eq:results.7}
\end{equation}
where $I_{\text{P}} = 6.61 \mbox{ eV}$ is the experimental gas phase ionization potential
of {\bf P} \cite{B79}, $A_{\text{F}} = 3.22 \mbox{ eV}$ is the experimental
gas phase electron affinity of {\bf F} 
\cite{LWW+13}, and $R = \mbox{6.26 {\AA}}$.
This gives an estimated CT energy of $\Delta E_{\text{CT}} = \mbox{1.09 eV}$.
It is only a rough estimate because the third term is the Coulomb attraction between 
point charges which is unlikely to be an accurate reflection of the charge distributions 
in {\bf P} and in {\bf F}.  An attempt to improve on this approximation by calculating the 
repulsion energy between the LUMO on {\bf F} and the HOMO on {\bf P} increased the estimated
CT energy to $\Delta E_{\text{CT}} = \mbox{1.35 eV}$. (See also Ref.~\cite{DCT+15} 
for more information about experimental and calculated ionization potentials and 
electron affinities for molecules of interest for organic solar cells.)

\begin{figure}
\includegraphics[width=\linewidth]{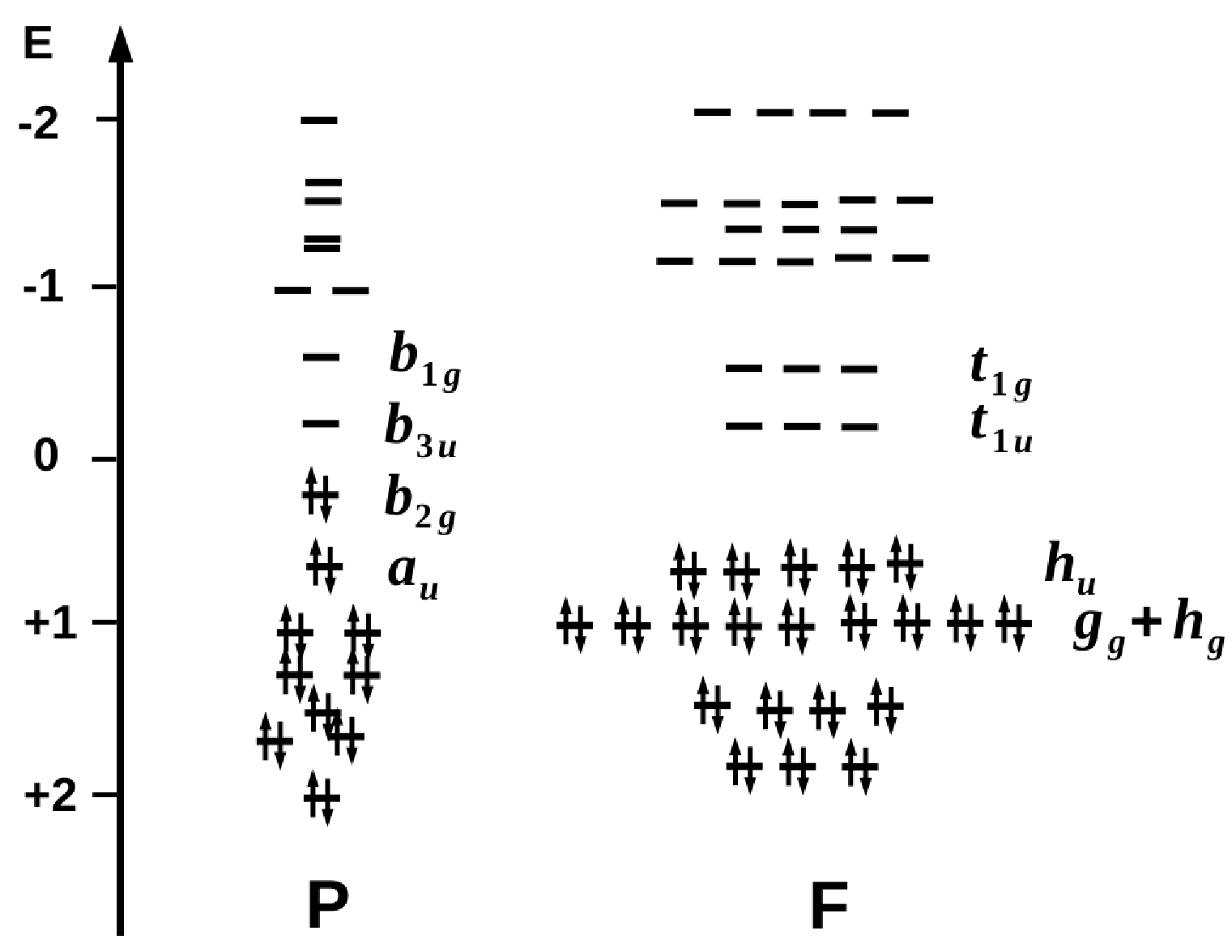} 
\caption{\label{fig:SHMO} Frontier molecular orbitals for {\bf P} and
for {\bf F} from simple H\"uckel molecular orbital theory.  Energies
are in H\"uckel units (i.e., the energy is $\alpha + \beta E$, where
both $\alpha$ and $\beta$ are negative numbers.)  Symmetry assignments
are for the point groups of the isolated molecules.
}
\end{figure}
As both {\bf F} and {\bf P} are $\pi$-systems, it is useful to take a brief look at the
results of simple H\"uckel molecular orbital theory.  The frontier orbitals are shown
in {\bf Fig.~\ref{fig:SHMO}}.  In this crude picture, the {\bf P} $b_{2g}$ HOMO is well
above the {\bf F} $h_u$ HOMO, but the {\bf P} $b_{3u}$ LUMO is quasi-degenerate with
the triply degenerate {\bf F} $t_{1u}$ LUMO.  Thus we should expect two types of low-lying
valence excitations, namely {\bf P}$^*$/{\bf F} and three quasi-degenerate 
{\bf P}$^+$/{\bf F}$^-$ states.  It is impossible from this level of theory to
predict which of these two types of excitations should be lowest in energy.


\begin{table}
\squeezetable
\begin{center}
\begin{tabular}{cccc}
\hline \hline
State & $\Delta E$ (eV) & $f$ & Character \\
\hline
  &\multicolumn{3}{c}{CC2} \\
  $S_{10}$ & 2.01 & 0.0005 &  {\bf P}/{\bf F}$^*$ \\ 
  $S_9$    & 2.00 & 0.0000 & {\bf P}/{\bf F}$^*$  \\
  $S_8$    & 1.98 & 0.0000 & {\bf P}/{\bf F}$^*$  \\
  $S_7$    & 1.90 & 0.0000 & {\bf P}/{\bf F}$^*$  \\
  $S_6$    & 1.90 & 0.0002 & {\bf P}/{\bf F}$^*$  \\
  $S_5$    & 1.88 & 0.0000 & {\bf P}/{\bf F}$^*$  \\
  $S_4$    & 1.87 & 0.0000 & {\bf P}/{\bf F}$^*$  \\ 
  $S_3$    & 1.70 & \underline{0.0355} & {\bf P}$^+$/{\bf F}$^-$  \\ 
  $S_2$    & 1.68 & \underline{0.0097} & {\bf P}$^+$/{\bf F}$^-$  \\ 
  $S_1$    & 1.67 & 0.0001 & {\bf P}$^+$/{\bf F}$^-$  \\ 
\hline \hline
\end{tabular}
\end{center}
\begin{tabular}{ccccccc}
\hline \hline
State & $\Delta E$ (eV) & $f$ & Character & $\Delta E$ (eV) & $f$ & Character\\
\hline
      & \multicolumn{3}{c}{SCS-CC2} 
      & \multicolumn{3}{c}{SOS-CC2} \\
$S_{10}$ & 2.37 & 0.0000 & {\bf P}/{\bf F}$^*$ &  
           2.54 & 0.0000 & {\bf P}/{\bf F}$^*$ \\ 
$S_9$    & 2.35 & 0.0000 & {\bf P}/{\bf F}$^*$ & 
           2.53 & 0.0000 & {\bf P}/{\bf F}$^*$ \\ 
$S_8$    & 2.34 & 0.0000 & {\bf P}/{\bf F}$^*$ & 
           2.51 & 0.0000 & {\bf P}/{\bf F}$^*$ \\ 
$S_7$    & 2.25 & 0.0000 & {\bf P}/{\bf F}$^*$ &  
           2.46 & 0.0001 & {\bf P}$^+$/{\bf F}$^-$\footnotemark[1] \\ 
$S_6$    & 2.24 & 0.0020 & {\bf P}/{\bf F}$^*$  & 
           2.45 & \underline{0.0089} & {\bf P}$^+$/{\bf F}$^-$ \\ 
$S_5$    & 2.23 & 0.0001 & {\bf P}/{\bf F}$^*$ &  
           2.41 & \underline{0.0246} & {\bf P}$^+$/{\bf F}$^-$ \\ 
$S_4$    & 2.21 & 0.0000 & {\bf P}/{\bf F}$^*$  &  
           2.39 & \underline{0.0044} & {\bf P}/{\bf F}$^*$ \\ 
$S_3$    & 2.17 & \underline{0.0380} & {\bf P}$^+$/{\bf F}$^-$ & 
           2.38 & 0.0000  & {\bf P}/{\bf F}$^*$ \\ 
$S_2$    & 2.14 & 0.0001 & {\bf P}$^+$/{\bf F}$^-$ & 
           2.36 & 0.0001 & {\bf P}/{\bf F}$^*$ \\ 
$S_1$    & 2.14 & 0.0011 & {\bf P}$^+$/{\bf F}$^-$ & 
           2.35 & 0.0000 & {\bf P}/{\bf F}$^*$ \\ 
\hline \hline
\end{tabular}
\footnotetext[1]{This state is actually about 50/50 {\bf P}$^+$/{\bf F}$^-$
and {\bf F}/{\bf P}$^*$ according to the transition density matrix analysis
shown in the SI.}
\caption{
\label{tab:CC2_C}
Characterization of {\bf P}/{\bf F} excited states using the CC2
{\em ab-initio} methods with the cc-pVDZ basis set for a model geometry {\bf C}.  
In all cases, 10 excited states have been requested.
Oscillator strengths near or larger than 0.01 have been underlined.
}
\end{table}

\begin{table}
\squeezetable
\begin{center}
\begin{tabular}{cccc}
\hline \hline
State & $\Delta E$ (eV) & $f$ & Character \\
\hline
      & \multicolumn{3}{c}{ADC(2)} \\ 
$S_{20}$ & 2.92 & 0.0000 & {\bf P}$^+$/{\bf F}$^-$ \\
$S_{19}$ & 2.40 & \underline{0.0415} & {\bf P}$^*$/{\bf F} \\
$S_{18}$ & 2.15 & 0.0002 & {\bf P}/{\bf F}$^*$ \\
$S_{17}$ & 2.14 & 0.0001 & {\bf P}/{\bf F}$^*$ \\
$S_{16}$ & 2.13 & 0.0002 & {\bf P}/{\bf F}$^*$ \\
$S_{15}$ & 2.11 & 0.0003 & {\bf P}/{\bf F}$^*$ \\
$S_{14}$ & 2.10 & 0.0001 & {\bf P}/{\bf F}$^*$ \\
$S_{13}$ & 2.10 & 0.0002 & {\bf P}/{\bf F}$^*$ \\
$S_{12}$ & 2.10 & 0.0001 & {\bf P}/{\bf F}$^*$ \\
$S_{11}$ & 2.08 & 0.0000 & {\bf P}/{\bf F}$^*$ \\
$S_{10}$ & 2.07 & 0.0003 & {\bf P}/{\bf F}$^*$ \\
$S_9$    & 2.05 & 0.0000 & {\bf P}/{\bf F}$^*$ \\
$S_8$    & 2.04 & 0.0000 & {\bf P}/{\bf F}$^*$ \\
$S_7$    & 1.96 & 0.0000 & {\bf P}/{\bf F}$^*$ \\
$S_6$    & 1.96 & 0.0001 & {\bf P}/{\bf F}$^*$ \\
$S_5$    & 1.94 & 0.0000 & {\bf P}/{\bf F}$^*$ \\
$S_4$    & 1.93 & 0.0000 & {\bf P}/{\bf F}$^*$ \\
$S_3$    & 1.73 & \underline{0.0179} & {\bf P}$^+$/{\bf F}$^-$ \\ 
$S_2$    & 1.71 & \underline{0.0072} & {\bf P}$^+$/{\bf F}$^-$ \\ 
$S_1$    & 1.70 & 0.0001 & {\bf P}$^+$/{\bf F}$^-$  \\
\hline \hline
\end{tabular}
\end{center}
\begin{tabular}{ccccccc}
\hline \hline
State & $\Delta E$ (eV) & $f$ & Character & $\Delta E$ (eV) & $f$ & Character\\
\hline
      & \multicolumn{3}{c}{SCS-ADC(2)} 
      & \multicolumn{3}{c}{SOS-ADC(2)} \\
$S_{20}$ & 3.34 & 0.0027 & {\bf P}$^*$/{\bf F}       &  
           3.32 & 0.0029 & {\bf P}$^*$/{\bf F}       \\ 
$S_{19}$ & 2.62 & \underline{0.0477} & {\bf P}$^*$/{\bf F} & 
           2.73 & \underline{0.0409} & {\bf P}$^*$/{\bf F} \\ 
$S_{18}$ & 2.53 & 0.0004 & {\bf P}/{\bf F}$^*$     & 
           2.72 & 0.0008 & {\bf P}/{\bf F}$^*$      \\ 
$S_{17}$ & 2.53 & 0.0004 & {\bf P}/{\bf F}$^*$     & 
           2.72 & 0.0042 & {\bf P}/{\bf F}$^*$      \\ 
$S_{16}$ & 2.51 & 0.0004 & {\bf P}/{\bf F}$^*$     & 
           2.70 & 0.0005 & {\bf P}/{\bf F}$^*$      \\ 
$S_{15}$ & 2.49 & 0.0013 & {\bf P}/{\bf F}$^*$     & 
           2.68 & \underline{0.0072} & {\bf P}/{\bf F}$^*$      \\ 
$S_{14}$ & 2.48 & 0.0000 & {\bf P}/{\bf F}$^*$     & 
           2.67 & 0.0000 & {\bf P}/{\bf F}$^*$      \\ 
$S_{13}$ & 2.43 & 0.0001 & {\bf P}/{\bf F}$^*$     & 
           2.59 & 0.0003 & {\bf P}/{\bf F}$^*$  \\ 
$S_{12}$ & 2.43 & 0.0011 & {\bf P}/{\bf F}$^*$     & 
           2.58 & 0.0032 & {\bf P}/{\bf F}$^*$      \\ 
$S_{11}$ & 2.41 & 0.0007 & {\bf P}/{\bf F}$^*$     &  
           2.58 & 0.0007 & {\bf P}/{\bf F}$^*$  \\ 
$S_{10}$ & 2.41 & 0.0000 & {\bf P}/{\bf F}$^*$     & 
           2.57 & 0.0000 & {\bf P}/{\bf F}$^*$     \\ 
$S_9$    & 2.39 & 0.0000 & {\bf P}/{\bf F}$^*$     & 
           2.55 & 0.0000 & {\bf P}/{\bf F}$^*$      \\ 
$S_8$    & 2.38 & 0.0000 & {\bf P}/{\bf F}$^*$      & 
           2.54 & 0.0000 & {\bf P}/{\bf F}$^*$      \\ 
$S_7$    & 2.29 & 0.0000 & {\bf P}/{\bf F}$^*$ & 
           2.47 & 0.0002 & {\bf P}/{\bf F}$^*$ \\ 
$S_6$    & 2.28 & 0.0009 & {\bf P}/{\bf F}$^*$  & 
           2.46 & 0.0049 & {\bf P}/{\bf F}$^*$ \\ 
$S_5$    & 2.27 & 0.0000 & {\bf P}/{\bf F}$^*$ & 
           2.43 & 0.0013 & {\bf P}/{\bf F}$^*$ \\ 
$S_4$    & 2.25 & 0.0000 & {\bf P}/{\bf F}$^*$ & 
           2.41 & 0.0000 & {\bf P}/{\bf F}$^*$ \\ 
$S_3$    & 2.18 & \underline{0.0318} & {\bf P}$^+$/{\bf F}$^-$ & 
           2.40 & \underline{0.0337} & {\bf P}$^+$/{\bf F}$^-$ \\ 
$S_2$    & 2.16 & 0.0002 & {\bf P}$^+$/{\bf F}$^-$ & 
           2.37 & 0.0002 & {\bf P}$^+$/{\bf F}$^-$\footnotemark[1] \\ 
$S_1$    & 2.15 & 0.0027 & {\bf P}$^+$/{\bf F}$^-$ & 
           2.36 & 0.0001 & {\bf P}$^+$/{\bf F}$^-$\footnotemark[1] \\ 
\hline \hline
\end{tabular}
\footnotetext[1]{This state is actually about 50/50 {\bf P}$^+$/{\bf F}$^-$
and {\bf F}/{\bf P}$^*$ according to the transition density matrix analysis
shown in the SI.}
\caption{
\label{tab:ADC2_C}
Characterization of {\bf P}/{\bf F} excited states using post-Hartree-Fock
ADC(2) {\em ab-initio} methods with the cc-pVDZ for a model geometry {\bf C}.  
In all cases, 20 excited states have been requested.
Oscillator strengths near or larger than 0.01 have been underlined.
}
\end{table}
{\bf Tables~\ref{tab:CC2_C}} and {\bf \ref{tab:ADC2_C}} shows the results of 
ADC(2) and CC2 calculations and 
their spin-scaled alternatives [SCS-ADC(2), SCS-CC2, SOS-ADC(2), and SOS-CC2].  
These latter might be important because ADC(2) and CC2 are only second-order methods 
and should not be considered to be the ultimate truth.  Indeed recent work has shown 
that improved results are obtained via spin scaling.  In particular, the SOS-ADC(2) 
method has been claimed to be an improvement over ADC(2) calculations
for describing CT excitations \cite{ABN+14}.  It should also be pointed out that these
are heavy calculations.  Calculating 20 states at the SCS-ADC(2) and SOS-ADC(2) levels 
took 13 days using 20 CPUs on our machines.  CC2 calculations are even more resource 
intensive.  Calculating 10 states at the SCS-CC2 and SOS-CC2 levels each took around 
20 days using 20 CPUs on our machines.  Thus these methods are {\em not} suitable for 
on-the-fly photodynamics simulations.  

Please note that the assignments given in Tables~\ref{tab:CC2_C} and 
\ref{tab:ADC2_C} and elsewhere in the present work are only given as a brief 
indication of the nature of the state.  
For reasons of space and brevity, we cannot 
give a detailed analysis of these assignments which therefore should be considered as 
only very approximate and subject to subjective bias.  
Nevertheless we do see the expected three {\bf P}$^+$/{\bf F}$^-$ states lying
lower in energy than the expected {\bf P}$^*$/{\bf F} state.
In the case of our high-quality {\em ab-initio} calculations a transition density 
analysis is given in the Supplementary Information (SI.) 
We have noticed that adding spin-scaling
increases the {\bf P}/{\bf F}$^*$ character relative to the {\bf P}$^+$/{\bf F}$^-$
character in the first three excited states.

Another important observation from these {\em ab-initio} calculations is that there 
is a dense manifold of singlet states (many of {\bf P}/{\bf F}$^*$ character)
which is well separated from the ground state.  
The precise ordering of these excited states can depend upon both the particular method 
used and upon small changes in the geometry of our {\bf P}/{\bf F} system.  
As semi-empirical methods are only intended to describe the valence orbitals and 
make use of a minimal-basis, we cannot expect to be able to describe all of these 
excited states with TD-lc-DFTB though we should be able to describe the most
important valence excitations.

\begin{figure}
\includegraphics[width=\linewidth]{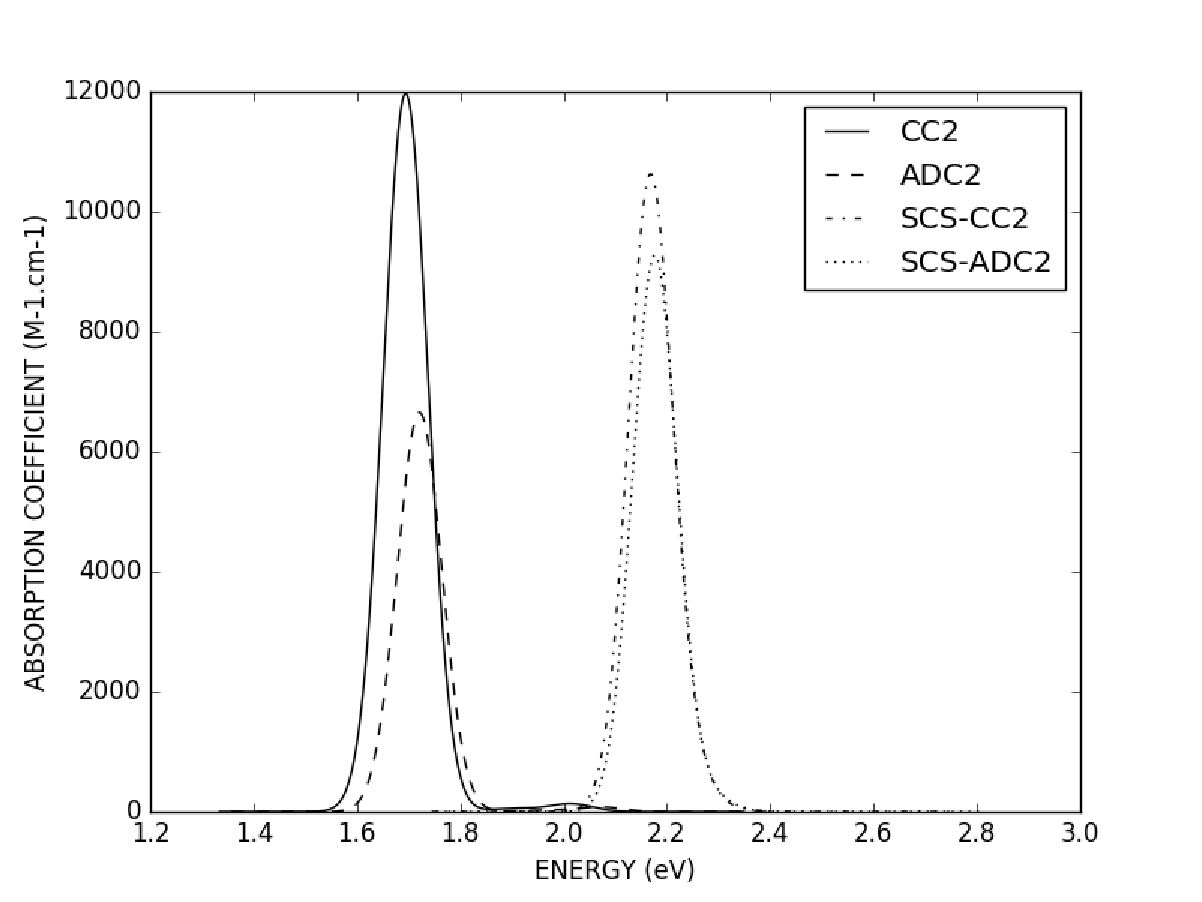} 
\caption{\label{fig:abinitio_C} Comparison of calculated spectra at 
geometry {\bf C} Gaussian convoluted with a 0.1 eV FWHM: ADC(2), CC2, SCS-ADC(2), and SCS-CC2.
All peaks correspond to the formation of the {\bf P}$^+$/{\bf F}$^-$ state.}
\end{figure}
\begin{figure}
\includegraphics[width=\linewidth]{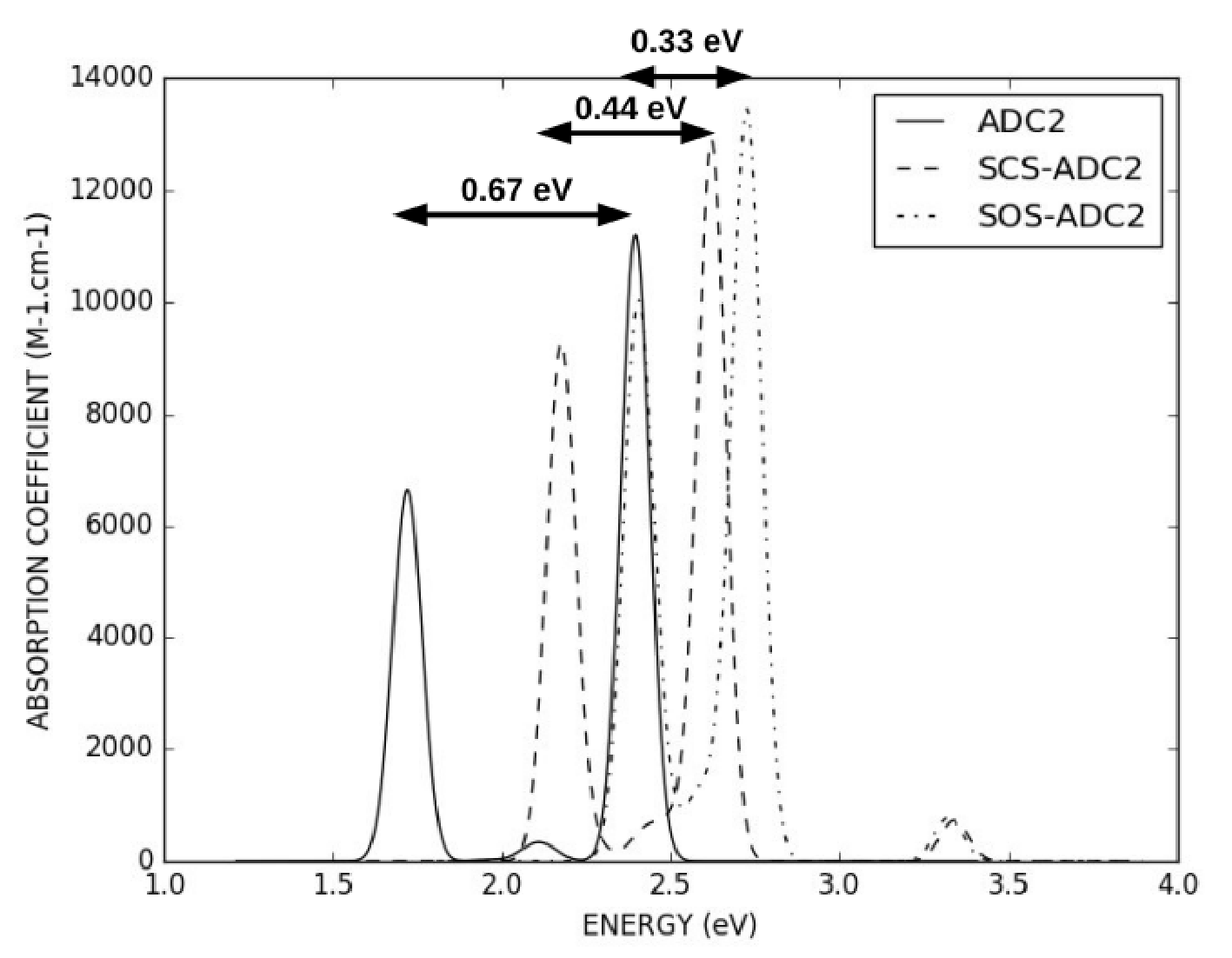} 
\caption{\label{fig:all_ADC_C_expanded} Comparison of calculated ADC(2), SCS-ADC(2), and SOS-ADC(2)
spectra Gaussian convoluted with a 0.1 eV FWHM at geometry {\bf C} over an expanded spectral 
range. The double-headed arrow indicates the energy difference between the lower {\bf P}$^+$/{\bf F}$^-$ CT peak and the upper {\bf P}$^*$/{\bf F} peak.}
\end{figure}

Let us return once more to the results of the {\em ab-initio} calculations shown
in Tables~\ref{tab:CC2_C} and \ref{tab:ADC2_C}.  
The lowest states are seen to be {\bf P}$^+$/{\bf F}$^-$ CT states with
some admixture of {\bf P}/{\bf F}$^*$ for some methods and are 
somewhat higher than our estimated {\bf P}$^+$/{\bf F}$^-$ CT energy of 
1.35 eV [1.70 eV for ADC(2), 1.67 eV for CC2, 2.15 eV for SCS-ADC(2), 2.14 eV 
for SCS-CC2, 2.36 eV for SOS-ADC(2), and 2.34 eV for SOS-CC2].  As illustrated 
in {\bf Fig.~\ref{fig:abinitio_C}},
CC2 and ADC(2) calculations without spin-scaling give very similar results.
As previously mentionned (Sec.~\ref{sec:TurboMole}), spin-scaling is thought to 
improve CT energies.  In this case we see that spin-scaling leads to blue-shifting
the energies.
{\bf Figure~\ref{fig:all_ADC_C_expanded}} shows spectra calculated with ADC(2) with and
without spin-scaling over a larger energy range.   In the present case, we see
that spin-scaling increases the energy of all states but increases the 
energy of CT states more than of the energy of a localized excitation on {\bf P}.
This is pretty much exactly what is usually seen when an lc functional is used
to improve CT energies in TD-lc-DFT, though we do not understand why such a qualitatively
similar observation should also hold for spin-corrected ADC(2) and CC2.

So which should we consider as ``exact''?  As there is less experience with the
spin-scaled methods we might be conservative and focus on the ADC(2) and CC2
results.  However, in the present context, the differences between the {\em ab-initio}
calculations with and without spin-scaling are much less important than the
simple result that there is an intense {\bf P}$^+$/{\bf F}$^-$ CT peak which is 
lower in energy than a still more intense peak corresponding to a local excitation 
on {\bf P} ({\bf P}$^*$/{\bf F}).  This is the ``exact'' behavior that we should 
hope to match with a semi-empirical method.

\begin{figure}
\includegraphics[width=\linewidth]{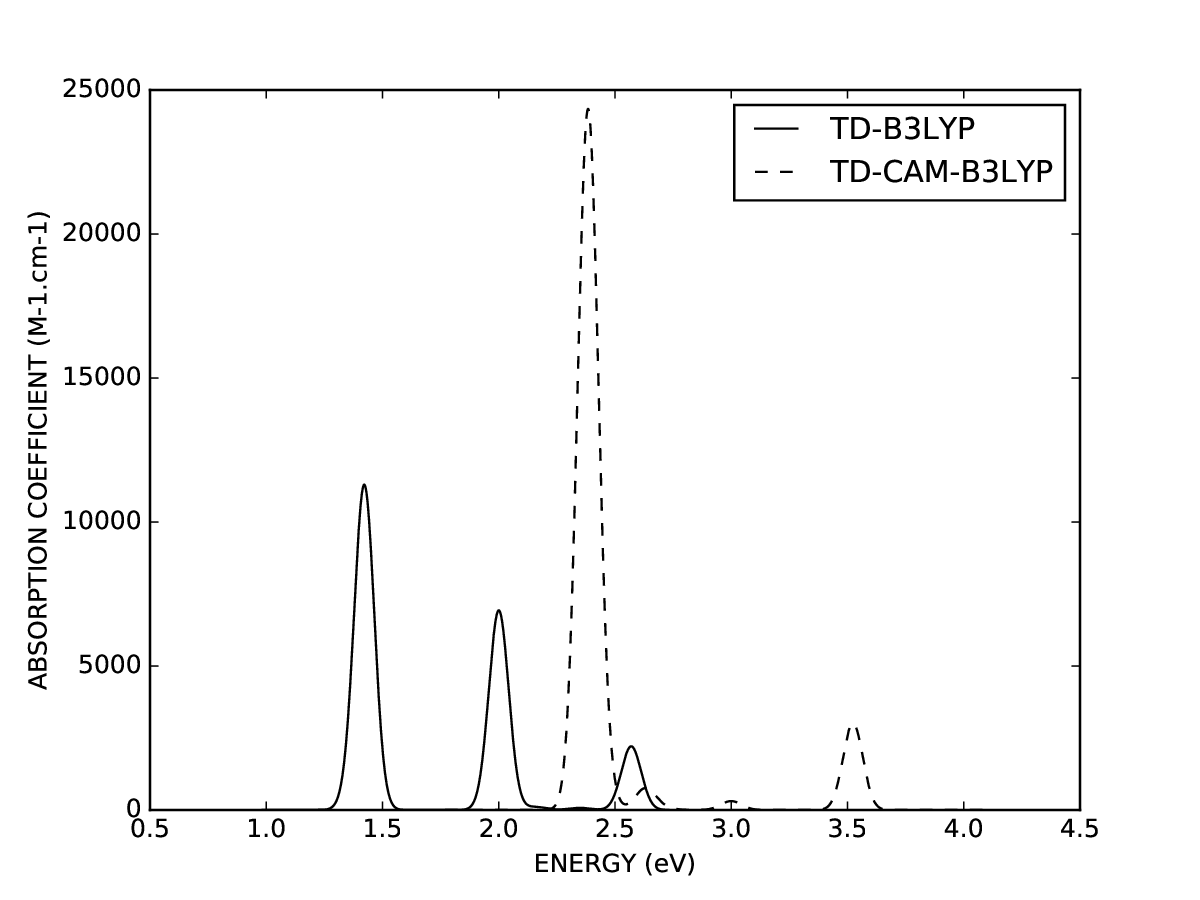} 
\caption{\label{fig:DFT_SCS_C} Comparison of 
TD-B3LYP and TD-CAM-B3LYP spectra,
Gaussian convoluted with a 0.1 eV FWHM at geometry {\bf C}.
The TD-B3LYP peak at around 1.5 eV corresponds to the {\bf P}$^+$/{\bf F}$^-$
CT state and the peak at around 2.0 eV corresponds to the local excitation 
{\bf P}$^*$/{\bf F}.  In contrast, these two peaks have merged into a single
peak at about 2.4 eV in the TD-CAM-B3YP spectrum.
}
\end{figure}

\begin{figure}
\includegraphics[width=\linewidth]{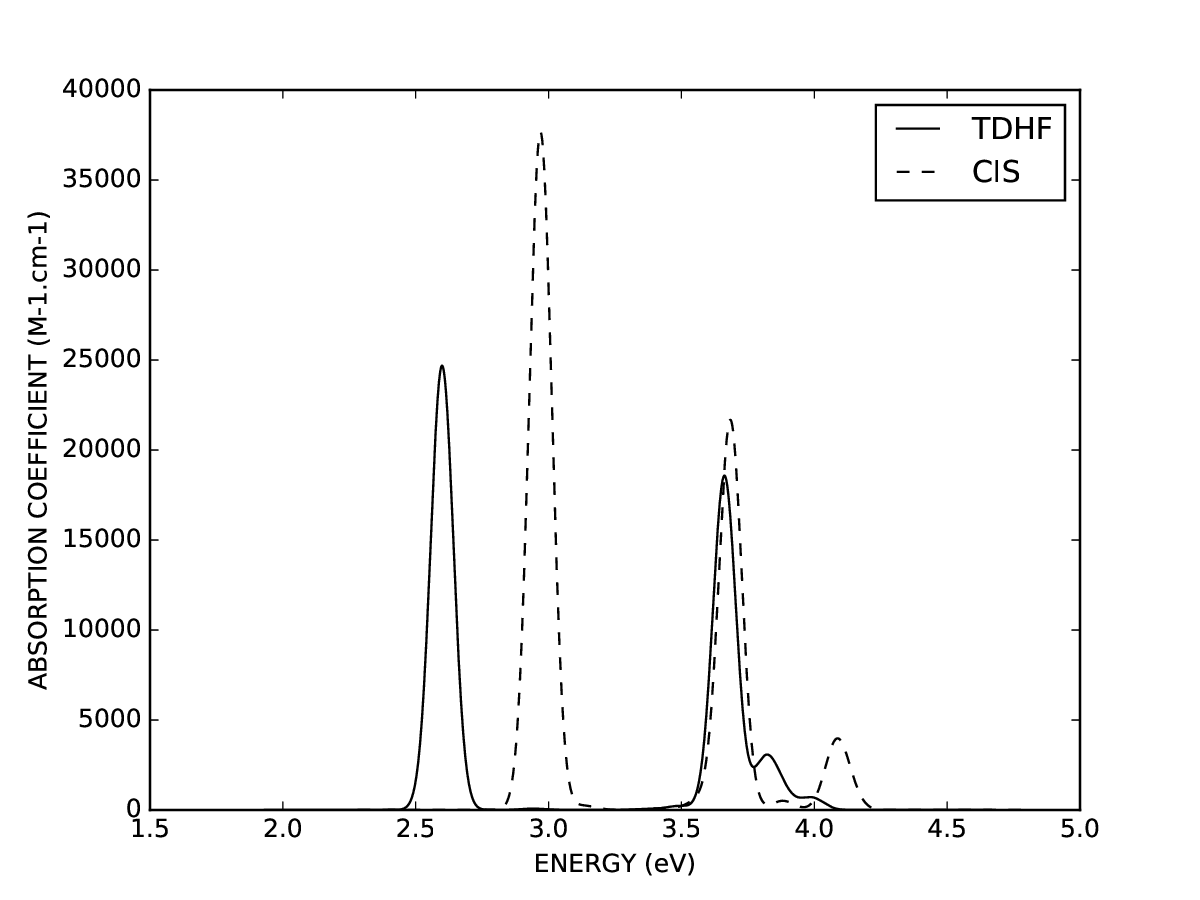} 
\caption{\label{fig:CIS_TDHF_C} Comparison of calculated CIS/6-31G(d,p) and 
TD-HF/6-31G(d,p) spectra Gaussian convoluted with a 0.1 eV FWHM at geometry 
{\bf C}. The lower peaks (at around 2.6 eV for TD-HF and at around 3.0 eV for CIS)
correspond to the local excitation {\bf P}$^*$/{\bf F}, while the higher peaks 
at around 3.6 eV for the two types of calculations correspond to the 
{\bf P}$^+$/{\bf F}$^-$ CT state.}
\end{figure}

As ``DFTB inherits the faults of DFT as well as some of its own,'' \cite{Mathias},
it is interesting to first see how well DFT works before examining DFTB.
We carried out TD-B3LYP/6-31G(d,p), TD-CAM-B3LYP/6-31G(d,p), TD-HF/6-31G(d,p), 
and CIS/6-31G(d,p) calculations of the spectra of our {\bf P}/{\bf F} complex. 
B3LYP is a well-known global hybrid functional with about 20\% exact exchange.
CAM-B3LYP adds a range-separated hybrid on top of the B3LYP global hybrid,
thus increasing the percentage of exact exchange at long-range.  Though pre-dating
Kohn-Sham DFT, Hartree-Fock (HF) may be considered to an extreme density functional
with 100\% exact exchange and no correlation.  Configuration interaction singles
(CIS) is TD-HF in the Tamm-Dancoff approximation. TD-DFT is known to underestimate
CT excitations by as much as 1-2 eV.  Range-separated functionals are known to
correct this problem at the cost of slightly increasing local excitation energies.
{\bf Figure~\ref{fig:DFT_SCS_C}} shows that TD-B3LYP gives a qualitatively correct
spectrum with the {\bf P}$^+$/{\bf F}$^-$ CT lower than the {\bf P}$^*$/{\bf F} 
locally excited state, but the intensities are wrong.  The {\bf P}$^*$/{\bf F} peak 
should be more intense than the {\bf P}$^+$/{\bf F}$^-$ peak but instead it is 
the {\bf P}$^+$/{\bf F}$^-$ peak which is more intense.  As expected the 
range-separated TD-CAM-B3LYP functional increases the energy of the 
{\bf P}$^+$/{\bf F}$^-$ CT peak relative to that of the {\bf P}$^*$/{\bf F} peak, but 
incorrectly makes them quasi-degenerate.  {\bf Figure~\ref{fig:CIS_TDHF_C}} shows 
that TD-HF and CIS calculations are qualitatively similar but that the 
{\bf P}$^+$/{\bf F}$^-$ CT peak is now higher in energy than the peak for 
the {\bf P}$^*$/{\bf F} locally excited state.
On the other hand, it is reassuring that the relative intensities of the
two peaks are close to what we expected from our high-quality calculations.

\begin{figure}
\includegraphics[width=\linewidth]{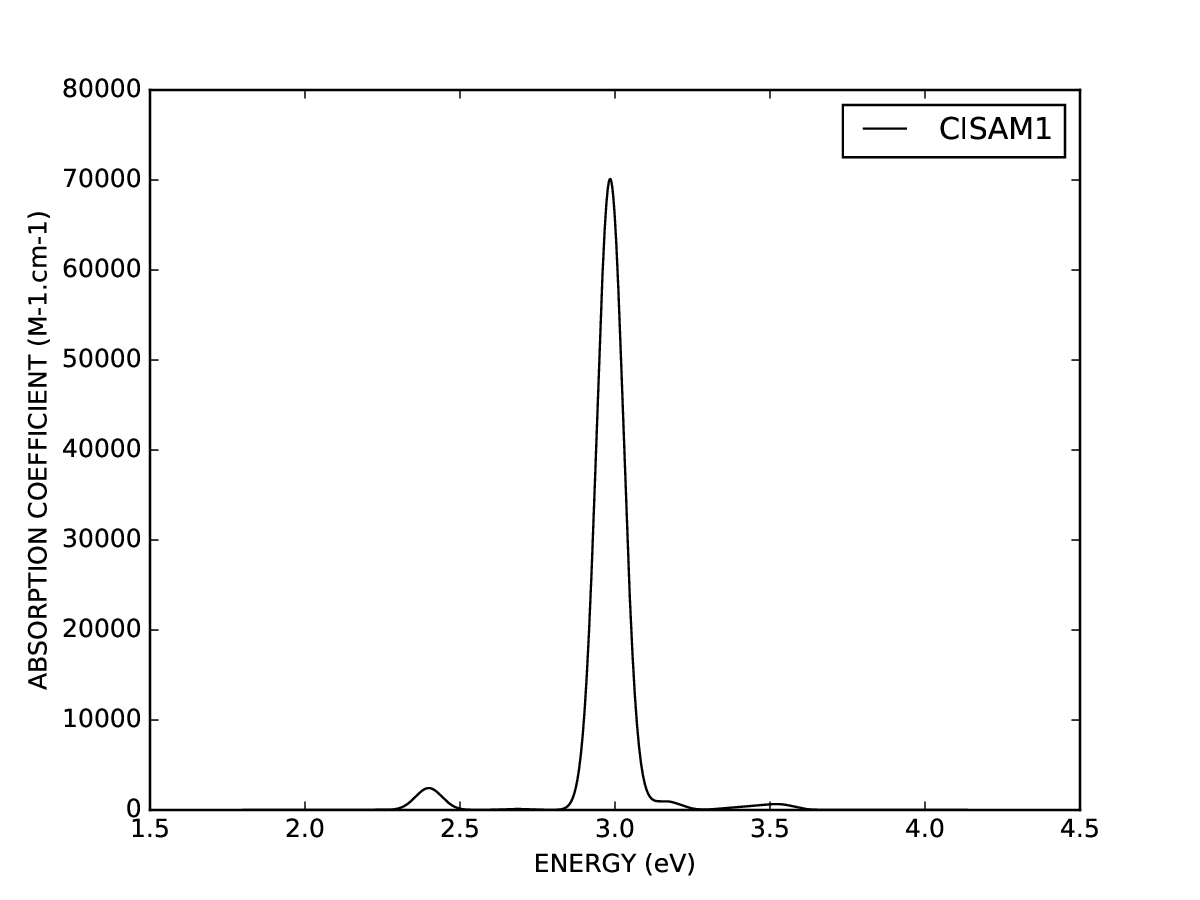} 
\caption{\label{fig:AM1CIS_C} CIS/AM1 spectrum,
Gaussian convoluted with a 0.1 eV FWHM at geometry {\bf C}.
The lower peak ($\sim$2.4 eV) correponds to the {\bf P}$^+$/{\bf F}$^-$
CT state while the upper peak ($\sim$3.0 eV) corresponds to the local
{\bf P}$^*$/{\bf F} excitation.
}
\end{figure}

Lastly, before turning to TD-lc-DFTB spectra, let us look at semi-empirical
CIS/AM1 spectra.  Note that this method has been parameterized using 
experimental data.  As such, though the structure of the calculations 
resemble a CIS calculation, the CIS/AM1 calculation actually
interpolates (or even extrapolates) experimental data by including implicit electron
correlation effects in its parameterization.  
As seen in {\bf Fig.~\ref{fig:AM1CIS_C}} the CIS/AM1 spectrum is very different
than the CIS spectrum in that AM1/CIS correctly places the {\bf P}$^+$/{\bf F}$^-$
CT peak at lower energy than the {\bf P}$^*$/{\bf F} peak while CIS places the 
{\bf P}$^+$/{\bf F}$^-$ CT peak at higher energy 
than the local {\bf P}$^*$/{\bf F} peak.  Moreover, the CIS/AM1 spectrum, though
shifted to higher energy than the ADC(2) spectrum, is in qualitative agreement
with the ADC(2) calculations in so far as the {\bf P}$^+$/{\bf F}$^-$ CT peak lies at 
lower energy than the {\bf P}$^*$/{\bf F} locally excitation peak and the energy 
differences between the {\bf P}$^+$/{\bf F}$^-$ and {\bf P}$^*$/{\bf F}  peaks are 
about the same in the CIS/AM1 and ADC(2) calculations.  We explain this by the implicit
inclusion of correlation effects in CIS/AM1 via fitting of parameters 
to experimental values.  On the other hand, the {\bf P}$^*$/{\bf F} peak is much 
more intense compared to the {\bf P}$^+$/{\bf F}$^-$ peak in the CIS/AM1 spectrum 
than is the case in the ADC(2) spectrum.

\begin{figure}
\includegraphics[width=\linewidth]{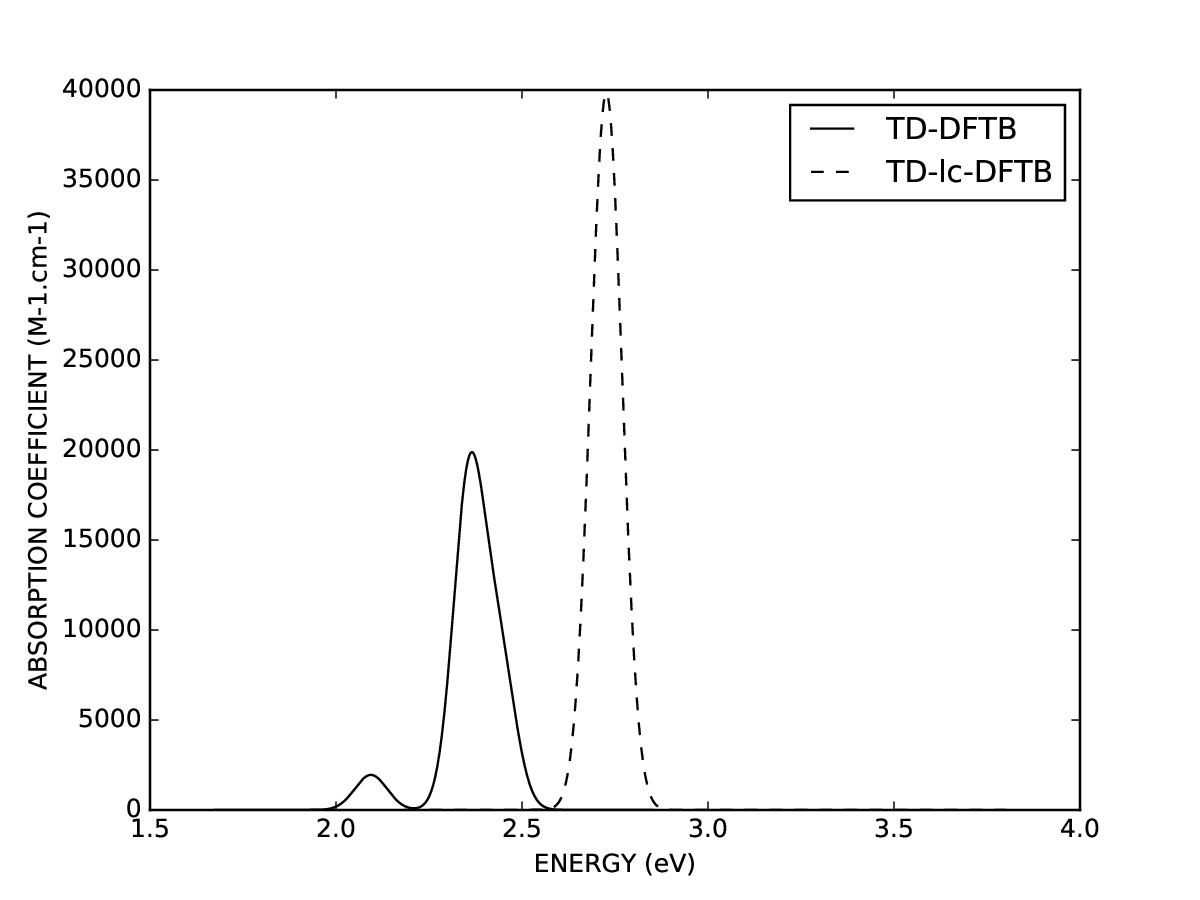} 
\caption{\label{fig:TD-DFTB_C} TD-DFTB ($R_{lc} = \infty$) and TD-lc-DFTB
with the default value of $R_{lc} = 3.03 \, a_0$
Gaussian convoluted with a 0.1 eV FWHM at geometry {\bf C}.
The TD-DFTB calculation is qualitatively correct with a lower energy 
{\bf P}$^+$/{\bf F}$^-$ CT peak ($\sim$2.1 eV) and a higher energy {\bf P}$^*$/{\bf F}
peak ($\sim$2.3 eV).  The TD-lc-DFTB calculation only shows the {\bf P}$^*$/{\bf F}
peak ($\sim$2.7 eV) in this figure.
}
\end{figure}
We now turn to TD-DFTB with and without lc.  {\bf Figure~\ref{fig:TD-DFTB_C}}
shows the TD-DFTB spectrum without lc and the TD-lc-DFTB spectrum with the 
default value of $R_{lc} = 3.03 \, a_0$.  It is immediately obvious that the
TD-DFTB spectrum is at least qualitatively correct with a lower energy 
{\bf P}$^+$/{\bf F}$^-$ CT peak and a higher energy {\bf P}$^*$/{\bf F}
peak.  The position of the {\bf P}$^*$/{\bf F} peak is even in good agreement
with that of the corresponding ADC(2) {\bf P}$^*$/{\bf F} peak, although we
note that this is the unexpected consequence of using the new 
$r_0(\mbox{C}) = 4.309 \, a_0$ confinement radius.  The older 
$r_0(\mbox{C}) = 2.657 \, a_0$ confinement radius results in a TD-DFTB spectrum
with the {\bf P}$^*$/{\bf F} peak at about 1.9 eV and a larger energy difference
between the {\bf P}$^*$/{\bf F} and {\bf P}$^+$/{\bf F}$^-$ peaks ($\sim$0.4 eV
as opposed to $\sim$0.2 eV with the new confinement radius.)  This highlights
how sensitive this particular calculation is to the choice of confinement radius.
Figure~\ref{fig:TD-DFTB_C} also shows that the TD-lc-DFTB spectrum with the
default $R_{lc} = 3.03 \, a_0$ is qualitatively incorrect as the lowest energy
peak corresponds to the {\bf P}$^*$/{\bf F} state irrespective of the choice
of confinement radius.  We conclude that something more profound is going on.

\begin{figure}
\includegraphics[width=0.8\linewidth]{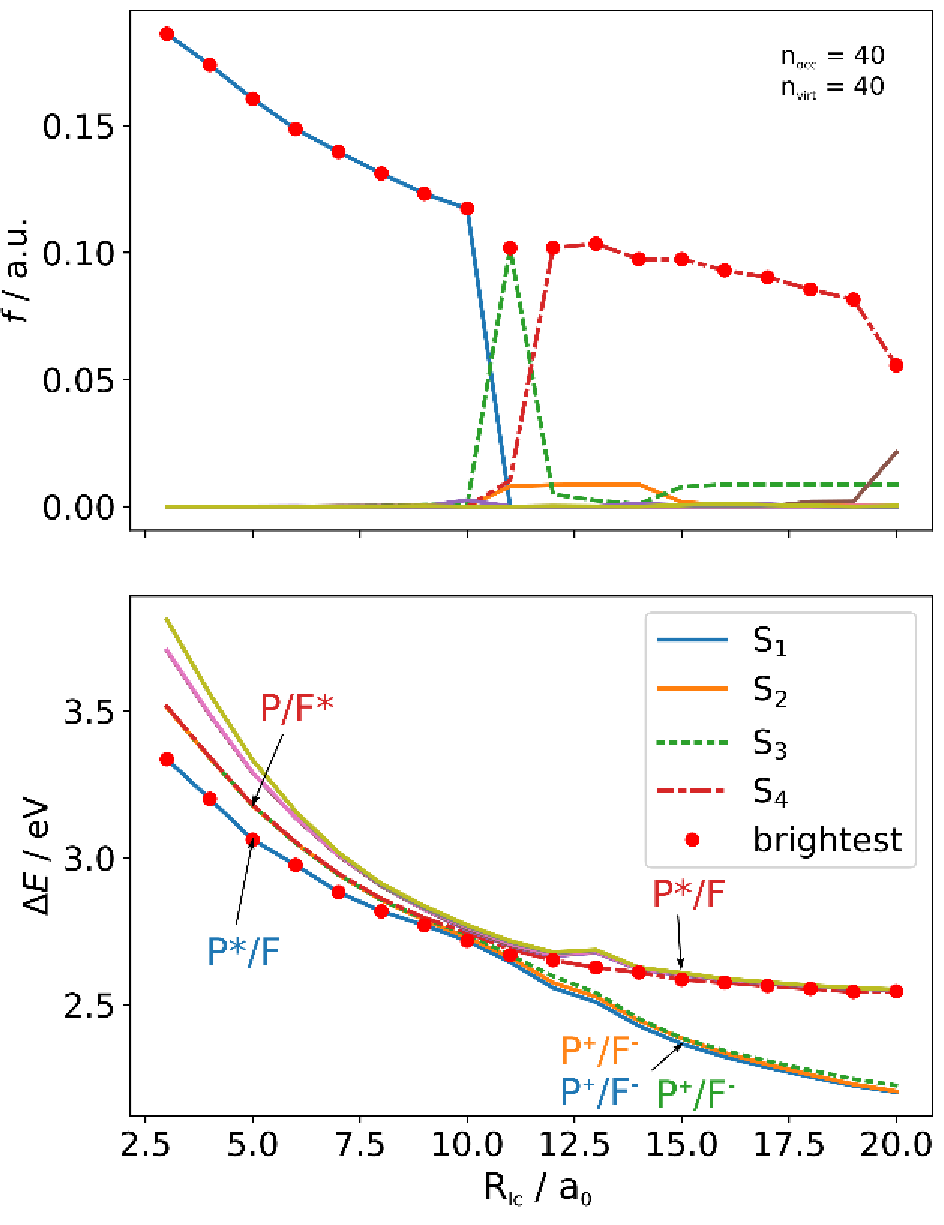} 
\caption{TD-lc-DFTB P/F excited-state energies as a function of $R_{lc}$:
upper, oscillator strengths; lower, excitation energies. 
\label{fig:StatesVSRcl}
}
\end{figure}
This is why we now turn to how the spectra of TD-lc-DFTB calculations vary as a 
function of $R_{lc}=1/\mu$.  {\bf Figure~\ref{fig:StatesVSRcl}} show the results 
of TD-lc-DFTB calculations for several values of $R_{lc}$.  
There is one excitation, identified as a local pentacene excitation, 
{\bf P}/{\bf F} $\rightarrow$  {\bf P}$^*$/{\bf F}, which has 
significantly greater oscillator strength than the other excitations.
It is useful to split the graph into regions:
\begin{itemize}
\item[I.] Above $R_{lc} > \, \sim 12.5 \, a_0$ this excited state 
is quasidegenerate with a locally excited buckminsterfullerene state 
{\bf P}/{\bf F}$^*$ and the two states can mix.  There are also three 
quasidegenerate {\bf P}$^+$/{\bf F}$^-$ CT excited states below 
the quasidegenerate {\bf P}$^*$/{\bf F} and {\bf P}/{\bf F}$^*$ states.  
The absorption spectrum in this region is qualitatively similar to 
what is seen in the high-quality {\em ab-initio} calculations.
\item[II.] The region $\sim 12.5 \, a_0 > R_{lc} > \, \sim 10 \, a_0$ is 
a crossing region where all of the states become quasidegenerate and considerable 
mixing occurs.  
\item[III.] Below $R_{lc} <  10 \, a_0$, the {\bf P}$^+$/{\bf F}$^-$ states 
move to higher energies and the two lower states have local excitation character 
with the {\bf P}$^*$/{\bf F} state being lower than the {\bf P}/{\bf F}$^*$ state.  
Inverse CT states of type {\bf P}$^-$/{\bf F}$^+$ are higher energy
states not shown on this graph.  The absorption spectrum in this region is
qualitatively similar to what is seen in the TD-HF and CIS calculations.
Note that this makes good sense as $R_{lc}$ is (roughly speaking) the cut-off 
beyond which exact (i.e., Hartree-Fock) exchange is used.  Thus a smaller value
of $R_{lc}$ corresponds to a more Hartree-Fock-like calculation.
\end{itemize}
On the basis of this picture, we might expect CT above $R_{lc} > 12.5 \, a_0$ 
and little or no CT below $R_{lc} < 10 \, a_0$.

We also tried one more idea that we call $\mu$-scanning.  This consists of finding the 
value of $R_{lc}$ which gives the best agreement with some other high-quality calculation.  
In our case, we found that $R_{lc} \approx 15 \, a_0$ allowed us to reproduce the 
spectra reported in Ref.~\cite{JRB17} at their geometry {\em using the older}
$r_0(\mbox{C})$ {\em confinement radius.}  We have not tried to repeat the procedure
with the new value of the confinement radius.

\subsection{Charge and Energy Transfer as a Function of $R_{lc}$}

\begin{figure}[b]
\includegraphics[width=1.0\linewidth]{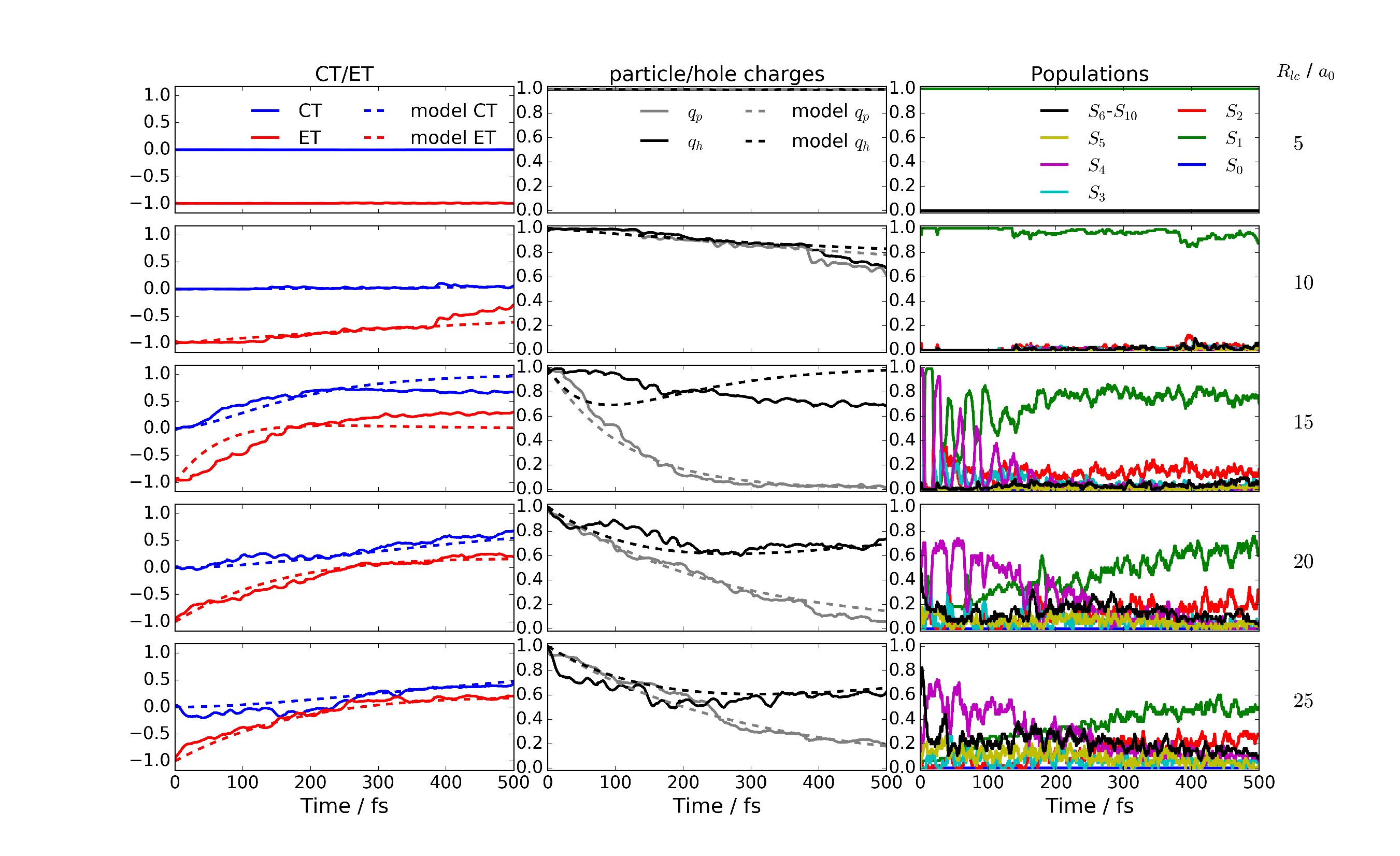} 
\caption{\label{fig:Alexander2} CT, ET, particle/hole populations, and state populations
for different values of the long-range parameter.}
\end{figure}

We now wish to see how {\bf P}/{\bf F} charge and energy transfer vary in TD-lc-DFTB FSSH 
calculations as a function of $R_{lc}$.  This will give us some idea of how
robust this model is with respect to $R_{lc}$ and will provide an estimate of
ET and CT times as a function of $R_{lc}$.  The basic procedure has already
been explained in Secs.~\ref{sec:theory} and \ref{sec:details}.
Ensemble averages
were over at least 50 trajectories for all values of $R_{lc}$ except for
$R_{lc} = 15 \, a_0$ where the average is over 100 trajectories.  Results are
shown in {\bf Fig.~\ref{fig:Alexander2}}.  Note that the range of $R_{lc}$ 
covers regions I, II, and III defined by  
the spectroscopic analysis of the last subsection. Thus we may expect 
substantial variation in the physics of CT and ET. 
\onecolumngrid

\twocolumngrid

From Fig.~\ref{fig:StatesVSRcl}, we see that
the initial excited state for $R_{lc} = 5 \, a_0$ is $S_1$ which has {\bf P}$^*$/{\bf F}
character.  There are no low-lying CT excited states so no CT is expected.  Furthermore
{\bf P}/{\bf F}$^*$ lies above $S_1$ so that ET also may not take place.  
The top row of Fig.~\ref{fig:Alexander2}
shows that neither CT nor ET takes place and that
the system remains in $S_1$ for the entire 
500 fs run.

From the upper part of Fig.~\ref{fig:StatesVSRcl} or from 
the second row of Fig.~\ref{fig:Alexander2} 
we see that the initial excited state is $S_1$ for $R_{lc} = 10 \, a_0$.  There are several
quasi-degenerate states near $S_1$ so that some mixing may occur with other states,
making it difficult to anticipate how it will decay.
The second row of Fig.~\ref{fig:Alexander2}
shows that CT is negligible
but that {\bf P}$^*$/{\bf F} $\rightarrow$ {\bf P}/{\bf F}$^*$ ET is 
important.  As $S_1$ remains by far the dominant state,
ET must be explained by variations in the geometry which lead to
changes in the nature of $S_1$.

From Fig.~\ref{fig:StatesVSRcl}, we see that the initial excited state 
for $R_{lc} = 15 \, a_0$ is $S_4$ which has {\bf P}$^*$/{\bf F} character.
There are three lower lying states of {\bf P}$^+$/{\bf F}$^-$ CT character,
so that we may expect to see CT dynamics.  This is confirmed by 
the third row of Fig.~\ref{fig:Alexander2}
which shows rapid population of 
{\bf P}$^+$/{\bf F}$^-$ states.  Interestingly the figure also shows 
significant ET.  

Examination of Fig.~\ref{fig:StatesVSRcl} might lead to the expectation that
the CT and ET dynamics for $R_{lc} = 20 \, a_0$ and $R_{lc} = 25 \, a_0$ should
be similar to that for $R_{lc} = 15 \, a_0$.  However, 
the last two rows of Fig.~\ref{fig:Alexander2}
show that this is not the case.  Instead ET dominates over CT for short times,
and CT dominates over ET for longer times.  
As we shall see, this switch-over may be explained by
the idea of that electron (particle) transfer from {\bf P} to {\bf F} begins 
by dragging the hole along with it but then nuclear motion kicks in and
restores the hole on {\bf P}.  This is similar to the mechanism of 
polaron formation in solids where the lattice distorts to create localized
charge defects.

This ``polaron formation'' is easiest to see after
recasting our results in terms of the particle $q_p = q_p^P$ and hole 
$q_h = q_h^P$ populations on {\bf P}. 
The results are also shown in Fig.~\ref{fig:Alexander2}.
Notice that there is a smooth
trend in going from 
the top to the bottom of Fig.~\ref{fig:Alexander2}.
As $R_{lc}$ increases, the $q_p$ curve which is always decreasing, changes its 
form from convex down to a nearly exponential decay.  Also as $R_{lc}$ increases,
the $q_h$ curve which is initially convex down (for $R_{lc} = 10 \, a_0$), 
straightens out (for $R_{lc} = 15 \, a_0$), and then becomes increasingly
sharply concave upward 
when $R_{lc} = 25 \, a_0$.
This gradual and seemingly smooth change is a little deceptive because, as we have seen with the
spectra, the underlying physics changes quite a bit with the value of $R_{lc}$.

Let us take a closer look.  For $R_{lc} = 10 \, a_0$
(second row of Fig.~\ref{fig:Alexander2})
we see that the particle and hole move together 
from {\bf P} to {\bf F} ({\bf P}$^*$/{\bf F} $\rightarrow$ {\bf P}/{\bf F}$^*$.)  
No significant CT happens, but the ET is almost exponential.  For $R_{lc} = 15 \, a_0$
(third row of Fig.~\ref{fig:Alexander2})
the hole transfer has slowed so that it is nearly linear.
On the other hand, 
the particle transfer has become nearly exponential.  For $R_{lc} = 20 \, a_0$
(fourth row of Fig.~\ref{fig:Alexander2})
the particle transfer is slowing and the hole, which
initially left {\bf P} to go to {\bf F}, starts to return to {\bf P} after some
time.  This phenomenon is even more marked for $R_{lc} = 25 \, a_0$
(last row of Fig.~\ref{fig:Alexander2}.)

\begin{figure}
\includegraphics[width=0.9\linewidth]{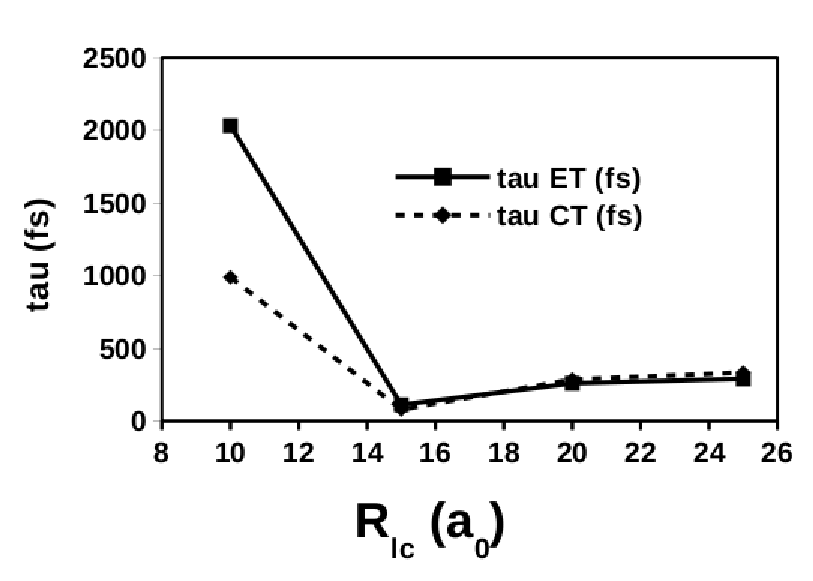} 
\caption{\label{fig:CTET} CT $\tau_{\text{CT}} = 1/k_{\text{CT}}$
and ET $\tau_{\text{ET}} = 1/k_{\text{ET}}$ times for different values of $R_{lc}$
obtained using the kinetic model shown in Eq.~(\ref{eq:results.1}).
}
\end{figure}

Figure~\ref{fig:Alexander2} 
shows that the overall trends can be captured very qualitatively
in the case of $R_{lc} = 10 \, a_0$ and $R_{lc} = 15 \, a_0$ and much more quantitatively
for $R_{lc} = 20 \, a_0$ and $R_{lc} = 25 \, a_0$ using the kinetic model 
shown in Eq.~\ref{eq:results.1} (see also the SI.)  This allows us to 
summarize CT and ET times in a single figure ({\bf Fig.~\ref{fig:CTET}}.)  
Using the three points on the right to extrapolate linearly backwards, then
somewhere around $R_{lc} = 12 \pm 2 \, a_0$, both the ET and CT times go to zero.  
This corresponds to our earlier conclusion that $R_{lc} \approx 10 \, a_0$ is 
the lower limit of region I.  At the value obtained by our
$\mu$-scan ($R_{lc} = 15 \, a_0$), $\tau_{\text{CT}} = \mbox{ 79 fs}$, in qualitative
agreement with the CIS/AM1 value of $\tau_{\text{CT}} = \mbox{ 164 fs}$
obtained from the single exponential fit or the 
$\tau_{\text{CT}}^{\text{slow}} \approx \mbox{ 91 fs}$
obtained from the double exponential fit.

\begin{figure}
\includegraphics[width=0.9\linewidth]{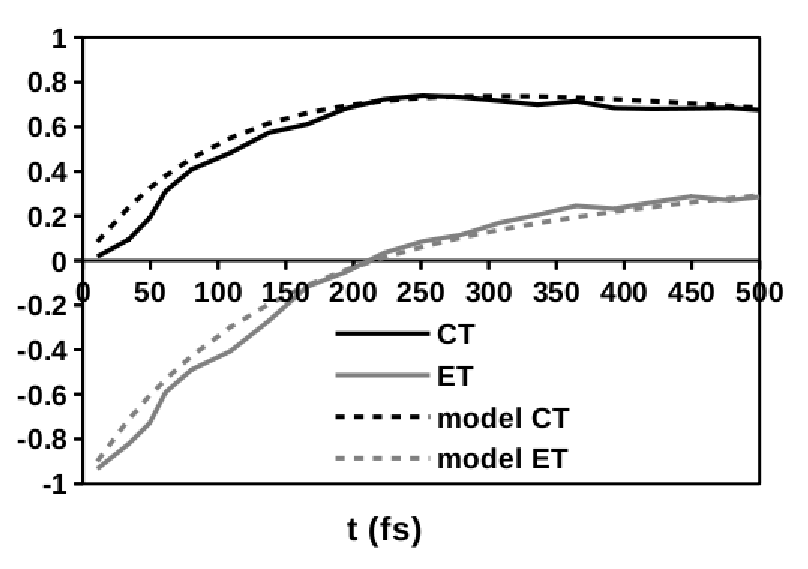} 
\caption{\label{fig:15a0fit} ET and CT curves for $R_{lc} = 15 \, a_0$
fit with Eq.~(\ref{eq:results.8}).
}
\end{figure}

However, there is another kinetics model which provides a significantly better fit
for the $R_{lc} = 15 \, a_0$ case.  This is two independent processes: hole migration,
from {\bf P} to {\bf F} ($k_h=1/\tau_h$) and particle migration from
{\bf P} to {\bf F} ($k_p=1/\tau_p$.)  The corresponding rate laws are
then,
\begin{eqnarray}
  q_h & = & e^{-t/\tau_h} \nonumber \\
  q_p & = & e^{-t/\tau_p} \, ,
  \label{eq:results.8}
\end{eqnarray}
which is the solution of the kinetic problem,
\begin{equation}
  \begin{array}{ccccc}  & k_p &   & k_h &   \\ 
  \mbox{{\bf P}$^*$/{\bf F}} & \longrightarrow & \mbox{{\bf P}$^+$/{\bf F}$^-$}
        & \longrightarrow & \mbox{{\bf P}/{\bf F}$^*$} 
  \end{array} \, ,
  \label{eq:results.9}
\end{equation}
when $k_p >\!\!> k_h$ ($\tau_h >\!\!> \tau_p$.)
A good fit (up to about 700 fs) is obtained with $\tau_p = \mbox{ 111 fs}$
and $\tau_h = \mbox{ 1380 fs}$ as shown in {\bf Fig.~\ref{fig:15a0fit}}.

We see that neither ET nor CT occur when $R_{lc} < 10 \, a_0$, but that both ET and CT
occur when $R_{lc} > 10 \, a_0$.  ET and CT relaxation times
are on the order of 100-300 fs in this range with values of around 100 fs being
our preferred best estimate based upon our $\mu$-scan value of $R_{lc} = 15 \, a_0$,
consistent with the CT relaxation time obtained from CIS/AM1.  At $R_{lc} = 20 \, a_0$
and $R_{lc} = 25 \, a_0$, the observed kinetics is in semi-quantitative agreement
with a mechanism where CT follows as a second step after an initial ET step, while
at $R_{lc} = 15 \, a_0$, the best agreement is found with a kinetic model involving
rapid particle transfer and slow hole transfer.  In terms of the ``polaron formation''
picture, rapid electron transfer is too rapid to delocalize the hole very much while
slower electron transfer gives the hole time to delocalize partially off of {\bf P}
before relocalizing back onto {\bf P} again.

\section{Conclusion}
\label{sec:conclude}

Our interest in TD-lc-DFTB is motivated by potential applications in the field
of organic electronics.  The present work may be regarded as a continuation
of our other work in this field \cite{DCT+15,TCC+17,DCJ+18,TCDA18,TZCA19}.  
In particular, this is the third in a series of papers \cite{DCT+15,DCJ+18} 
aimed at exploring the use of density-functional tight-binding (DFTB) for 
investigating ET and CT for molecules and assemblies of molecules typical 
of organic electronics.

The choice of TD-lc-DFTB is governed, in part, by practicality.  FSSH photodynamics 
calculations easily become very resource intensive, so computational efficiency is 
important even when dealing with only moderately large systems.  
DFTB has recently gained an immense popularity as demonstrated by 
DFTB options in increasingly many major quantum chemistry 
codes.  This is because DFTB is designed to behave like DFT, the dominant workhorse for 
routine quantum chemistry calculations these days, but DFTB has better scalability than 
DFT with respect to the number of atoms in our system.  
DFTB accomplishes this by
borrowing approximation methods from the toolbox of semi-empirical methods and adding
some of its own.  

We emphasize that technically DFTB is an approximate form of DFT rather
than a semi-empirical theory because it is parameterized to fit DFT rather than 
parameterized using experimental data.  However many of the approximations made in 
DFTB are inherited from earlier work on semi-empirical methods.  This is even more 
true once a long-range correction is introduced because of the need to include 
Hartree-Fock exchange for which older semi-empirical methodology is well-developed.

The conventional six-step model for the physics of organic solar cells was 
presented in the introduction.  We would like to explore phenomena which happen
on the scale of hundreds of femtoseconds.  From Table~\ref{tab:SixStepModel},
this is mainly step (iii) CT.  However we might also get some insight into 
the initial physics of longer steps such as (ii) exciton diffusion 
and (iv) charge separation.
We know from our knowledge of DFT that we will need a dispersion 
correction and that we should  use an lc functional, especially for TD-DFT
calculations of CT excitations.  Not every computer code has the DFTB 
version of all of these options and can carry out FSSH calculations.  
However the {\sc DFTBaby} TD-lc-DFTB computer code \cite{HM17} that we 
used here is one of the few codes 
to have all of these options (but see also Ref.~\cite{KYNN20}.)

Our objective has been to test the TD-lc-DFTB FSSH method for its description
of CT at a model heterojunction of an organic solar cell.  For this purpose,
we chose a particularly well-studied system.  The 
pentacene ({\bf P})/buckminsterfullerene ({\bf F}) solar cell is not the most
efficient of all the organic solar cells, but it is probably the most studied
organic solar cell both experimentally and theoretically.  It should probably
be emphasized that, while a good deal is already known about this solar cell,
it is also frequently used as a model system for a deeper investigation of
new or older insufficiently understood phenomena.  Thus we should not discount
the possibility of learning something new about this system.
Our testing of TD-lc-DFT focused on a bimolecular model of a {\bf P}/{\bf F}
heterojunction  with an eye to going to yet larger systems with several 
{\bf P} and several {\bf F} molecules.  

Given the reliance of DFTB on semi-empirical technology, we might ask what
we should expect to obtain from TD-lc-DFTB?  
The answer to the question depends, in
part, upon whether it is possible to use a more sophisticated and rigorous
method for the application of interest.  If the answer is ``yes,'' then 
semi-empirical theories have a well-established place as tools for building
understanding by showing what phenomena follow from simple models.
If the answer is ``no,'' then semi-empirical theories provide a way
to extend the more sophisticated and rigorous method beyond its normal
range of applicability.  DFT does this with more resource intensive 
{\em ab-initio} quantum-chemistry methods, and DFTB does this with DFT.
Nevertheless, given the number of approximations which are made, our emphasis
should be on qualitative phenomena and upon trends.  Here we have emphasized
trends in ET and CT times as a function of the range-separation parameter 
$R_{lc}$.

Although {\sc DFTBaby} seemed the aptest starting point for our study, we
still found it necessary to add three improvements.  The first improvement
is the choice of the initial state.  In photochemical applications, the 
initial state is usually thought of as the excitation of a wave packet to
the Franck-Condon point of one of the adiabatic potential energy surface
of one of the excited electronic states.  This would be appropriate if we
believed that the excitation occurred at the heterojunction of the organic
solar cell.  However, the conventional model prescribes that the 
exciton is most probably formed away from the heterojunction and diffuses
to the heterojunction.  We have therefore modified the program to model
the arrival of the {\bf P}$^*$ exciton at the interface by the projection
of the {\bf P}$^*$ excitation onto the {\bf P}$^*$/{\bf F} model system.
Note that, while this may be a rare choice of FSSH initial condition,  it is 
also completely in line with applications anticipated by Tully who allows
an arbitrary initial state.

The second improvement has been the implementation of the decoherence
correction of Ref.~\cite{GP07}.  This solves (or at least reduces the
effects of) several known problems caused by overcoherence in 
FSSH \cite{T90,LS11,LS12,LFS13}.  
In particular, we have shown by explicit CIS/AM1 FSSH calculations that
the surface and the electronic wavevector methods become essentially
indistinguishable when it comes to calculating ensemble-averaged
electronic expectation values.  This is important for the third 
improvement.

This third improvement is the implementation of a way to automatically
calculate the charge of excited electrons (particles) and holes on each
of the fragments.  This gives us a direct definition of CT and ET.
We note that it is not the only definition as another definition is 
possible on the basis of Kasha's exciton model \cite{KRE65,PL12,DCJ+18}
which can account for the exchange of charge between neighboring molecules
even when no net charge transfer occurs.  However our direct approach
seems better adapted to the needs of organic electronics where the primary
concern is the movement of real net physical charges.

Our {\bf P}/{\bf F} model is already beginning to be rather large (96 atoms) for 
studying CT and ET by the mixed quantum/classical trajectory-based fewest-switches 
surface hopping (FSSH) method which requires many thousands of electronic structure 
calculations.  Nevertheless both the CIS/AM1 FSSH and TD-lc-DFTB FSSH methods
showed themselves to be suitable for meeting this challenge.  Thus we were able
to calculate ensemble averages for up to 500 to 800 fs with reasonably good ensemble
statistics over 50 to 100 trajectories.  This may be compared with the two
previous FSSH studies that we know for {\bf P}/{\bf F} systems: The TD-$\omega$B97X-D FSSH
study from the Br\'edas group \cite{JRB17} used a more accurate electronic 
structure method than the one used here but was limited to 100 fs and 7 trajectories.
The periodic TD-DFT FSSH calculations by the  Prezhdo group \cite{AP14} treated
a larger, more realistic, heterojunction model but made several severe approximations,
including the assumption that the excited-state trajectories follow the same paths
as ground-state thermal trajectories and several uses of the independent particle
approximation.  The present study represents a compromise between the accuracy of 
the electronic structure method and the need to run enough trajectories to get reasonably
good ensemble averages, while still maintaining the basic structure of a rigorous
TD-DFT FSSH calculation.

As we have emphasized, lc-DFTB is an active area of methods development.  As it stands,
the structure of the TD-lc-DFTB differs enough from that of TD-lc-DFT that, while
$R_{lc}$ is often similar for the two methods, it need not always be similar.  Indeed
the present work shows that a much larger value of $R_{lc}$ is needed in our 
TD-lc-DFTB calculations than might have been expected based upon experience with
TD-lc-DFT.  Examination of calculated spectra show that small values of $R_{lc}$
behave like TD-HF or CIS in that the {\bf P}$^+$/{\bf F}$^-$ peak is higher in energy
than the {\bf P}$^*$/{\bf F} peak.  Increasing $R_{lc}$ beyond 10 $a_0$ leads to
the lowering of the energy of the {\bf P}$^+$/{\bf F}$^-$ peak to below that of the
{\bf P}$^*$/{\bf F} peak, in agreement with the spectra of high-quality {\em ab-initio} 
calculations.  A simple chemical kinetics model allowed us to extract ET and CT
times as a function of $R_{lc}$, hence anticipating that the choice of $R_{lc}$
might be tuned to achieve ET and CT times obtained from high-quality {\em ab-initio} 
FSSH calculations.  Unfortunately, such calculations are much too resource intensive
to be practical at the present time.  Nevertheless we did carry out CIS/AM1 FSSH
calculations, a method whose parameters are fit to experiment, whose spectra show 
the same qualitatively correct ordering of 
{\bf P}$^+$/{\bf F}$^-$ and {\bf P}$^*$/{\bf F} peaks as in the 
high-quality {\em ab-intio} calculations.  TD-lc-DFTB FSSH CT times were found to
be similar to CIS/AM1 FSSH times for $R_{lc} = 15 \, a0$.

The fact that we can calculate not only net charges but also electron (particle) and
hole charges allowed us to make a remarkable observation of an unexpected yet
physically reasonable phenomenon. 
During the initial {\bf P}$^*$ $\rightarrow$ {\bf F} transfer of an electron, the
positively-charged hole also starts to delocalize, following the negatively-charged
electron.  This delocalization does not last long before the hole relocalizes back
onto {\bf P}, presumably because of electronic and nuclear relaxation effects.
As such, this resembles polaron formation in solids.  This phenomenon only lasts
about 5 fs in our CIS/AM1 calculations but is present.  It is also present in our
TD-lc-DFTB calculations and is of longer duration with the ``polaron formation time''
increasing as $R_{lc}$ increases.

Looking forward to the future, we think that the experience gained through the
present work should allow us to design TD-lc-DFTB FSSH {\em in silico} experiments
for larger cluster models of organic heterojunctions.  Not only could this be
a step towards more realistic models of charge transfer at organic heterojunctions
but it may also allow us to investigate at least the initial steps of exciton
diffusion and charge separation.

\section{Acknowledgements}
\label{sec:thanks}
This work has been supported in part by the French National Research Agency
({\em Agence Nationale de la Recherche}, ANR) ORGAVOLT (ORGAnic solar cell VOLTage)
project ANR-12-MONU-0014-02. Ala Aldin M.\ H.\ M.\ Darghouth acknowledges a 
Franco-Iraqi PhD scholarship administered via the French agency {\em Campus France}  
during the period of his doctoral research and support from the University of Mosul
for funding that allowed him to continue working on this project after taking up
his current position in Iraq.
The authors wish to acknowledge the support from Grenoble Alps University's
ICMG ({\em Institut de Chimie Mol\'eculaire de Grenoble}) Chemistry Nanobio Platform
PCECIC ({\em Plateau du Centre d'Exp\'erimentation et de Calcul Intensif en Chimie})
on which this work has been performed.  Pierre Girard is gratefully acknowledged
for his help and support using this platform.
Alexander Humeniuk and Roland Mitri\'c acknowledge financial support within
the European Research Council (ERC) Consolidator Grant DYNAMO (grant number 646737).
Some of the early work on this project was work done in Singapore where it was
supported, in part, by the Society of Interdisciplinary Research (SOIR\'EE).
Mark E.\ Casida and Ala Aldin M.\ H.\ M.\ Darghouth would also like to acknowledge
a useful trip to W\"urzburg funded by the German GRK 2112 Project ``Biradicals,''
as well as useful discussions with Drs.\ Mathias Rapacioli 
and Hemanadhan Myneni.  Ala Aldin M.\ H.\ M.\ Darghouth acknowledges
having followed an advanced course taught by Prof.\ Mario Barbatti on 
``Theoretical Aspects of Organic Femtochemistry'' together with a tutorial 
on the {\sc Newton-X} program.  We are grateful to Mario Barbatti for 
his extensive comments on an early version of this manuscript.
Felix Plasser is acknowledged for having drawn our attention to Ref.~\cite{PL12}.
Mark E.\ Casida acknowledges an inspiring discussion with Lucia Reining regarding
the place of semi-empirical methods in the theorist's toolbox.
\appendix
\section{Brief Review of DFTB and TD-DFTB}
\label{sec:DFTB}

This appendix contains a very concise review of DFTB and of TD-DFTB as an aid for
understanding the still-rather-new implementation of TD-lc-DFTB used in this work \cite{HM15}.
We give a brief review of
DFTB \cite{KM09,OSHD09,ES14}
and 
TD-DFTB \cite{NSD+01,FSE+02,HNWF07,N09,DAF+13}
and of their relation to 
DFT \cite{PY89,DG90,KH00}
and 
TD-DFT \cite{C09,CH12,U12}.  DFT and TD-DFT are now so well
established that it seems that little needs to be said about them.  
DFTB and TD-DFTB are more recent but would still be 
familiar to experts.  
Here we will concentrate on a concise review of just enough of the basic methodology to 
be able to explain why it is difficult to introduce lc into TD-lc-DFTB and what compromises
have been made in the current methodology. 

In both DFT and DFTB, the total energy is expressed as,
\begin{equation}
  E = E_{\text{BS}} + E_{\text{rep}} \, ,
  \label{eq:theory.1}
\end{equation}
where the band-structure term is the occupation-number weighted sum of Kohn-Sham orbital energies,
\begin{equation}
  E_{\text{BS}} = \sum_i n_i \epsilon_i \, ,
  \label{eq:theory.2}
\end{equation}
and the repulsion potential term in DFT is,
\begin{equation}
  E_{\text{rep}} = E_H[\rho] + E_{xc}[\rho] - \int v_{xc}[\rho](\vec{r}) \, d\vec{r} + V_{n,n} \, ,
  \label{eq:theory.3}
\end{equation}
where $\rho$ is the electron density, $H$ refers to the Hartree energy expression,
$xc$ stands for exchange-correlation terms, and $V_{n,n}$ stands for the nuclear repulsion terms.  
DFTB makes use of several ideas from semi-empirical quantum
chemistry as well as some of its own.  Like semi-empirical methods, DFTB assumes a minimal basis
description of the valence electrons and all the other electrons are treated implicitly as part 
of ionic cores.  In the original form of DFTB \cite{PFK+95}, the total charge density was simply
assumed to be the sum of unperturbed atomic charge densities, 
$\rho^0 = \sum_I \rho_I^0$ and the  Hartree plus xc potential is assumed separable,
$v_{Hxc}[\rho] = \sum_I v_{Hxc}[\rho_I]$.  These two approximations are used,
together with neglect of any three-center terms, to construct the matrix of the Kohn-Sham operator.  
Solving the matrix form of the DFTB Kohn-Sham equation then gives the orbital energies and hence 
the band-structure term in the total energy.  The repulsion potential is assumed to be the sum 
of pairwise interatomic repulsion potentials, $E_{\text{rep}} = \sum_{I<J} V_{I,J}(R_{I,J})$.   
In principle, a new set of pairwise potentials is needed for each new approximate 
density-functional.  In practice, generating these new pairwise potentials for each new
density functional is impractical.

Modern self-consistent charge (SCC) DFTB adds a Coulomb term,
\begin{eqnarray}
  E_{\text{coul}} & = & \frac{1}{2} \int \int \delta \rho(\vec{r}_1)
  \left( f_H(\vec{r}_1,\vec{r}_2) + f_{xc}(\vec{r}_1,\vec{r}_2) \right)
  \nonumber \\
  & \times & \delta \rho(\vec{r}_2) \, d\vec{r}_1 d\vec{r}_2 \, ,
  \label{eq:theory.4}
\end{eqnarray}
to the energy expression of Eq.~(\ref{eq:theory.1}) to account for the fact 
that the charge density is not simply the sum of the unperturbed charge densities
but rather,
\begin{equation}
  \rho(\vec{r}) = \rho^0(\vec{r}) + \delta \rho(\vec{r}) \, .
  \label{eq:theory.5}
\end{equation}
Here $f_H$ and $f_{xc}$ are the second functional derivatives with respect to $\rho$ of
the classical Coulomb (Hartree) repulsion and $E_{xc}$ respectively.  
Mulliken's integral approximation and a monopole
expansion of products of molecular orbitals in terms of atom-centered $s$-type functions 
$g_I(\vec{r})$, then allows $\delta \rho$ to be replaced by Mulliken charges 
and the essential Coulomb integral in Eq.~(\ref{eq:theory.4}) becomes,
\begin{equation}
  E_{\text{coul}} = \frac{1}{2} \sum_{I,J} \Delta q_I \gamma_{I,J} \Delta q_J \, ,
  \label{eq:theory.6}
\end{equation}
where,
\begin{equation}
  \gamma_{I,J} = \int \int g_I(\vec{r}_1) \left( f_H(\vec{r}_1,\vec{r}_2) + f_{xc}(\vec{r}_1,\vec{r}_2) \right) g_J(\vec{r}_2) \, d\vec{r}_1 d\vec{r}_2 \, ,
  \label{eq:theory.7}
\end{equation}
and the $\Delta q_I$ are Mulliken charge differences.
This has a couple of interesting consequences.  The first is that variational
minimization of the three term energy expression results in a Kohn-Sham orbital
equation which must now be solved self-consistently because the Kohn-Sham
operator depends upon the Mulliken charges which are themselves calculated
from the molecular orbital coefficients (hence SCC-DFTB.)

The second interesting consequence is that it is now possible to set up and solve the TD-DFTB
analogue \cite{NSD+01} of Casida's equation \cite{C95},
\begin{equation}
  \left[ \begin{array}{cc} {\bf A} & {\bf B} \\ {\bf B}^* & {\bf A}^*
  \end{array} \right] \left( \begin{array}{c} \vec{X}_I \\ \vec{Y}_I
  \end{array} \right) = \omega_I 
   \left[ \begin{array}{cc} {\bf 1} & {\bf 0} \\ {\bf 0} & -{\bf 1}
  \end{array} \right] \left( \begin{array}{c} \vec{X}_I \\ \vec{Y}_I
  \end{array} \right),
  \label{eq:theory.8}
\end{equation}
where,
\begin{eqnarray}
  A_{ia\sigma,jb\tau} & = & \delta_{i,j} \delta_{a,b} \delta_{\sigma,\tau}
  \left( \epsilon_{a\sigma} - \epsilon_{i\sigma} \right) 
   + K_{ia\sigma,jb\tau} \nonumber \\
  B_{ia\sigma,jb\tau} & = & K_{ia\sigma,bj\tau} \, .
  \label{eq:theory.9}
\end{eqnarray}
Here $\omega_I$ is the electronic excitation energy and the $(\vec{X}_I,\vec{Y}_I)$
are used for calculating spectra oscillator strengths and nonadiabatic coupling
elements.  In TD-DFTB,
\begin{equation}
  K_{ia\sigma,jb\tau} = \sum_{I,J} q_I^{ia} \gamma^{\sigma,\tau}_{I,J} q_I^{jb} 
  \, ,
  \label{eq:theory.10}
\end{equation}
where the $q_I^{ia}$ are Mulliken transition charges.

\section{Brief Review of TD-DFT FSSH}
\label{sec:FSSH}

This appendix provides a very brief review of how TD-DFT is usually combined
with Tully's molecular dynamics with quantum transitions (nowadays often
simply referred to as FSSH) \cite{T90,HT94}.  As FSSH is now widely known, it is
not our intention to review that method here.  Instead, the reader is referred 
to a recent review \cite{CB18}.

The first implementation of TD-DFT FSSH was due to Tapavicza, Tavernelli,
and R\"othlisberger in 2007 \cite{TTR07} in a development version of the
{\sc CPMD} code.  They proposed that the nonadiabatic coupling be calculated
using Casida's {\em Ansatz} which was originally intended as an aid for
assigning TD-DFT excited states \cite{C95}.  Specifically, an excited-state
wave function
\begin{equation}
  \Psi_I = \sum_{i,a,\sigma} \Phi_{i\sigma}^{a\sigma} C_{ia\sigma} \, ,
  \label{eq:theory.16}
\end{equation}
made up of singly excited determinants $\Phi_{i\sigma}^{a\sigma}$
(corresponding to the $i\sigma \rightarrow a\sigma$ excitation)
is postulated and it is argued that
\begin{equation}
  C_{ia\sigma}^I = \sqrt{\frac{\epsilon_{a\sigma}-\epsilon_{i\sigma}}{\omega_I}}
   F_{ia\sigma}^I \, ,
  \label{eq:theory.17}
\end{equation}
where
\begin{equation}
  \vec{F}_{I} \propto \left( {\bf A} - {\bf B} \right)^{-1/2}
          \left( \vec{X}_I + \vec{Y}_I \right) \, ,
  \label{eq:theory.18}
\end{equation}
is renormalized so that,
\begin{equation}
  \vec{F}_I^\dagger \vec{F}_I = 1 \, .
  \label{eq:theory.19}
\end{equation}
Then Eqs.~(\ref{eq:theory.16}) and (\ref{eq:theory.17}) are combined with 
\begin{eqnarray}
  d_{I,J}\left({\bf R}(t+\frac{\Delta}{2})\right) & = & \frac{1}{2\Delta} \left[
  \langle \Psi_I({\bf R}(t)) \vert \Psi_J({\bf R}(t+\Delta) \rangle \right. \nonumber \\
  & - & \left . \langle \Psi_I({\bf R}(t+\Delta) \vert \Psi_J({\bf R}(t) \rangle \right]
  \, ,
  \label{eq:theory.19b}
\end{eqnarray}
to obtain the nonadiabatic coupling (NAC) elements.
This leads to a linear combination of overlap terms between two Slater determinants
at different times which is evaluated using the observation 
that the overlap of two Slater determinants is the determinant of overlap integrals \cite{L55}.
Their implementation was followed by an application
to the photochemical ring opening of oxirane \cite{TTR+08} which showed that
the nonexistence of a proper conical intersection in conventional TD-DFT
\cite{LKQM06} was not a serious practical problem for TD-DFT FSSH.  TD-DFT
FSSH has also been implemented in a version of {\sc TurboMol} capable of
calculating nonadiabatic coupling elements analytically \cite{SF10} and
this was applied early on to study the photochemistry of vitamin-D \cite{TMF10}.

\section{Algorithm for Calculating $p$ and $h$ Charges}
\label{sec:charges}

The particle density matrix ${\bf \gamma}^p$ and the hole 
density matrix ${\bf \gamma}^h$ are easy to calculate
using second quantization.  We will use the common molecular orbital (MO) index
convention,
\begin{equation}
   \underbrace{a, b, c, \cdots, g, h}_{\mbox{unoccupied}}
   \underbrace{i, j, k, l, m, n}_{\mbox{occupied}}
   \underbrace{o, p, q, \cdots, x, y, z}_{\mbox{free}} \, .
   \label{eq:theory.24} 
\end{equation}
The physical vacuum $\vert 0 \rangle$ is taken to be the ground state single determinant.
Casida's {\em Ansatz} takes the form of a singles configuration interaction (CIS) wave 
function \cite{C95},
\begin{equation}
  \vert I \rangle = \sum_{i,a} a^\dagger i \vert 0 \rangle C_{i,a} \, . 
  \label{eq:theory.25} 
\end{equation}
Specifically,
\begin{equation}
  C_{i,a} = {\cal N} \left(X_{i,a}+Y_{i,a}\right) \, ,
  \label{eq:theory.25b} 
\end{equation}
and the normalization constant ${\cal N}$ is chosen to that
\begin{equation}
  \sum_{i,a} \vert C_{i,a} \vert^2 = 1 \, .
  \label{eq:theory.26} 
\end{equation}
The hole density matrix in the MO representation is,
\begin{eqnarray}
  \gamma^h_{k,l} & = & \langle I \vert l k^\dagger \vert I \rangle
                 = \sum_a C^*_{l,a} C_{k,a} \nonumber \\
  {\bf \gamma}^h & = & {\bf C} {\bf C}^\dagger 
  \, ,
  \label{eq:theory.27} 
\end{eqnarray}
and the particle density matrix in the MO representation is,
\begin{eqnarray}
  \gamma^p_{c,d} & = & \langle I \vert d^\dagger c \vert I \rangle 
                 = \sum_i C^*_{i,d} C_{i,c} \nonumber \\
  {\bf \gamma}^p & = & {\bf C}^\dagger {\bf C} 
  \, .
  \label{eq:theory.28} 
\end{eqnarray}

Finally we must calculate the particle and hole charges on each fragment.
This is done in the usual way using Mulliken charges.  Atom-centered
basis functions are labeled with lower case Greek indices.  The overlap matrix is,
\begin{equation}
  S_{\mu,\nu} = \langle \mu \vert \nu \rangle \, .
  \label{eq:theory.29} 
\end{equation}
The density matrices $\gamma$ are converted into ``charge and bond order''
density matrices $P$ (capital $\rho$) by the usual transformation,
\begin{eqnarray}
  P_{\mu,\nu} & = & \sum_{r,s} c_{\mu,r} \gamma_{r,s} c_{\nu,s}^* \nonumber \\
  {\bf P}  & = & {\bf c} {\bf \gamma} {\bf c}^\dagger \, .
  \label{eq:theory.30} 
\end{eqnarray}
The final equations that we seek are then,
\begin{eqnarray}
   q_h^P & = & \sum_{\mu \in P} \sum_\nu S_{\mu,\nu} P_{\mu,\nu}^h \nonumber \\
   q_p^P & = & \sum_{\mu \in P} \sum_\nu S_{\mu,\nu} P_{\mu,\nu}^p \nonumber \\
   q_h^F & = & \sum_{\mu \in F} \sum_\nu S_{\mu,\nu} P_{\mu,\nu}^h \nonumber \\
   q_p^F & = & \sum_{\mu \in F} \sum_\nu S_{\mu,\nu} P_{\mu,\nu}^p \, .
   \label{eq:theory.31} 
\end{eqnarray}

\section*{Conflicts of Interest}
The authors declare no conflict of interest.
\section*{Supplemental Information}
\label{sec:SI}

\begin{figure}
\includegraphics[width=0.6\textwidth]{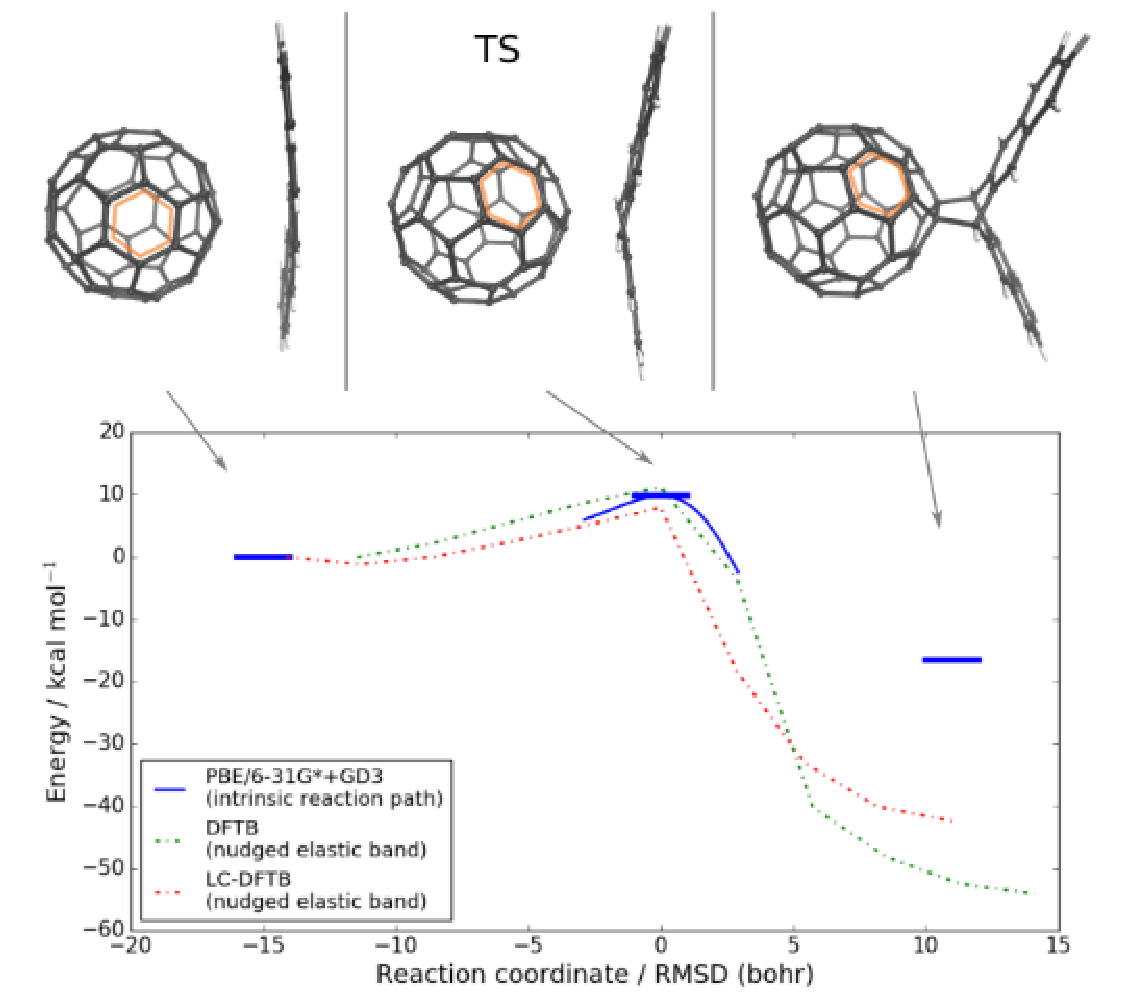} 
\caption{\label{fig:PES}
This figure shows that to get from the reactant minimum to the transition 
state the C$_{60}$ molecule has to rotate around the axis perpendicular to the 
pentacene.  This rotation can be seen by concentrating on the orange hexagon.   
The  barrier  height  is  reproduced  reasonably  well  by  DFTB  (both  with  
and  without lc), but the reaction is predicted to be far too exoergic.  The 
reason why the reaction is observed in the simulations is probably that the 
initial geometry is already close to the transition state, as the C$_{60}$
molecule has the correct orientation for the ring closing reaction to happen.
{\bf Computational details}: {\em PBE/6-31G*+GD3}.  Optimization of the reactant
and the product of the electrocyclic addition reaction was carried out at the AM1 level.
Further optimization of educt and product at the PBE/6-31G* + GD3 (Grimme's dispersion 
correction).  A frequency calculation confirmed that both stationary points are 
minima (all frequencies positive). Search for transition state connecting the
reactant and product, a frequency calculation produced exactly one negative frequency.
The intrinsic reaction coordinate was followed starting from the transition state 
in both directions. {\em {\sc DFTB} with and without lc.}  To  locate  the  
transition  state  with  tight-binding  DFT  the  nudged  elastic  band  
algorithm  
was employed. The  initial  guess  for  the  
path  was  generated  by  interpolating  10  geometries  linearly  between  the
PBE-optimized educt and product geometries.  The nudged elastic band was then 
allowed to relax until the total forces on all images were below 0.02 (atomic units).
}
\end{figure}

\begin{itemize}
\item[$\bullet$] C$_{60}$ + pentacene electrocyclic addition reaction.  Figure~\ref{fig:PES}.
\item[$\bullet$] Geometry C for {\bf P}/{\bf F}.  
\begin{table}
\begin{tabular}{rcccc}
\hline \hline
Atom & Type & $x$ & $y$ & $z$ \\
\hline
 1 & C &  6.389931 &  3.367151 &  1.602947 \\
 2 & C &  6.380509 &  3.717252 &  0.199539 \\
 3 & C &  6.271531 &  4.421802 &  2.571059 \\
 4 & C &  6.497451 &  2.042099 &  1.971424 \\
 5 & C &  6.252930 &  5.102553 & -0.158302 \\
 6 & C &  6.480776 &  2.720966 & -0.749282 \\
 7 & C &  6.141263 &  6.061830 &  0.792373 \\
 8 & C &  6.150713 &  5.714740 &  2.183638 \\
 9 & C &  6.578158 &  1.008483 &  1.012835 \\
10 & C &  6.573813 &  1.358384 & -0.389437 \\
11 & C &  6.595313 & -0.340195 &  1.375643 \\
12 & C &  6.603895 &  0.339694 & -1.343713 \\
13 & C &  6.574315 & -1.358957 &  0.421169 \\
14 & C &  6.587548 & -1.009050 & -0.981023 \\
15 & C &  6.482189 & -2.721740 &  0.780395 \\
16 & C &  6.516248 & -2.042847 & -1.940144 \\
17 & C &  6.391222 & -3.718300 & -0.169084 \\
18 & C &  6.409622 & -3.368179 & -1.572400 \\
19 & C &  6.265020 & -5.103950 &  0.187888 \\
20 & C &  6.301012 & -4.423167 & -2.541291 \\
21 & C &  6.162846 & -6.063553 & -0.763529 \\
22 & C &  6.181147 & -5.716445 & -2.154696 \\
23 & H &  6.277117 &  4.155348 &  3.623724 \\
24 & H &  6.494281 &  1.778112 &  3.025515 \\
25 & H &  6.244366 &  5.361629 & -1.212805 \\
26 & H &  6.466770 &  2.982955 & -1.803701 \\
27 & H &  6.042557 &  7.104209 &  0.507990 \\
28 & H &  6.058911 &  6.501105 &  2.925532 \\
29 & H &  6.576440 & -0.603747 &  2.429472 \\
30 & H &  6.591655 &  0.603192 & -2.397651 \\
31 & H &  6.461488 & -2.983749 &  1.834699 \\
32 & H &  6.519808 & -1.778877 & -2.994237 \\
33 & H &  6.249754 & -5.363039 &  1.242312 \\
34 & H &  6.313308 & -4.156715 & -3.593900 \\
35 & H &  6.065140 & -7.106210 & -0.479820 \\
36 & H &  6.096859 & -6.503072 & -2.897203 \\
37 & C &  3.555957 &  0.239775 &  0.650973 \\
38 & C &  3.118066 &  1.595476 &  0.913217 \\
39 & C &  2.390413 &  1.593252 &  2.166840 \\
40 & C &  2.385538 &  0.235533 &  2.679753 \\
41 & C &  3.110446 & -0.600480 &  1.742993 \\
42 & C &  2.687868 & -1.896011 &  1.492289 \\
43 & C &  1.509958 & -2.411476 &  2.162283 \\
44 & C &  0.814186 & -1.612660 &  3.056325 \\
45 & C &  1.262343 & -0.259492 &  3.321751 \\
46 & C &  0.089981 &  0.579658 &  3.479296 \\
47 & C &  0.094278 &  1.875677 &  2.989020 \\
48 & C &  1.271099 &  2.395233 &  2.318892 \\
49 & C &  0.829409 &  3.236018 &  1.223271 \\
50 & C &  1.527676 &  3.238879 &  0.025756 \\
\hline \hline
\end{tabular}
\caption{
\label{tab:geomCa}
Cartesian coordinates in {\AA}ngstr\"om for geometry C, part I.
}
\end{table}
\begin{table}
\begin{tabular}{rcccc}
\hline \hline
Atom & Type & $x$ & $y$ & $z$ \\
\hline
51 & C &  2.700948 &  2.402154 & -0.132490 \\
52 & C &  2.705504 &  1.888765 & -1.489874 \\
53 & C &  1.534526 &  2.407810 & -2.169215 \\
54 & C &  0.806748 &  3.242683 & -1.232778 \\
55 & C & -0.578874 &  3.243890 & -1.238286 \\
56 & C & -1.308974 &  3.241331 &  0.014240 \\
57 & C & -0.620533 &  3.236268 &  1.216594 \\
58 & C & -1.074838 &  2.396430 &  2.308071 \\
59 & C & -2.196428 &  1.598707 &  2.148315 \\
60 & C & -2.200633 &  0.242566 &  2.661161 \\
61 & C & -1.082925 & -0.255376 &  3.310897 \\
62 & C & -0.635441 & -1.609747 &  3.049361 \\
63 & C & -1.326072 & -2.406075 &  2.150397 \\
64 & C & -0.598519 & -3.241101 &  1.214602 \\
65 & C &  0.787100 & -3.244115 &  1.220078 \\
66 & C &  1.518034 & -3.242556 & -0.032678 \\
67 & C &  0.829308 & -3.237556 & -1.235699 \\
68 & C & -0.620637 & -3.233349 & -1.240526 \\
69 & C & -1.318620 & -3.236292 & -0.043688 \\
70 & C & -2.490858 & -2.397619 &  0.114094 \\
71 & C & -2.495631 & -1.884682 &  1.470312 \\
72 & C & -2.923016 & -0.590605 &  1.719909 \\
73 & C & -3.365380 &  0.250785 &  0.625010 \\
74 & C & -2.916148 &  1.603591 &  0.889648 \\
75 & C & -2.482467 &  2.406215 & -0.152883 \\
76 & C & -2.478003 &  1.893252 & -1.509100 \\
77 & C & -1.301513 &  2.411077 & -2.179849 \\
78 & C & -0.606203 &  1.612625 & -3.073315 \\
79 & C &  0.843441 &  1.611129 & -3.068777 \\
80 & C &  1.289575 &  0.256595 & -3.330623 \\
81 & C &  0.115960 & -0.578975 & -3.497492 \\
82 & C & -1.055715 &  0.259632 & -3.338428 \\
83 & C & -2.180082 & -0.234889 & -2.697611 \\
84 & C & -2.907394 &  0.600502 & -1.762135 \\
85 & C & -3.360991 & -0.239522 & -0.670789 \\
86 & C & -2.913796 & -1.593706 & -0.931849 \\
87 & C & -2.184068 & -1.591018 & -2.184749 \\
88 & C & -1.063674 & -2.392158 & -2.335588 \\
89 & C &  0.112406 & -1.874989 & -3.007208 \\
90 & C &  1.282274 & -2.398119 & -2.327741 \\
91 & C &  2.402809 & -1.599591 & -2.166810 \\
92 & C &  2.406110 & -0.241853 & -2.679710 \\
93 & C &  3.126068 &  0.591934 & -1.737197 \\
94 & C &  3.560386 & -0.249678 & -0.641684 \\
95 & C &  3.120513 & -1.604030 & -0.907438 \\
96 & C &  2.692600 & -2.409435 &  0.134922 \\
\hline \hline
\end{tabular}
\caption{
\label{tab:geomCb}
Cartesian coordinates in {\AA}ngstr\"om for geometry C, part II.
}
\end{table}
\item[$\bullet$] Transition density matrix analysis of some of our {\em ab-initio} calculations.
                 Following
                 \begin{quote}
                 \noindent
                 S.\ Tretiak and S.\ Mukamel, ``Density Matrix Analysis and Simulation of 
                 Electronic Excitations in Conjugated and Aggregated Molecules'', {\em 
                 Chem. Rev.} {\bf 9}, 3171 (2002)
                 \end{quote}
                 and
                 \begin{quote}
                 \noindent
                 P.\ Plasser and H.\ Lischka, ``Analysis of Excitonic and Charge Transfer 
                 Interactions from Quantum Chemical Calculations'', {\em J. Chem. Theory
                 Comput.} {\bf 8}, 2777 (2012),
                 \end{quote}
                 suppose that we have a CIS-type expansion for the $I$th excited state in the
                 molecular orbital (MO) representation,
\begin{equation}
  \vert I \rangle = \sum_{i,a} C^I_{i,a} a^{\dagger} i \vert 0 \rangle \, ,
  \label{eq:SI.2}
\end{equation}
                 then the transition density matrix (TDM) in the atomic orbital (AO) 
                 representation is,
\begin{equation}
  P_{\mu,\nu} = \sum_{i,a} c_{\mu,i} C^{I,*}_{i,a} c_{\nu,a}^* \, ,
  \label{eq:SI.3}
\end{equation}
                 where ${\bf c}$ is the matrix of MO coefficients.  It is actually the 
                 ``fraction of transition density matrix'' (FTDM) which is displayed,
\begin{equation}
  \mbox{FTDM}_{X,Y} = \frac{\sum_{\mu \in X} \sum_{\nu \in Y} P_{\mu,\nu}^2}
     {\sum_{\mu \in \text{dimer}} \sum_{\nu \in \text{dimer}} P_{\mu,\nu}^2}
   \, ,
  \label{eq:SI.4}
\end{equation}
where $X$ and $Y$ refer to fragments ({\bf P}, {\bf F}).  
The FTDM is a $2 \times 2$ matrix whose diagonal elements
correspond to local excitations and whose off-diagonal elements refer to charge-transfer
excitation in the sense of Kasha's exciton model.  In the figures, the matrix elements
are printed if they are larger than 0.09.  Figures~\ref{fig:CC_TMA} and \ref{fig:ADC_TMA}.
\item[$\bullet$] More elaborate kinetics model:
\begin{equation}
   \mbox{{\bf A}} 
   \begin{array}[c]{c} k_{\text{\bf AB}} \\ \longrightarrow \\ \longleftarrow \\ k_{\text{\bf BA}} \end{array}
   \mbox{{\bf B}} 
   \begin{array}[c]{c} k_{\text{\bf BC}} \\ \longrightarrow \\ \longleftarrow \\ k_{\text{\bf BC}} \end{array}
   \mbox{{\bf C}}
   \, ,
  \label{eq:SI.1}
\end{equation}
with {\bf A} = {\bf P}$^*$/{\bf F}, {\bf B} = {\bf P}/{\bf F}$^*$, and {\bf C} = {\bf P}$^+$/{\bf F}$^-$.
Fitting results are given in Figures~\ref{fig:CISAM1}, \ref{fig:10a0}, \ref{fig:15a0}, \ref{fig:20a0},
and \ref{fig:25a0}.
\end{itemize}
\begin{figure}
\includegraphics[width=0.3\textwidth]{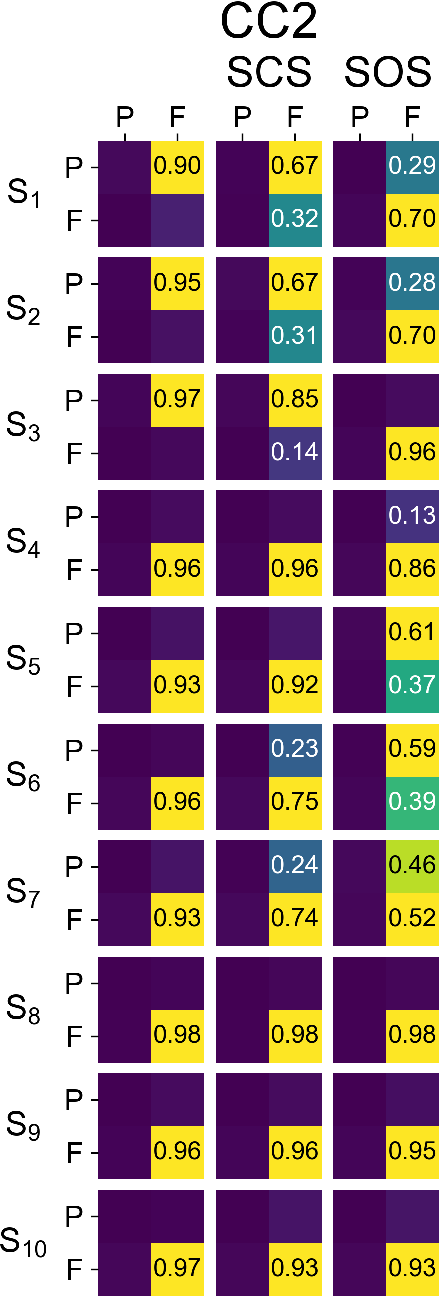} 
\caption{\label{fig:CC_TMA}
Transition density matrix analysis of the first 10 excited states of our CC2, SCS-CC2,
and SOS-CC2 calculations for geometry {\bf C}.
}
\end{figure}
\begin{figure}
\includegraphics[width=0.5\textwidth]{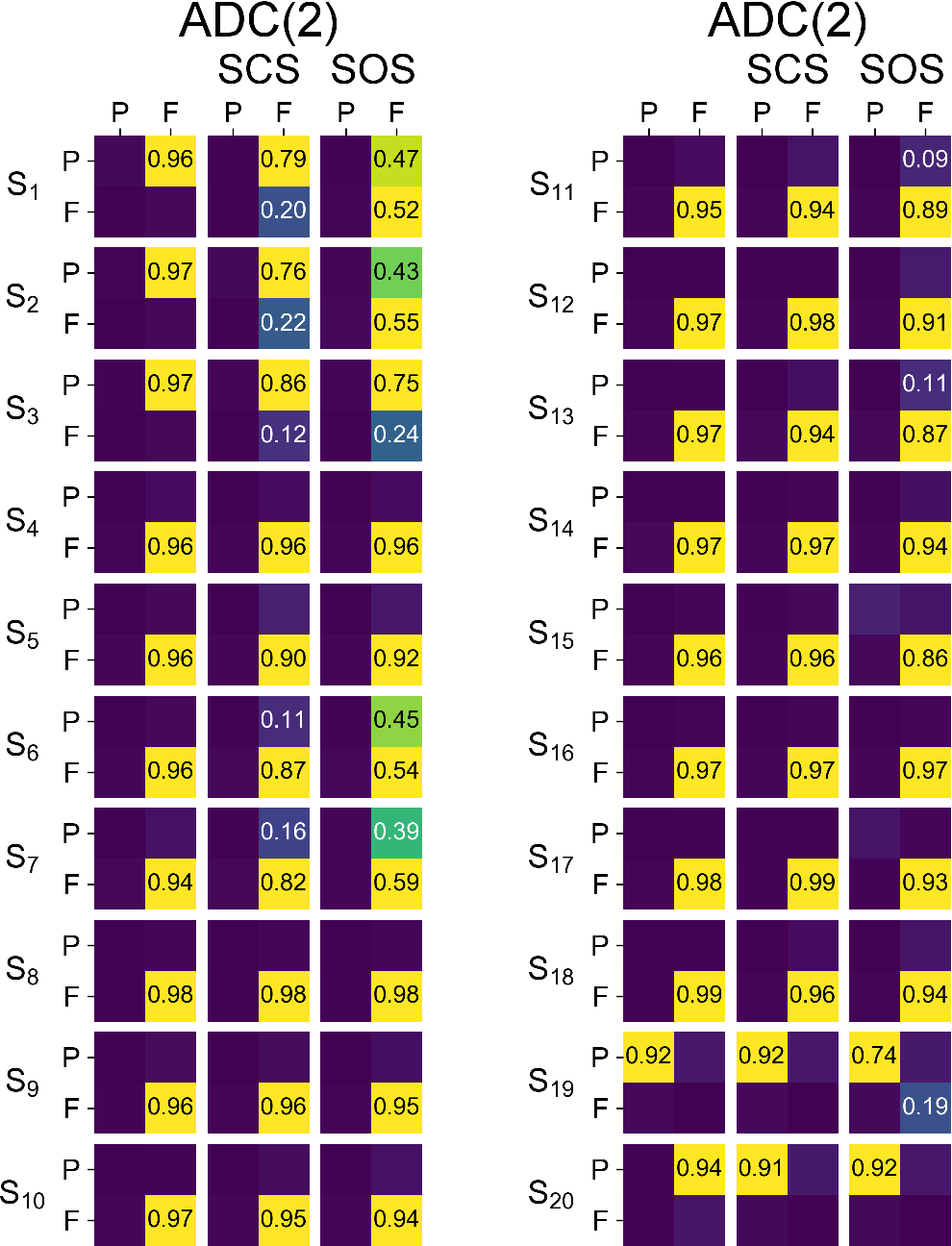}
\caption{\label{fig:ADC_TMA}
Transition density matrix analysis of the first 20 excited states of our ADC(2), SCS-ADC(2),
and SOS-ADC(2) calculations for geometry {\bf C}.
}
\end{figure}
\onecolumngrid

\newpage
\begin{figure}
\includegraphics[width=1.0\textwidth]{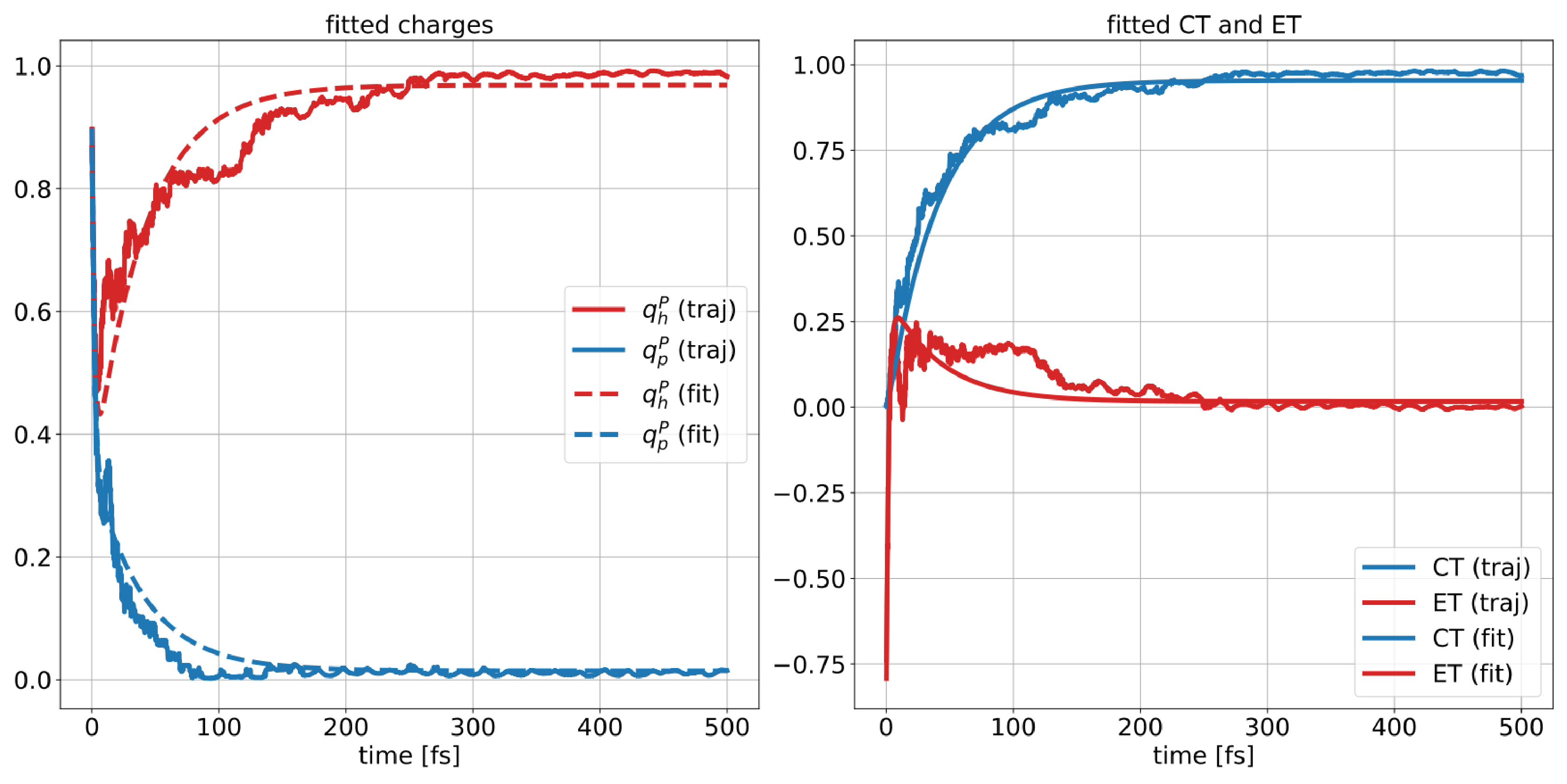}
\caption{\label{fig:CISAM1}
Kinetics fit to the CIS/AM1 FSSH results: 
$k_{\text{\bf AB}} = \mbox{0.320 fs$^{-1}$}$ ($\tau_{\text{\bf AB}} = \mbox{3.12 fs}$),
$k_{\text{\bf BA}} = \mbox{0.150 fs$^{-1}$}$ ($\tau_{\text{\bf BA}} = \mbox{6.67 fs}$),
$k_{\text{\bf BC}} = \mbox{0.0357 fs$^{-1}$}$ ($\tau_{\text{\bf BC}} = \mbox{28.0 fs}$), and
$k_{\text{\bf CB}} = \mbox{0.00117 fs$^{-1}$}$ ($\tau_{\text{\bf CB}} = \mbox{855 fs}$).
}
\end{figure}
\begin{figure}
\includegraphics[width=1.0\textwidth]{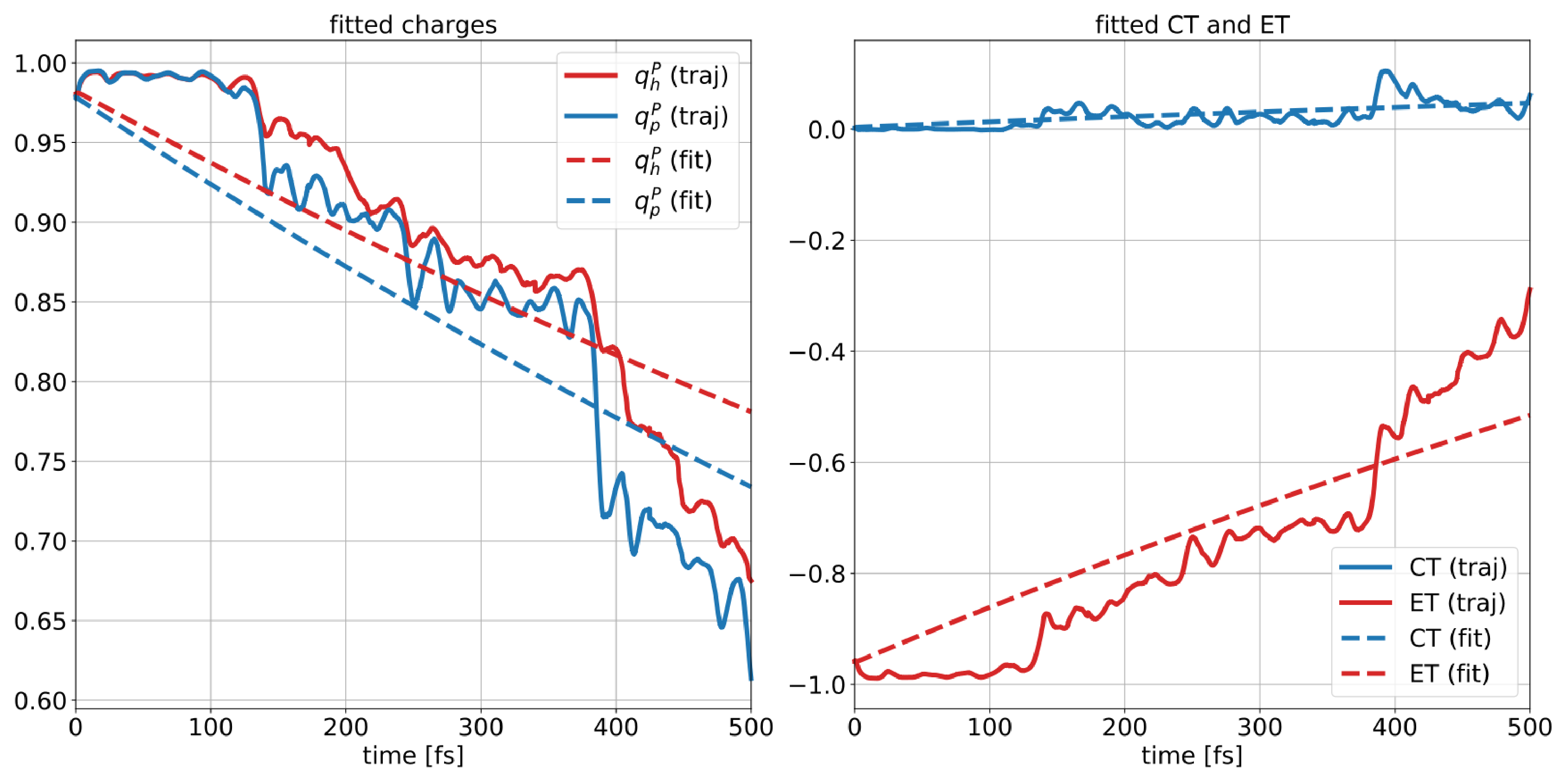} 
\caption{\label{fig:10a0}
Kinetics fit to the TD-lc-DFTB FSSH results with $R_{lc} = 10 \, a_0$: 
$k_{\text{\bf AB}} = \mbox{0.000576 fs$^{-1}$}$ ($\tau_{\text{\bf AB}} = \mbox{1 740 fs}$),
$k_{\text{\bf BA}} = \mbox{1. $\times$ 10$^{-10}$ fs$^{-1}$}$ 
($\tau_{\text{\bf BA}} = \mbox{1. $\times$ 10$^{+10}$ fs}$),
$k_{\text{\bf BC}} = \mbox{0.185 fs$^{-1}$}$ ($\tau_{\text{\bf BC}} = \mbox{5.41 fs}$), and
$k_{\text{\bf CB}} = \mbox{0.856 fs$^{-1}$}$ ($\tau_{\text{\bf CB}} = \mbox{1.17 fs}$).
}
\end{figure}
\begin{figure}
\includegraphics[width=1.0\textwidth]{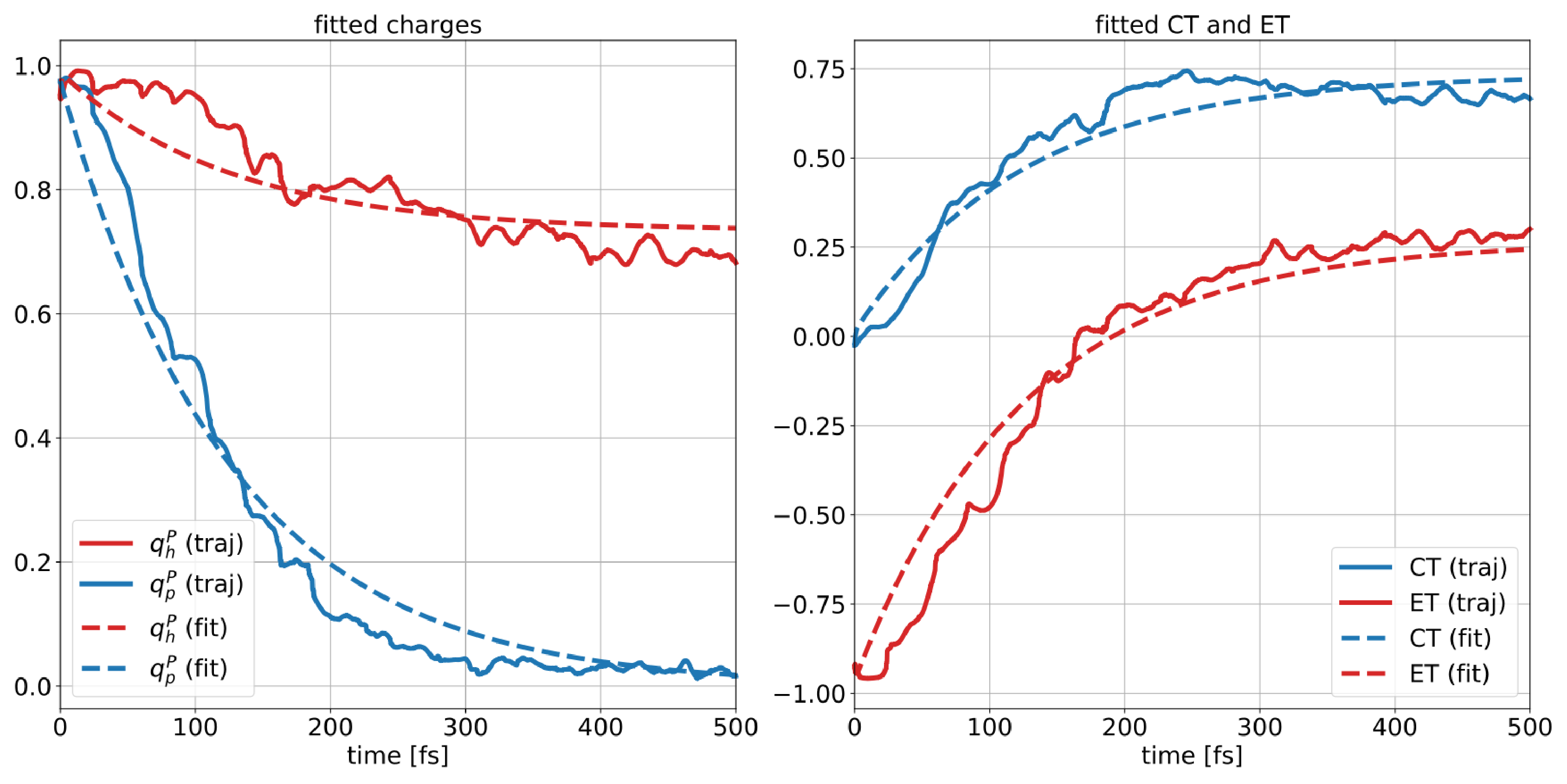} 
\caption{\label{fig:15a0}
Kinetics fit to the TD-lc-DFTB FSSH results with $R_{lc} = 15 \, a_0$: 
$k_{\text{\bf AB}} = \mbox{0.00800 fs$^{-1}$}$ ($\tau_{\text{\bf AB}} = \mbox{125 fs}$),
$k_{\text{\bf BA}} = \mbox{1.347 $\times$ 10$^{-6}$ fs$^{-1}$}$ 
($\tau_{\text{\bf BA}} = \mbox{472 000 fs}$),
$k_{\text{\bf BC}} = \mbox{0.971 fs$^{-1}$}$ ($\tau_{\text{\bf BC}} = \mbox{1.03 fs}$), and
$k_{\text{\bf CB}} = \mbox{0.353 fs$^{-1}$}$ ($\tau_{\text{\bf CB}} = \mbox{2.83 fs}$).
}
\end{figure}
\begin{figure}
\includegraphics[width=1.0\textwidth]{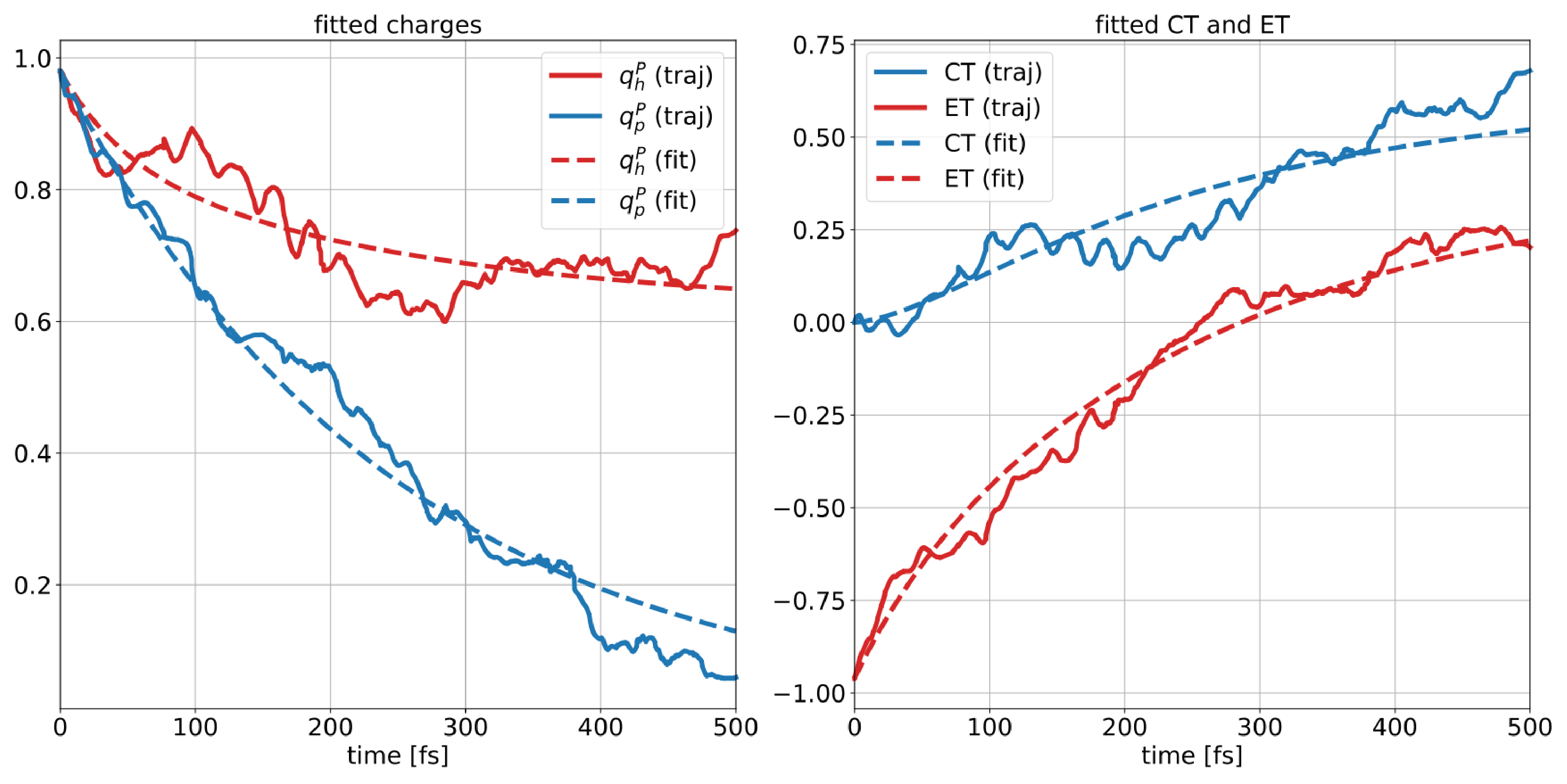} 
\caption{\label{fig:20a0}
Kinetics fit to the TD-lc-DFTB FSSH results with $R_{lc} = 20 \, a_0$: 
$k_{\text{\bf AB}} = \mbox{0.00405 fs$^{-1}$}$ ($\tau_{\text{\bf AB}} = \mbox{247 fs}$),
$k_{\text{\bf BA}} = \mbox{1. $\times$ 10$^{-10}$ fs$^{-1}$}$ 
($\tau_{\text{\bf BA}} = \mbox{1. $\times$ 10$^{+10}$ fs}$),
$k_{\text{\bf BC}} = \mbox{0.0133 fs$^{-1}$}$ ($\tau_{\text{\bf BC}} = \mbox{75.2 fs}$), and
$k_{\text{\bf CB}} = \mbox{0.0082 fs$^{-1}$}$ ($\tau_{\text{\bf CB}} = \mbox{122 fs}$).
}
\end{figure}
\begin{figure}
\includegraphics[width=1.0\textwidth]{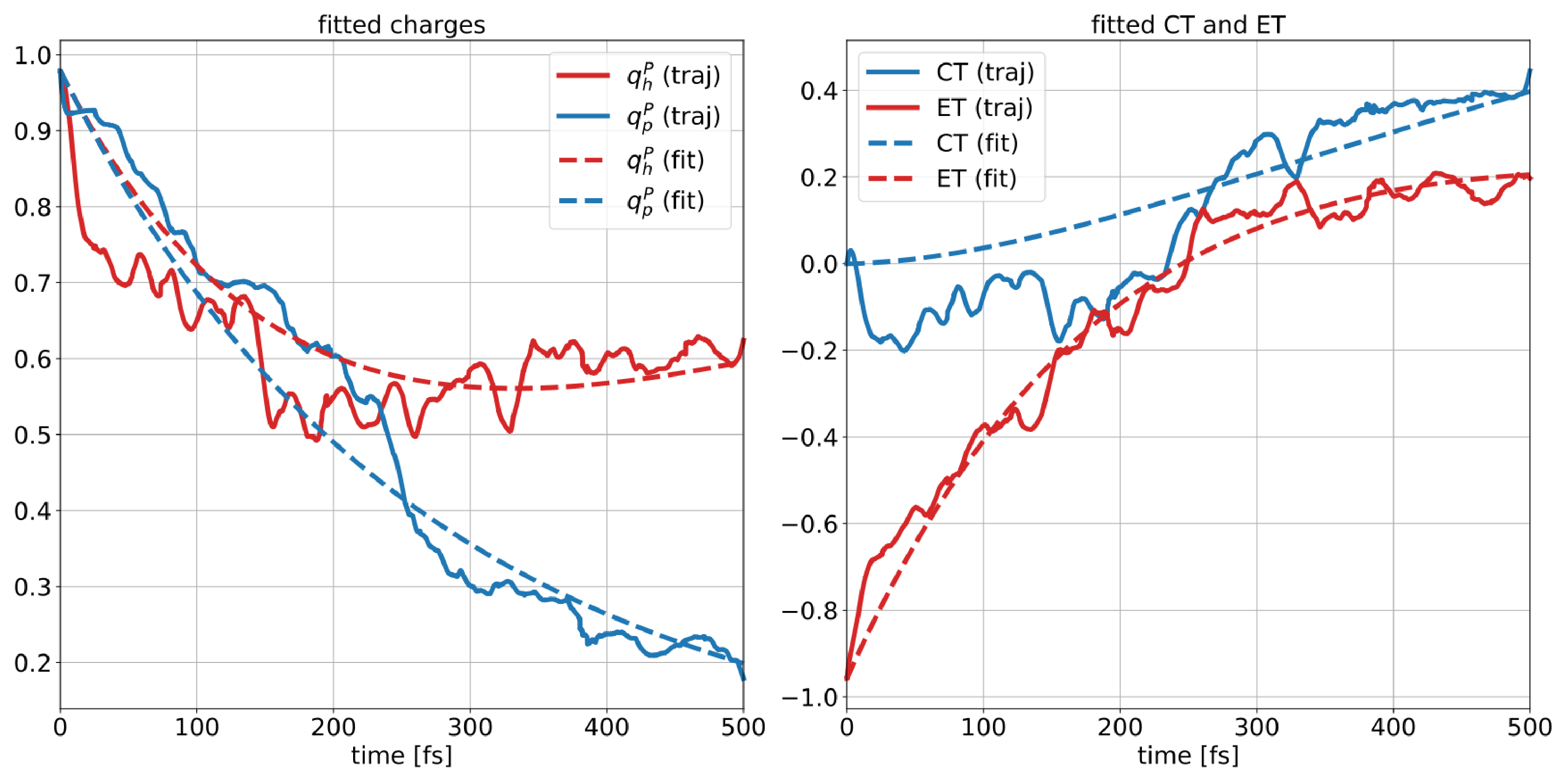} 
\caption{\label{fig:25a0}
Kinetics fit to the TD-lc-DFTB FSSH results with $R_{lc} = 20 \, a_0$: 
$k_{\text{\bf AB}} = \mbox{0.00364 fs$^{-1}$}$ ($\tau_{\text{\bf AB}} = \mbox{275 fs}$),
$k_{\text{\bf BA}} = \mbox{0.0000444 fs$^{-1}$}$ ($\tau_{\text{\bf BA}} = \mbox{22 500 fs}$), 
$k_{\text{\bf BC}} = \mbox{0.00222 fs$^{-1}$}$ ($\tau_{\text{\bf BC}} = \mbox{450 fs}$), and
$k_{\text{\bf CB}} = \mbox{1.0 $\times$ 10$^{-10}$ fs$^{-1}$}$, 
($\tau_{\text{\bf CB}} = \mbox{1. $\times$ 10$^{+10}$ fs}$). 
}
\end{figure}
\twocolumngrid

\end{document}